\documentclass[12pt,preprint2]{aastex}

\newcommand{\hii}{H\,{\sc ii}}
\newcommand{\sixiii}{Si\,{\sc xiii}}
\newcommand{\medE}{\langle E \rangle}
\begin{document}

\title{An X-ray Census of Young Stars in the Massive Southern Star-Forming Complex NGC 6357}

\author{Junfeng Wang,\altaffilmark{1} Leisa K.\ Townsley,\altaffilmark{1}
Eric D.\ Feigelson,\altaffilmark{1} Konstantin V.\ Getman,\altaffilmark{1}
Patrick S.\ Broos,\altaffilmark{1} Gordon P.\ Garmire,\altaffilmark{1} and
Masahiro Tsujimoto \altaffilmark{2}}

\altaffiltext{1}{Department of Astronomy \& Astrophysics, The
Pennsylvania State University, 525 Davey Lab, University Park, PA
16802; {\sf jwang@astro.psu.edu, townsley@astro.psu.edu, edf@astro.psu.edu}}

\altaffiltext{2}{Department of Physics, Rikkyo University, 3-34-1, Nishi-Ikebukuro, Toshima, Tokyo, 171-8501, Japan}

\begin{abstract}

We present the first high spatial resolution X-ray study of the
massive star forming region NGC 6357, obtained in a 38 ks {\it
Chandra}/ACIS observation. Inside the brightest constituent of this
large HII region complex is the massive open cluster Pismis 24. It
contains two of the brightest and bluest stars known, yet remains
poorly studied; only a handful of optically bright stellar members
have been identified. We investigate the cluster extent and Initial
Mass Function and detect $\sim$800 X-ray sources with a limiting 
sensitivity of $\sim 10^{30}$ ergs s$^{-1}$;
this provides the first reliable probe of the rich intermediate-mass
and low-mass population of this massive cluster, increasing the number
of known members from optical study by a factor of $\sim 50$. The high
luminosity end ($\log L_h$[2-8 keV]$\ge 30.3$ ergs s$^{-1}$) of the observed 
X-ray luminosity function in NGC 6357 is clearly consistent with a
power law relation as seen in the Orion Nebula Cluster and Cepheus B,
yielding the first estimate of NGC 6357's total cluster population, a
few times the known Orion population.  We investigate the structure of the
cluster, finding small-scale substructures superposed on a spherical
cluster with 6 pc extent, and discuss its relationship to the nebular
morphology. The long-standing $L_X - 10^{-7}~L_{bol}$ correlation for
O stars is confirmed.  Twenty-four candidate O stars and one
possible new obscured massive YSO or Wolf-Rayet star are presented.
Many cluster members are estimated to be intermediate-mass stars from
available infrared photometry (assuming an age of $\sim 1$ Myr), but
only a few exhibit $K$-band excess.  We report the first detection of
X-ray emission from an Evaporating Gaseous Globule at the tip of a
molecular pillar; this source is likely a B0-B2 protostar.

\end{abstract}

\keywords{ISM: individual (NGC 6357) - open clusters and
associations: individual (Pismis 24) - stars: formation -
stars: mass function - stars: pre-main sequence - X-Rays: stars}

\section{Introduction}

\object{NGC 6357} (= W 22 = RCW 131 = S 11) is a large \hii\ region complex
in the southern sky excited by the OB association Pismis 24, at a
distance of $\sim 2$ kpc \citep{Neckel78,Felli90,Massi97,Bohigas04}.
This young stellar cluster and its surrounding gaseous environment are
relatively poorly studied; only a handful of stellar members have been
identified and characterized, and the molecular cloud has not been
well-mapped.  In optical and near-infrared images, the exciting
cluster is not prominent against the Galactic field population except
for a few bright members (Figure \ref{fig:dss2mass}).  Yet, two of
these stars are remarkably luminous and massive with spectral types O3
If (M$\sim 200-300$ M$_\odot$) and O3 III \citep[M$\sim 100$
M$_\odot$;][]{Massey01, Walborn02}.  Such massive stars are usually
found in only the richest Galactic clusters and the giant starburst
regions such as Carina, 30 Doradus, and NGC 3603.  This raises the
question is Pismis 24 much richer than it appears, or is it a poor cluster with a top-heavy Initial Mass Function (IMF)?
This would be highly unusual, as top-heavy IMFs are seen only in
extreme starburst environments such as the super-star clusters of M 82
\citep{Elmegreen05}.

We investigate this question by measuring, for the first time, the low-mass
pre-main sequence (pre-MS) population of the Pismis 24 cluster and its
environs.  This is achieved with a sensitive X-ray observation with
the {\it Chandra} X-ray Observatory.  {\it Chandra} observations can
penetrate heavy absorption up to $A_V \simeq 500$ \citep{Grosso05} and
resolve point sources with sub-arcsecond resolution.  In most cases, the
X-ray detected stars appear on existing near-infrared (NIR) images;
the advantage is that the strong X-ray emission of pre-MS stars
\citep[elevated $10^1-10^4$ above MS stars][]{Preibisch05a} traces
magnetic activity rather than photospheric or circumstellar disk
emission, therefore effectively discriminating cluster members from
older Galactic field stars.  We estimate that the cluster has $\simeq
10,000$ stars, $\simeq 5$ times richer than the known Orion Nebula Cluster.
The presence of two O3 stars is compatible with a cluster of this
population.  The known massive stars and the large number of low mass
X-ray stars suggest that Pismis 24 is a very rich cluster with a
standard IMF.

Our study also gives considerable new information about the cluster
and its X-ray properties.  We provide a census of $\sim$750 cluster
members with sub-arcsecond positions, 613 of which have NIR
counterparts.  Although not complete, the survey includes pre-MS stars
with masses extending down to $\sim 0.3 M_{\odot}$; most do not have
circumstellar disks revealed by $K$-band excess.  We study the
spatial distribution of the cluster members and identify a possible
subcluster. We examine the ringlike morphology of the NGC 6357 nebula
and investigate the possible feedback from the massive stars to their
environs. We measure X-ray properties of a dozen known O stars, and
list over 20 new candidate O stars.  A faint X-ray source is found
associated with a protostar in an Evaporating Gaseous Globule (EGG)
for the first time.  A few pre-MS stars are seen with very powerful
X-ray flares with peak $L_x \sim 10^{32}$ ergs s$^{-1}$.

Section \ref{overview.sec} reviews past study of the Pismis 24 cluster
and NGC 6357 region.  The {\it Chandra} observation and its analysis
are described in \S \ref{obs.sec}, and a NIR observation using the
Simultaneous 3 color InfraRed Imager for Unbiased Surveys (SIRIUS)
camera on the Infrared Survey Facility (IRSF) telescope is summarized
in \S \ref{sirius.sec}. The association between X-ray sources and
optical and near-infrared (ONIR) stars, as well as the infrared colors of X-ray
counterparts are presented in \S \ref{counterparts.sec}.  The cluster
population inferred from the X-ray emitting stars, their X-ray
luminosity function (XLF), the spatial distribution of the cluster
members, and the morphology of the nebula are discussed in \S
\ref{XLF.sec}.  X-ray properties of O stars and intermediate- to
low-mass pre-MS stars are presented in \S \ref{ximf.sec}. The possibility
of diffuse X-ray emission is examined in \S \ref{diffuse.sec}. A brief
summary of our findings is given in \S \ref{summary.sec}. This study
is part of a series of X-ray investigations of high-mass star
formation regions that includes the Rosette Nebula (Townsley et al.\
2003; Wang et al., in preparation), M 17 (Townsley et al.\ 2003; Broos
et al., in preparation), the Orion Nebula (Feigelson et al.\ 2003; Getman
et al.\ 2005 and associated articles), 30 Doradus (Townsley et al.\
2006a, 2006b), and Cepheus B/OB3b (Getman et al.\ 2006).

\section{Observational Overview of NGC 6357 and Pismis 24
\label{overview.sec}}

NGC 6357, a large \hii\ region complex showing an annular morphology
in the radio and optical \citep{Haynes78,Lortet84}, is located in the
Sagittarius spiral arm and spans $60^{\prime}\times 40^{\prime}$ in
the southern sky.  NGC 6334, another prominent star-forming complex
likely associated with NGC 6357, lies $\sim 60^{\prime}$ away
\citep{Neckel78,Ezoe06}.  The massive open cluster Pismis 24 (= C
1722-343 = OCl 1016) lies in NGC 6357's central cavity
\citep{Pismis59, Alter70}.  The kinematic distance to NGC 6357,
$d=1.0\pm 2.3$ kpc, is given by \citet{Wilson70}. \citet{Neckel78}
derives $d=1.74\pm 0.31$ kpc with improved accuracy based on $UBV$ and
H$\beta$ data. Spectroscopic parallax of 10 high-mass members in
Pismis 24 gives a larger distance of $2.56\pm 0.10$ kpc
\citep{Massey01}, which we adopt in this paper.  We should note that
recent work \citep{Kharchenko05,Prisinzano05,Arias06} has found
smaller distances to other high mass star forming regions (e.g., M8)
than previous studies; we will discuss the uncertainties in
distance-related stellar properties wherever applicable.

\begin{figure}
 \centering
 
 \includegraphics[width=0.8\textwidth]{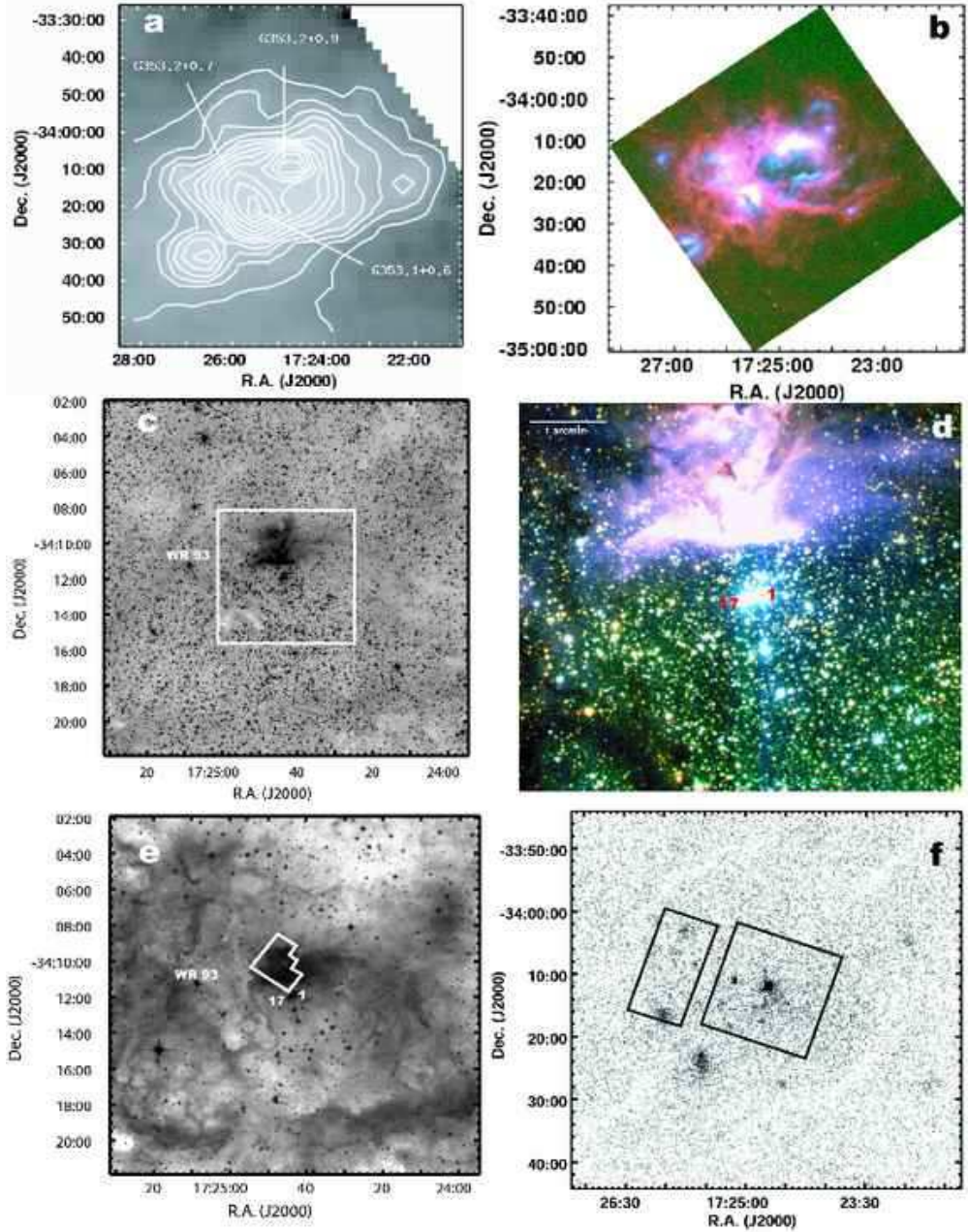}
 \caption{\small Multiwavelength images of the NGC 6357 region from
 long to short wavelengths. (a) $1.5^\circ \times 1.5^\circ$ view of 6
 cm radio continuum contours from Parkes survey \citep{Haynes78} with
 the three \hii\ regions labeled.  (b) The same region shown as a
 3-color composite mid-infrared image from the MSX survey: red is 8.28
 $\mu$m, green is 12.13 $\mu$m, and blue is 21.3 $\mu$m.  (c)
 $20\arcmin \times 20\arcmin$ K-band image from the 2MASS survey
 showing the \hii\ region G353.2+0.9, dark clouds, and the stellar
 field.  The box shows the SIRIUS FOV.  (d) $\sim 6\arcmin
 \times 6\arcmin$ $JHK$ composite image from the SIRIUS
 observation. $J$ is shown in blue, $H$ is green, and $K$ is red. (e)
 $20\arcmin \times 20\arcmin$ R-band image from the Digitized Sky
 Survey emphasizing H$\alpha$ emission. The box shows HST/WFPC2
 coverage. The brightest members WR 93, Pis 24-1, and Pis 24-17 are
 marked. (f) An unpublished X-ray image from a short ROSAT PSPC
 observation. The three brightest regions are associated with the
 NGC~6357 \hii\ regions. The ACIS-I FOV is shown as a box, with the off-axis S2+S3
 chips represented as a rectangle to the east. Note that S3 captures
 G353.2+0.7 at its southern edge.
 \label{fig:dss2mass}}
\end{figure}

Figure~\ref{fig:dss2mass}b shows the large-scale morphology in the
mid-infrared, observed with the Midcourse Space Experiment (MSX)
satellite. A large ring-like structure extends $\sim 60^{\prime}$ in
diameter and bright nebulosity is seen at the northern rim. The
central cavity appears free of large amounts of warm
dust. Far-infrared (FIR) continuum revealed several luminous (L$\sim
10^5$ L$_\odot$) embedded sources coincident with CO line and radio
continuum emission peaks \citep{McBreen83}. G353.2+0.9, which we study
here, is the youngest ($t\sim 10^5$ yr) and brightest of NGC 6357's
three \hii\ regions. It lies at the northern interface between the
cluster and molecular cloud and contains a number of embedded sources
\citep[Figure~\ref{fig:dss2mass}a;][]{Frogel74,Sakellis84,Persi86,
Massi97,Bohigas04}. In contrast with NGC 6334, almost no water maser emission
is found in NGC 6357 \citep{Healy04a}.  This may either indicate that
massive star formation has ceased in the region \citep{Persi86} or
that it has been overrun by the ionization front \citep{Healy04a}.
G353.1+0.6 is a more evolved \hii\ region viewed edge-on, bounded on
the northern side by a different molecular cloud. It contains a single
O5 star, several B stars and embedded IR sources, but no indication of
current star formation can be found
\citep{Felli90,Massi97}. G353.2+0.7, which is also a FIR peak, does
not contain early type stars \citep{Felli90,Massi97}, but it hosts the
only water maser found in NGC 6357 \citep{Sakellis84}. However,
\citet{Persi86} identifies this water maser source as an evolved late
M star.

Despite a number of optical, infrared, and radio studies on NGC 6357,
no X-ray observation of this field has been reported.
Figure~\ref{fig:dss2mass}f shows an unpublished 9 ks {\em ROSAT} PSPC
observation revealing a few strong X-ray sources, including NGC 6357's
three \hii\ regions.  The brightest X-ray emission coincides with the
core of the Pismis 24 cluster, but the {\em ROSAT} instrumentation was
unable to resolve the source population.

The OB cluster Pismis 24 is centered $\sim 1^{\prime}$ south of the
G353.2+0.9 ionization front.  Early optical study of NGC 6357 and
Pismis 24 revealed $\sim 20$ O-type and early B-type stars
\citep{Moffat73,Neckel78,Neckel84,Lortet84}, including the unusally
late-type WC7+O7-9 Wolf-Rayet (WR)/O star binary HD 157504 \citep[= WR
93;][]{Hucht01}. Two of the cluster members, namely Pismis 24-1 (=HDE
319718) and Pismis 24-17, were recently classified as spectral type
O3.5, some of the brightest and bluest stars known \citep{Massey01,Walborn02}.
Only a dozen other O3 stars are known in the Galaxy
\citep{Massey93,Walborn94,Maiz-Apellaniz04}. The two luminous O3 stars
are easily identified close to the sharp boundary of the bright
optical nebula in the optical image, accompanied by several other
bright cluster members extending $\sim 4^{\prime}$ north-south
(Figure~\ref{fig:dss2mass}e), but are not as apparent in the Two
Micron All Sky Survey (2MASS) K-band image
(Figure~\ref{fig:dss2mass}c). Figure~\ref{fig:dss2mass}d shows a
SIRIUS $JHK$ composite image, which has higher resolution and
penetrates deeper into the cloud than 2MASS. The two luminous O stars
are evident in the center of the SIRIUS image, which contains
thousands of stars. Other noticeable features are the bright nebula
with tenuous diffuse filaments in the north and the infrared dark
column occupying the south-east corner of the image, previously noted as
the south eastern complex (S.E.C.) in CO mapping
\citep{Massi97}. Several heavily reddened stars can be seen in the
S.E.C. region. However, most NIR stars are foreground and background
objects, not cluster members.

Although most researchers believe G353.2+0.9 is excited by the hot
stars in Pismis 24 \citep[e.g.,][]{Lortet84}, \citet{Felli90} argued
that Pismis 24 is projected onto the \hii\ region by chance, and the
embedded massive stars \citep[IRS 1, 2, and 4;][]{Persi86} supply the
ionizing radiation.  Recent studies do not support this claim.
\citet{Massi97} have shown that most molecular gas lies behind the
\hii\ region and the cluster location is consistent with a face-on
blister-type \hii\ region morphology. Optical spectroscopic study by
\citet{Bohigas04} indicates that the hot O stars of Pismis 24 emit
$10^{50}$ UV photons s$^{-1}$, sufficient to ionize the \hii\ region.

\citet{Massey01} consider the W-R/O binary WR 93, lying 4$^\prime$ away
from the cluster core, to be a likely member of Pismis 24 and caution
that the cluster might be larger than the central concentration of O
stars.  Spectroscopic study indicates a mass loss rate for WR 93 of
$\dot{M} = 2.5 \times 10 ^{-5}$ M$_{\odot}$ yr$^{-1}$ using a clumped
wind model with terminal wind velocity $v_{\infty}=2290$ km s$^{-1}$
\citep{Prinja90,Nugis98,Hucht01}. The nearby radio continuum and CO peaks
suggest that WR 93 ionizes and heats the surrounding cloud
\citep{McBreen83}.  The visual extinction towards WR 93 is high, with
$E(B-V)=1.82$ or $A_V \simeq 5.8$ mag \citep{Nugis98,Hucht01}.

\citet{Healy04b} and \citet{Hester05} present an HST/WFPC2 image of
the G353.2+0.9 \hii\ region in NGC 6357, illustrating a sequence
of formation of low-mass stars and describing their evolution from molecular cores to
EGGs to proplyds in the vicinity of massive stars. Very recently,
\citet{DeMarco06} analyze HST observations of eight HII regions,
including Pismis 24, to look for protoplanetary disks. In Pismis 24,
they find a jet-like feature, similar to a Herbig-Haro object in their
source \# 14. However, no central star is seen in the jet candidate.

\section{Observations and Data Reduction \label{obs.sec}}

\subsection{{\it Chandra} Observation and Data Selection}

NGC 6357 was observed on July 9, 2004 with the Imaging Array of the
Advanced CCD Imaging Spectrometer (ACIS-I) on board {\it Chandra}.
Detailed description of the instrument can be found in
\citet{Weisskopf02}. Four front illuminated (FI) CCDs form the ACIS-I
which covers a field-of-view (FOV) of $\sim 17^{\prime}\times
17^{\prime}$. Two CCD chips on the ACIS Spectroscopic Array (ACIS-S)
were also set to be functioning during the observation, although the
mirror point spread function (PSF) degrades significantly far
($>15^{\prime}$) off-axis.  Here the ACIS-S data will not be
quantitatively discussed.  The observation was made in the standard
Timed Exposure, Very Faint mode, with 3.2-second integration time and
5 pixel $\times$ 5 pixel event islands. The total exposure time is 38
ks and the satellite roll angle is 289 degrees. The aim point is
centered on the O3 If star Pis 24-1, the heart of OB association
Pismis 24. The observation ID is \dataset[ADS/Sa.CXO#obs/4477]{4477}.

Appendix B of \citet{Townsley03} and \citet{Getman05a} provide detailed
explanations of the Penn State ACIS data reduction procedure. Data
reduction starts with filtering the Level 1 event list processed by
the {\it Chandra} X-ray Center pipeline to recover an improved Level 2
event list.  To improve absolute astrometry, X-ray positions of ACIS-I
sources were obtained by running the {\it wavdetect} wavelet-based
source detection algorithm \citep{Freeman02} within the {\it Chandra}
Interactive Analysis of Observations (CIAO) package on the original
Level 2 event list, using only the central $8^{\prime}\times
8^{\prime}$ of the field. The resulting X-ray sources were matched to
the 2MASS point source catalog. We calculated the position offsets
between 277 X-ray sources and their NIR counterparts, and applied an
offset of $+0^{\prime \prime}.02$ in right ascension (RA) and
$-0^{\prime \prime}.33$ in declination (Dec) to the X-ray
coordinates. The CCD charge transfer inefficiency (CTI) was corrected
using the Penn State CTI corrector (version 1.16) developed by
\citet{Townsley00,Townsley02}.

\subsection{Image Reconstruction and Source Finding}

A multifaceted source finding procedure is summarized here, which is
designed to locate all potential sources, even in the presence of
crowding or a broad PSF, with the recognition that features due to
detector noise may also enter the source list. First, we assemble a
large number of candidate sources using a variety of techniques and
criteria (wavdetect, reconstruction, visual selection, etc., as
described below), which includes a number of possible false detections. These
spurious sources are identified and removed after the photon
extraction step (\S 3.3), where our customized tool evaluates the
statistical significance of each detection above background, rejecting
likely false detections when significances fail to meet a chosen threshold.
These photons are then returned to the background and we re-evaluate the source
significance for the surviving candidates, finally defining reliable
detections and marginal detections for inclusion in tables of source properties.  The details of this procedure are described below.

\begin{figure}
 \centering
\includegraphics[width=1.0\textwidth]{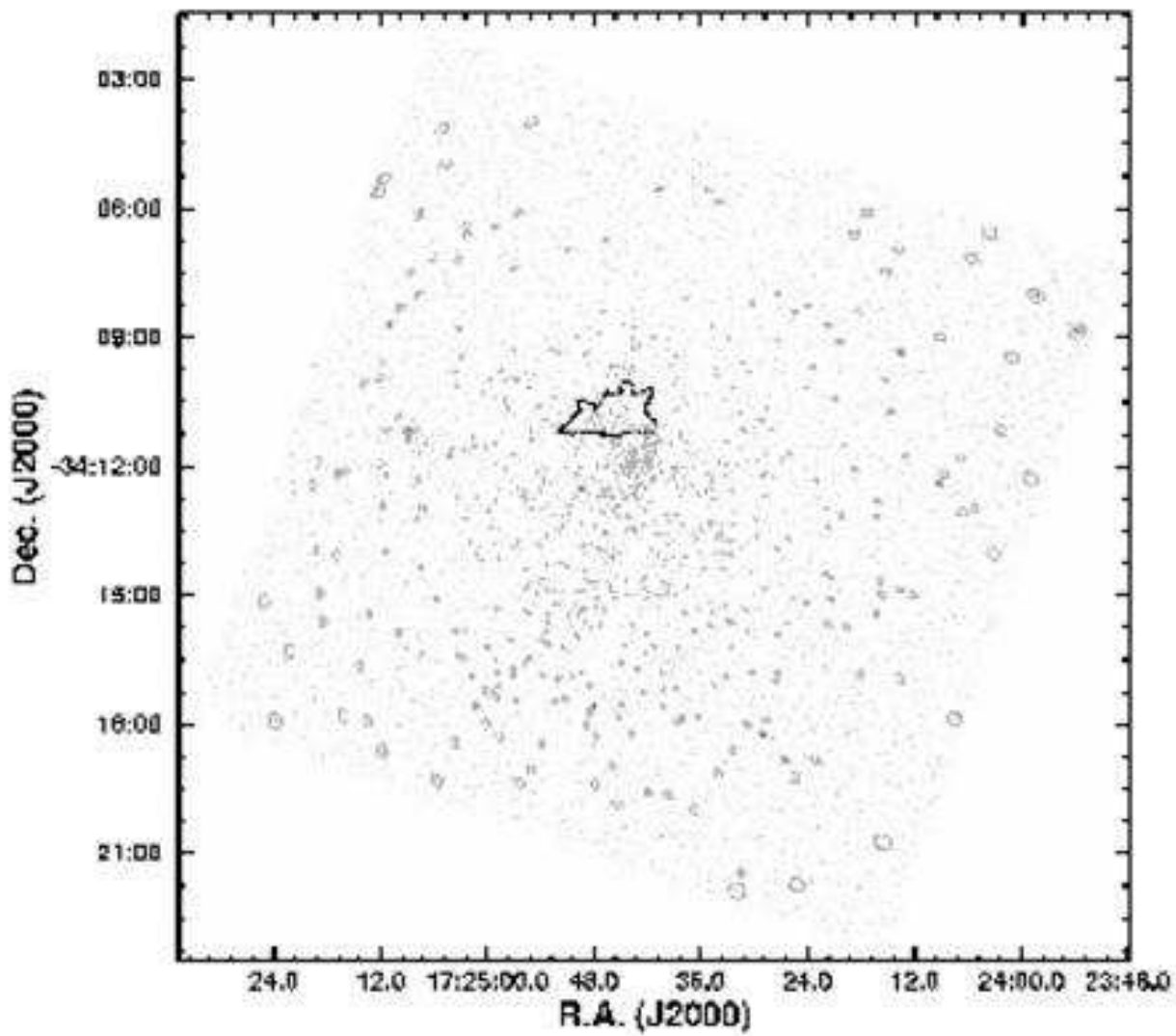}
\caption{(a) The $17^{\prime}\times 17^{\prime}$ ACIS-I image (0.5--7.0
keV) binned by $2^{\prime\prime}\times 2^{\prime\prime}$. The grey
polygons show the source extraction regions that are matched to the
local {\it Chandra} point spread function. The black contour in the
center outlines the ionization front seen in the optical image. (b)
The ACIS full-field adaptively smoothed image showing all four I-chips
and two S-chips. Red represents the soft band (0.5--2.0 keV) X-ray
emission and blue represents the hard band (2.0--7.0 keV) X-ray
emission. The cyan and white polygons outline the extraction regions
for possible diffuse emission in the central region, and in a larger
area, respectively. The yellow polygon outlines the region used for
diffuse background estimation. See \S \ref{diffuse.sec}. (c)
Zooming-in on the inner $\sim 4^{\prime}\times 4^{\prime}$ of the
ACIS-I data, binned by $0.25^{\prime\prime}\times
0.25^{\prime\prime}$, then adaptively smoothed.
\label{fig:ACIS_I}}
\end{figure}

Source detection was carried out with the improved Level 2 event list
that is band-limited to 0.5--7.0 keV with cosmic ray afterglows
removed. Effective exposure maps and images for the full field, the
$17^{\prime} \times 17^{\prime}$ I array, the central $8^{\prime}$ of the I
array, and the inner $4^{\prime}$ of the field were made with 4-pixel, 2.6-pixel, 1-pixel, and
half-pixel binning, respectively (an ACIS sky pixel is $0.5^{\prime\prime}$ on a side).  
Figure~\ref{fig:ACIS_I}a shows the
raw ACIS I-array image at reduced resolution.  Adaptive-kernel
smoothed flux images were created with the CIAO tool {\it csmooth}
\citep{Ebeling06} to help identify additional potential faint sources and any possible diffuse emission.  The resulting full field two-color smoothed image and a zoom-in on the
cluster core are shown in Figure~\ref{fig:ACIS_I}; red and blue
represent the soft band emission (0.5--2.0 keV) and hard band emission
(2.0--7.0 keV), respectively.  

The {\it wavdetect} program was run
with source significance threshold $1 \times 10^{-5}$ on each of the four binned images
described above, in three energy bands (0.5--2.0 keV, 2.0--7.0 keV, and
0.5--7.0 keV).  This source significance threshold may give some false detections, but experience has shown us that this threshold is appropriate for crowded fields that might contain complex backgrounds due to diffuse emission.  Merging the detections from the 12 runs resulted in a list
of 614 potential point sources on ACIS-I. Careful visual inspection
revealed that {\it wavdetect} still missed several apparent sources in the
central region of the cluster.  To take advantage of the sub-arcsecond
PSF at positions around the aimpoint, we applied a subpixel
positioning code \citep{Mori01} to improve spatial resolution in the
inner part of the field and performed an image reconstruction with the
Lucy-Richardson maximum likelihood algorithm \citep{Lucy74}.

\begin{figure}
 \centering
\includegraphics[width=1.0\textwidth]{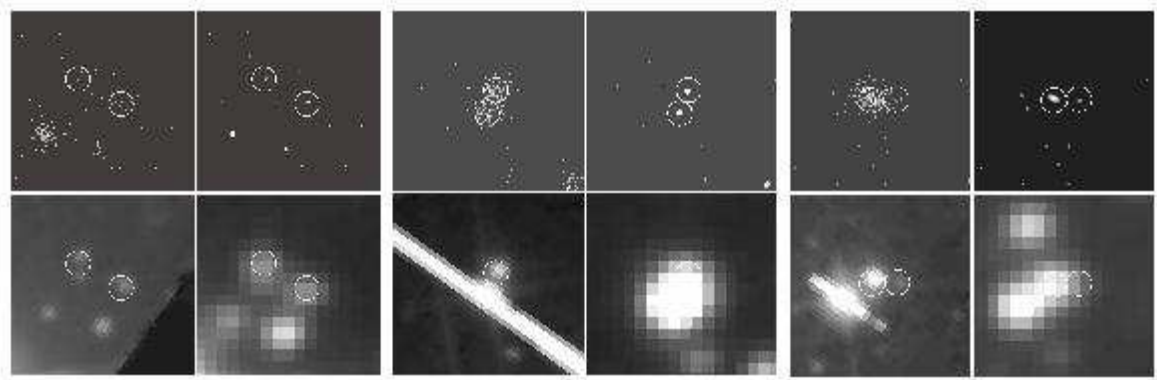}
 \caption{\small Examples of close X-ray source pairs (ACIS \#313,
 317; \#402, 406; \#427, 428; shown as dashed circles) recovered from
 image reconstruction. Each panel includes: $6^{\prime\prime}\times
 6^{\prime\prime}$ raw ACIS image binned by $0.25^{\prime\prime}$ (top left),
 reconstructed image (top right), HST image (bottom left), and SIRIUS
 image (bottom right). The white stripe in the HST image is the
 diffraction spike from a nearby bright star.  \normalsize
 \label{fig:recon}}
\end{figure}

Our image reconstruction technique for crowded stellar clusters is
described in detail in \citet{Townsley06} and a similar procedure is
followed here. The maximum likelihood image reconstruction code
\footnote{From the IDL Astronomy User's Library maintained by Wayne
Landsman at \url{http://idlastro.gsfc.nasa.gov/homepage.html}.} was
run in the crowded central region using a $\sim 50^{\prime\prime}
\times 50^{\prime\prime}$ image made with 0.2-pixel binning and centered on Pis
24-1, using the 1.0 keV PSF at that on-axis
location from the PSF library in the {\it CIAO} calibration
database. We chose the 200th iteration of the algorithm as the most
appropriate image for source searching, as it showed good PSF removal without over-resolving the data. We adopted the {\it find}
procedure in the {\it DAOPHOT} package \citep{Stetson87} to obtain the
centroid, shape, and brightness parameters of the resolved
peaks. Several combinations of flux ($f$), roundness ($r$), and
sharpness ($s$) limits were tested to minimize the number of peaks
from one photon, and the adopted parameters to reject spurious peaks
were: $r>1.0$ or $r<-1.0$; or $f < 0.8$; or $f < 1.6$ and $s >0.98$.
Using SAOImage DS9 \citep{Joye03}, world coordinate system (WCS)
region files were made, centered on the unrejected peaks, based on
90$\%$ PSF contours. The final reconstructed source list was made with
the aid of the I-array image overlaid with these regions where
any remaining reconstructed peaks that contain only single photon events within
90$\%$ PSF contours were rejected.  This image reconstruction
procedure was performed separately using another image centered on Pis
24-17 because {\it Chandra}'s PSF varies at different locations and an
appropriate local PSF must be used for the reconstruction. Examples of
valid sources from image reconstruction are shown in
Figure~\ref{fig:recon}. The list of potential sources from the maximum
likelihood reconstruction was merged with the {\it wavdetect} source
catalog. By overlaying regions representing the potential sources onto
the original and smoothed ACIS-I images, an additional 40 sources were
added based on spatial concentrations of $\ge 3$ photons and proximity
of 2MASS near-IR sources ($\le 1^{\prime\prime}$).  The source finding
procedure described here results in a total of 910 potential sources
identified on the ACIS-I array.

\subsection{Photon Extraction and Limiting Sensitivity}

A preliminary event extraction for the 910 potential X-ray sources was
made with {\it ACIS Extract} (version 3.79; hereafter {\it AE}), a
versatile script written in IDL that performs source extraction, fits
X-ray spectra, creates light curves, and computes a wide variety of
statistical properties of the sources
\citep{Broos02}\footnote{\url{http://www.astro.psu.edu/xray/docs/TARA/ae\_users\_guide.html}}.
Using the {\it AE}-calculated probability $P_B$ that the extracted
events are solely due to Poisson fluctuations in the local background,
source validity can be statistically evaluated while taking account of
the large distorted PSF at far off-axis locations and spatial
variations in the background.  The traditional source significance, defined as
the photometric signal to noise ratio, is calculated for every source
as well. We rejected sources with $P_B > 1\%$ likelihood of being a
background fluctuation.  The trimmed source list includes 779 sources,
with full-band (0.5--8.0 keV) net (background-subtracted) counts
ranging from 1.7 to 1837 counts. We performed source extraction with
this list using {\it AE} as described in detail by \citet{Townsley06}.

The 779 valid sources are purposely divided into two lists: the 665
sources with $P_B<0.1\%$ make up our primary source list of highly
reliable sources (Table~\ref{tbl:primary}), and the remaining 114
sources with $P_B \ge 0.1\%$ likelihood of being spurious background
fluctuations\footnote{A few $P_B$ values are greater than 0.01, our
rejection threshold in the preliminary run. This is due to the
feedback to the background from the rejected sources when we
re-extract the trimmed source list. As described in detail in \S 2.1
of \citet{Townsley06}, our catalog cutting scheme is iterative: the
initial set of 910 potential sources is extracted and then cut at the
threshold of 1\%. The rejected sources are allowed to contribute to
the background and the trimmed list of 779 valid sources is then
re-extracted, resulting in different local background estimates and
different source validity metrics $P_B$. Thus, some $P_B$ values in
Table 2 are $>1$\%.} are listed as tentative sources in
Table~\ref{tbl:tentative}. We believe that most of those tentative
sources are likely real detections\footnote{The fraction of tentative
sources (Table 2) that have optical and near-infrared counterparts is
50\%, suggesting that they are not random spurious detections (\S 4.1). For
the 64 weakest X-ray sources that have 3 counts or less, 10 of them
have been cautioned with a "U" flag (unreliable sources that have large
probability of being background, \S 4.5) in the table, while 27 of them have optical and
near-infrared identifications. Most of these faint sources are
clustered within the central $2^{\prime}$ of the field and are likely low-mass cluster
members.}.

We should note that the separation of reliable sources, tentative
sources, and invalid sources is based on a single simple statistic,
the Poisson probability of observing the extracted counts given the
local background density.  This permits scientists to evaluate the
reliability of each source individually.  A high cutoff (around 4
counts for a typical {\it Chandra} exposure on-axis) reduces spurious
noise-based sources ({\it false positives}) but misses some real
sources ({\it false negatives}).  A low cutoff reduces the false
negatives while increasing the false positives. In the absence of a
consensus statistical procedure for the Poisson
weak-signal-in-background problem, there is no way to clean the list
in an objective fashion without losing real sources.  Since different
scientists make different decisions on the balance between false
positives and false negatives, and there are good scientific reasons
to believe many astronomically interesting faint sources lie in {\it Chandra}
images of rich stellar clusters, we have opted to report marginal
sources in Table~2.  The reader is clearly warned that the reality of
some of these weakest sources is suspect, and each source's $P_B$ can
be examined individually.  These choices concerning how we choose to
report faint sources do not affect our science results.  The weakest
X-ray sources are only used for X-ray luminosity function estimation
(\S 5.1), contributing to the faintest two bins of the XLF, which do not
impact on our conclusions regarding the NGC 6357 stellar population.  We further
note that {\it every} detection scheme must -- explicitly or
implicitly -- estimate the local background, and every such estimate
must implicitly identify the data that are guessed not to be
background.  For crowded young stellar clusters, this can never be
accomplished in a unique, objective manner as it depends on the unknown
distribution of true faint sources (including thousands of stars which
lie below any detection threshold) and possibly diffuse emission from
O star winds \citep{Townsley03}.

Table~\ref{tbl:primary} and Table~\ref{tbl:tentative} have a format
that closely follows the tables in \citet{Getman06} and are identical
to Tables 1 and 2 in \citet{Townsley06}.  Column 1 gives the ACIS
running sequence number, and column 2 provides the IAU
designation. Columns 3--6 show RA and Dec in degrees, positional error
in arcseconds, and off-axis angle in arcminutes. Column 7--11 give net
counts in the full band, the associated $1\sigma$-equivalent
uncertainty based on Poisson statistics, background counts in the
extraction region, net counts in the hard band, and the fraction of
the PSF used for source extraction (reduced values for this quantity indicate source crowding). Columns 12 and 13 present two
measures of the source validity: the photometric significance, and the
probability that the source is a spurious background fluctuation $P_B$,
used for establishing source validity as described above. Column 14 lists possible
anomalies in the observation and Column 15 checks the source
variability. Table footnotes give details for Column 13--15. Column 16
gives the ``effective'' exposure time, defined as the amount of
exposure time needed for the source to accumulate the current number
of counts if placed at the {\it Chandra} aimpoint.  Column 17 gives
the source's median energy in the full spectral band after background
subtraction.

The on-axis limiting sensitivity of this 38 ks observation can be
estimated using the Portable Interactive Multi-Mission Simulator
(PIMMS
\footnote{\url{http://heasarc.gsfc.nasa.gov/Tools/w3pimms.html}}). Assuming
an on-axis detection of 3 counts (0.5--8.0 keV) in 38170 seconds and
an absorbing column $N_H=1.0\times 10^{22}$ cm$^{-2}$ \citep[estimated
from an average $A_V \sim 6.0$ mag to Pismis 24 in Bohigas et~al.\ 2004,
adopting the conversion between $N_H$ and $A_V$ from][]{Vuong03}, the
observed X-ray flux for a thermal plasma with temperature of 1 keV
(appropriate for a pre-MS star's X-ray spectrum) is $5.6 \times
10^{-16}$ ergs s$^{-1}$ cm$^{-2}$. Correcting for absorption and a
distance of 2.56 kpc to Pismis 24, this gives a limiting luminosity of
$\log L_{t,c} = 30.2$ erg s$^{-1}$, where $L_{t,c}$ is the intrinsic
(absorption-corrected) luminosity in the total {\it Chandra} band
0.5--8.0 keV.

\subsection{Source Variability and Spectral Fitting}
\label{specfit.sec}

A band-limited, adaptively smoothed light curve for each source was
generated by {\it AE} together with a median energy time-series. The
variability of the source was evaluated by the significance of a
one-sided Kolmogorov-Smirnov statistic $P_{KS}$, comparing the source
events arrival times to that of a uniform light curve model. If
$P_{KS}>0.05$, the null hypothesis of a constant source was
accepted. If $P_{KS}<0.005$, the source was classified as
``variable.'' Otherwise, the source was ``possibly variable.''

For the 40 sources with $>$80 net counts, the extracted spectra were
fit to models of optically thin thermal plasmas using the {\it XSPEC}
package \citep{Arnaud96}, based on source spectra, background spectra,
ancillary response functions (ARFs) and redistribution matrix
functions (RMFs) constructed by the {\it AE} script. The data were
grouped in bins with $\ge 10$ counts and the $\chi^2$ statistic was
adopted to evaluate the goodness-of-fit. A single {\it apec} \citep{Smith01} thermal
plasma model with a single absorption component was considered.
Interactive fitting with alternative models (e.g., {\it vapec} thermal
plasma with variable abundances; two-temperature {\it apec} thermal
plasma model) was performed for a few strong sources that did not
achieve a satisfying fit with the automated process.
Table~\ref{tbl:apec} gives the best fit results using the thermal
plasma model; its format closely follows Table~3 in
\citet{Townsley06}. Uncertainties representing 90\% confidence
intervals are given; incomplete confidence levels imply the parameters
are not well constrained, or {\it XSPEC} may have encountered some
abnormality in the error calculations. For the 144 sources that have
net counts between 20 and 80, we fit ungrouped spectra with a single
temperature {\it apec} thermal plasma model using the likelihood ratio
test ($C$ statistic); the fitting results are presented in
Table~\ref{tbl:apec2}.

None of the NGC 6357 sources are bright enough to warrant correction
for photon pile-up.   The brightest source in the field is the W-R/O
binary WR 93 which has 1837 ACIS counts in the 38 ks exposure, or a
count rate of 0.15 counts per CCD frame.  As it lies $5^{\prime}.4$
off-axis where the PSF is somewhat degraded, the photon pile-up
effect will not be significant enough to distort the spectral
fitting, according to previous simulation work \citep{Townsley02}.

\subsection{SIRIUS NIR Observation}
\label{sirius.sec}

To identify infrared counterparts with images that are deeper and of
higher spatial resolution than 2MASS, $J$, $H$, and $K$ band images
were obtained on 30 September 2005 using the wide-field NIR SIRIUS
camera on the 1.4 m IRSF telescope at the South Africa Astronomical
Observatory.  Details of the instrument can be found in
\citet{Nagayama03}. The pixel scale of the detector is 0.45 arcsec
pixel$^{-1}$ with a $7^{\prime}.7 \times 7^{\prime}.7$
field-of-view. Three second exposures were made at each of 20 dither
positions for a total exposure of 60 seconds. A typical limiting
magnitude for such SIRIUS observations is $K_s$=15.7.  Unfortunately,
the observing conditions were non-photometric and the seeing was
1.1\arcsec\/ FWHM in the $K$ band. We nonetheless found these images
to be very valuable for identifying infrared counterparts (as described
below) because of the higher spatial resolution compared to 2MASS.

\section{Identification of the Chandra Sources}
\label{counterparts.sec}

\subsection{X-ray Sources with Stellar Counterparts}

Identifying stellar counterparts for X-ray sources in massive star
forming regions beyond 1 kpc is challenging.  In regions such as the
Chamaeleon I cloud ($0.2$ kpc) or Cepheus B region ($0.75$ kpc), the
cluster members are sufficiently bright and well-separated that the
2MASS catalog provides adequate identifiers for nearly all {\it
Chandra} stellar sources \citep{Feigelson04, Getman06}.  However, for
NGC 6357 at $2.5$ kpc, deeper and higher resolution observations such
as the SIRIUS images are needed for source identification.  In massive
star forming regions such as NGC 6357, optical and NIR imaging have
the additional challenge of detecting faint stars amidst highly
structured background emission from the \hii\ region or deeply
embedded in the molecular cloud.  Thus we cannot arrive at a
quantitative statement regarding the completeness of the search for
stellar counterparts of {\it Chandra} sources.  But we can say that,
with very few exceptions, e.g., infrared luminous active galactic
nuclei (AGNs), the counterparts which we do identify are indeed
cluster members.

For sources with off-axis angle $\theta \le 3^{\prime}$, the {\it
Chandra} sourcelist was matched to the 2MASS Catalog of Point Sources
with a 1$^{\prime\prime}$ matching radius. The search radius was
enlarged to 2$^{\prime\prime}$ for $\theta > 3^{\prime}$ in
consideration of the degraded off-axis {\it Chandra} PSF
\citep{Getman05a}.  We find that 445 of our X-ray sources ($58\%$)
have 2MASS counterparts. The matching results are presented in
Table~\ref{tbl:class}. Four of the ACIS sources (\#88, 517, 688, 774)
have two counterparts within $2^{\prime\prime}$ and we assign the star
with the smaller separation as the real counterpart.  Three pairs of
sources -- 252 and 257, 401 and 407, 404 and 406 -- share a single
2MASS counterpart within $1^{\prime\prime}$. Given the high density of
both X-ray sources and 2MASS objects around cluster center, we follow
the simple Monte Carlo technique from \citet{Alexander03} to evaluate
the number of chance superpositions between 2MASS sources and X-ray
detections. A random offset between $5^{\prime\prime}$ and
$10^{\prime\prime}$ is applied to the catalog to remove the true
associations, and then the same cross-identification radii are used to
match X-ray positions to 2MASS positions. This procedure is repeated
many times and the resulting false detection rate is $2-3$\%.

For the 564 X-ray sources covered by SIRIUS imaging, we visually
identified 476 infrared counterparts with separations $\le
1^{\prime\prime}$ from X-ray sources. Of these, 308 are known 2MASS
counterparts that have been matched to ACIS sources. Therefore the
combined matches from 2MASS and SIRIUS yield 613 NIR counterparts for
the 779 X-ray point sources.  Of the 166 X-ray sources without NIR
counterparts, 89 lie in the SIRIUS FOV.

\begin{figure}
 \centering
 \plottwo{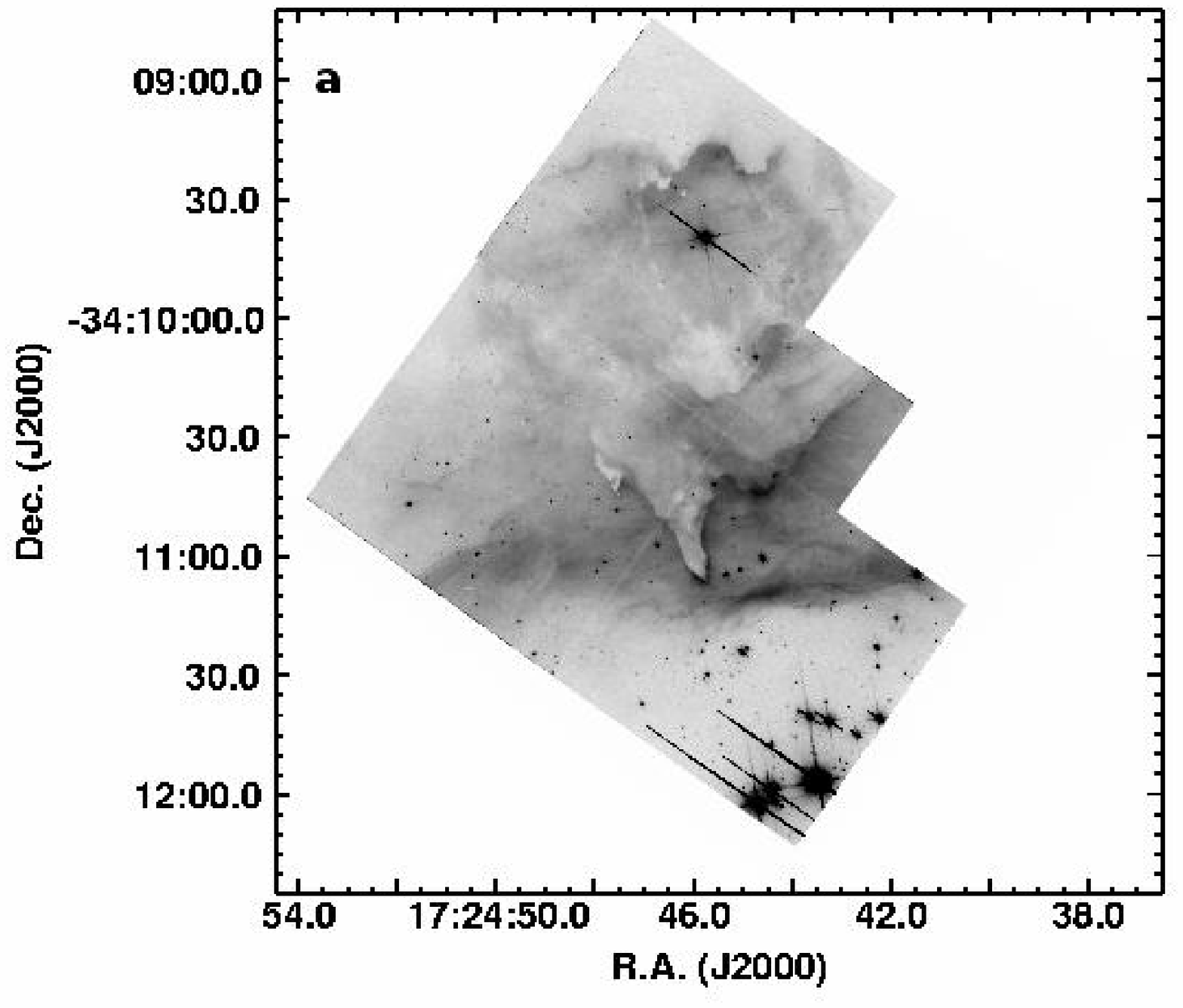}{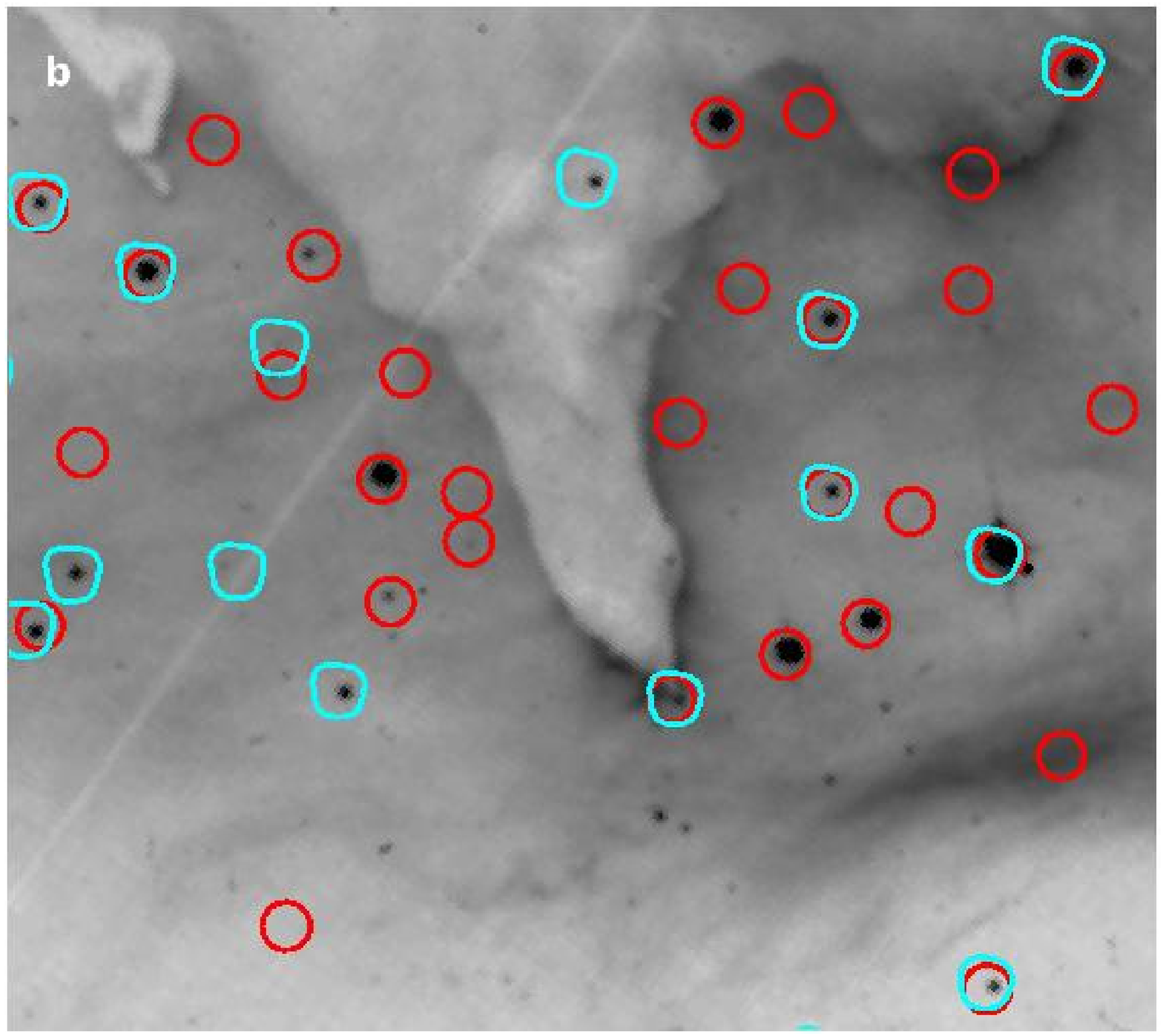}
 \caption{(a) HST/WFPC2 image (F814W) revealing
 details of the interface between the massive stars, the \hii\ region, and
 the molecular cloud; (b) zoom in on the elephant trunk with 2MASS
 near-IR sources (red circles) and X-ray sources (cyan polygons)
 overlaid. The cyan polygons are not circles; they
 represent the X-ray source extraction polygons based on the PSF. \label{fig:hst}}
\end{figure}

In spite of its small FOV, an archival {\it HST} WFPC2 (F814W) image
of the \hii\ region G353.2+0.9 in NGC 6357 (Figure~\ref{fig:hst};
Healy et~al.\ 2004b) was visually examined for optical
counterparts. By registering to a 2MASS image, an astrometric
correction of $\Delta RA=0.9^{\prime\prime}$ and $\Delta
Dec=-0.4^{\prime\prime}$ was applied to the HST image.  There are 154
X-ray sources covered by the WFPC2 imaging, 99 of which have HST
counterparts. The WFPC2 image demonstrates details of this interface
between the hot tenuous \hii\ region and the cold structured molecular
cloud. Some of the optically bright sources are not X-ray emitters and
are probably foreground stars; a few stars appear very faint and are
probably deeply embedded in the molecular cloud. All but three of the
HST counterparts (corresponding to ACIS \#359, 375, 390) have matched
infrared counterparts. This increases the number of X-ray sources with
ONIR counterparts to 616. Thanks to HST's high resolution imaging,
those ACIS sources are resolved from the crowded field where infrared
images fail to detect them. Even though optical imaging suffers
heavily from extinction, we speculate that $\sim 10-20$ more HST
counterparts for the less-embedded ACIS sources with separations $\le
1^{\prime\prime}$ from other sources would have been recovered due to
the high spatial resolution, if the HST spatial coverage of Pismis 24
were complete.

\subsection{Extragalactic Contamination}

We estimate the contamination from the extragalactic population
following the method described in \citet{Getman06}. Monte-Carlo
simulations of the expected extragalactic source contamination are
constructed by placing fake AGNs randomly across the ACIS-I field
superposed on the ACIS background.  The X-ray flux distribution
follows the log$N$--log$S$ distribution from \citet{Moretti03} and the
corresponding power law photon index for individual sources follows
the flux dependence found by \citet{Brandt01}. Synthetic source
spectra were generated with the XSPEC $fakeit$ function and convolved
with an absorption column $N_H$. We ran simulations with three typical
$N_H$: the total Galactic absorption towards NGC 6357 \citep[$N_H =
1.2 \times 10^{22}$ cm$^{-2}$,][]{Dickey90}; Galactic absorption plus
an additional $N_H = 1.0 \times 10^{22}$ cm$^{-2}$ typical of Pismis
24 OB stars; and Galactic absorption plus an additional $N_H = 3.0
\times 10^{22}$ cm$^{-2}$ typical of reddened 2MASS stars.  Then we identified the closest real
X-ray detection to the fake source. Source significance for each fake AGN was then
calculated using the local background of the nearby real source in the NGC
6357 background map. We set a detection threshold of $Signif \ge 2.3$,
where $Signif$ is the quantity given in column 12 of
Table~\ref{tbl:primary}.  This limiting source significance is the
median $Signif$ of those hard X-ray sources without infrared
counterparts and lying far off-axis, which are the most likely candidates
for AGNs in the field.

The number of detectable AGNs from typical simulations with the three
absorption column densities ranges from 14 to 29.  The level of AGN
contamination for the NGC 6357 region (Galactic coordinates $l=353.2$,
$b=+0.9$) is similar to the contamination calculated for the Cepheus B
($l=110.2$, $b=+2.6$) observation which had a similar {\it Chandra}
exposure \citep{Getman06}. The true extragalactic contamination may be
lower due to localized regions of very high absorption by the
molecular cloud behind the OB association.  We conclude that no more
than 4\%, and more likely $\simeq 2$\%, of the 779 X-ray sources are
extragalactic contaminants.  Among possible AGN candidates, the
brightest in X-rays is source \#1. Based on its flat X-ray spectrum,
large separation from the central cluster, non-variable lightcurve,
and lack of optical or near-IR counterpart, ACIS \#1 is considered to
be a likely AGN. Using a power law model with a single absorption
component, its spectrum can be adequately fit using a power law photon
index $\Gamma=1.4$ and an absorption column $N_H=8.0\times 10^{21}$
cm$^{-2}$.

\subsection{Field Star Contamination}

To estimate possible contamination from foreground field stars, we
follow the techniques described in the {\it Chandra} Orion Ultradeep Project (COUP) membership study and the {\it
Chandra} study of the Cep B star forming region
\citep{Getman05b,Getman06}. We use simulations based on the stellar
population synthesis model of Galactic disk stars by the Besan\c{c}on
group \citep{Robin03}. The models are calculated using their on-line
service\footnote{\url{http://bison.obs-besancon.fr/modele/}}.
Synthetic catalogues of stars generated from these models predict
$\sim 5600$ field stars with $V < 22$ mag in a solid angle of $\sim
0.08$ deg$^2$ (ACIS-I FOV) to the distance of NGC 6357. About 10\% are
F-type main sequence (MS) stars, $\sim 20\%$ are G-type MS stars,
$\sim 45\%$ are K-type MS stars, $\sim 20\%$ are M-type dwarfs; the
rest are giants with a few subdwarfs.

We then adopt the X-ray luminosity functions (XLFs) for MS K-M stars
\citep{Schmitt95}, F-G stars \citep{Schmitt97}, and late type giants
\citep{Huensch96} derived from a complete volume-limited sample in the
solar neighborhood. The conversion of intrinsic X-ray flux in the
ROSAT PSPC $0.1 - 2.4$ keV band to {\it Chandra} ACIS-I $0.5-8.0$ keV
count rate is from PIMMS, using a thermal plasma model assuming
$kT \approx (4\pi d^2 F_X \times 5.5 \times 10^{-26})^{0.2}/11.34$ keV
from \citet{Gudel98}, where $d$ is the stellar distance and $F_X$ is
the X-ray flux in ROSAT PSPC $0.1 - 2.4$ keV band.  Only Galactic
absorption is assumed to be present for stars in the foreground of the
NGC 6357 cloud and the amount of absorption to the fake star is
calculated based on its distance when the simulated catalog is
generated.

Following the same procedures used in the simulations for
extragalactic contamination, a simulated field of $\sim 5600$ X-ray
field stars was added to the ACIS background map, and the star
locations were examined for source significance.  For detection
thresholds ranging from $Signif \sim 1.5$ to $Signif \sim 3.0$, 10000
simulation runs give a typical number of 6 to 16 detectable foreground
field stars.  These sources should be readily distinguished from
extragalactic AGN and most young cluster members by the brightness of
their stellar counterparts and (when sufficient X-ray counts are
present to measure median energies) low absorption in their spectra.
Eleven individual stars are identified as likely foregound stars in
the field (\S \ref{classification.sec}).  Those contaminants will be
removed from further consideration of the young cluster
population. The likelihood of background star contamination is small
because of the heavy absorption of the molecular cloud and the large
distance. 

The contamination from Galactic disk stars is thus $1-2$\%
of the observed source population, about half that of the
extragalactic contamination. A complete removal of all non-members is
impossible without further information, thus we ignore the remaining
foreground star and AGN contamination in our further analysis given
that the fraction is statistically very small.

\subsection{Likely New Stellar Members}

For the 163 ACIS sources without ONIR counterparts, at most 20--30 of
them are possible extragalactic contaminants (\S 4.2). The rest of the 
$\sim$140 sources are probably new members of Pismis 24, in addition to
the 605 likely members (excluding the 11 candidate foreground stars) with
ONIR counterparts identified from the thousands of ONIR stars in the field (\S
4.1).

Sixty-two ACIS sources without ONIR counterparts are clustered within
$2^{\prime}$ of the O3 star Pis 24-1.  The rest are widely
distributed throughout the field, although a few are located to the
northwest of the ionization front where there is much less optical
nebulosity. Five of them are clustered around the WR star. Most of
these likely members are faint X-ray sources; only two of them have
more than 10 net counts. More than 100 of them have median photon
energy $\medE \ge 2.0$ keV and $\sim 50$ have $\medE \ge 3.0$ keV,
implying heavy obscuration (see Figure~8 from Feigelson et~al.\
2005). We speculate that, due to the face-on geometric configuration,
these represent the more distant cluster members that are located
behind those optically visible members, deeply immersed in the
molecular cloud.  Some may be very young, embedded in molecular cores
or protostellar envelopes that have not yet been identified. But many
are likely to be older Class II and III pre-MS stars that have
counterparts in {\it Spitzer Space Telescope} surveys. The properties
and spatial distribution of these likely new members will be further
discussed in \S \ref{spatial.sec} and \S \ref{embedded.sec}, after we
classify the X-ray sources.

\subsection{Classification of the X-ray Sample}
\label{classification.sec}

From the results of the previous sections, we can establish a rough
classification of the X-ray sources.  These class designations appear
in Table~\ref{tbl:class}. The numbers in parentheses after the class
names are the number of sources in each category.

\begin{description}

\item[R (600)--] Reliable X-ray sources with ONIR stellar
identifications. They represent the more massive and less obscured
members of the stellar population.

\item[C (590)--] X-ray sources with off-axis angle $\Theta
\le 5.0 \arcmin$. They are likely members of the central cluster.

\item[H (109)--] Hard X-ray sources with $\langle E \rangle \ge 3.0$
keV. Most of these are heavily obscured members of the embedded stellar
population. About 20-30 of these are probably extragalactic AGNs.

\item[U (18)--] Unreliable X-ray detections. We consider the 18 X-ray
sources in Table~2 with large probability of being
background ($\log P_B \ge -1.8$) to be possible spurious detections.

\item[F (11)--] Foreground field stars. Seven soft X-ray sources
($\langle E \rangle <1.3$ keV, corresponding to $\log N_H \le 22.0$
cm$^{-2}$) with more than 20 net counts in 0.5--8.0 keV (to have
reliable median photon energy measurements) are identified as
foreground sources, excluding any known O stars.  One X-ray emitting
A8 subgiant, HD 157528, was recovered using this criterion. The
limitation for this criterion is that unidentified massive stars with
soft X-ray emission may be included. Two of these candidate foreground
stars, ACIS \#9, \#539, are included as candiate O stars (see \S
\ref{newO.sec}). In addition, we select 4 soft X-ray sources ($\langle
E \rangle <1.3$ keV) as likely field stars since their $JHK$ colors
are consistent with the locus of Besan\c{c}on simulated field
stars. All these sources have visual photometry in the UCAC2 catalogue
\citep{Zacharias04}, which provides additional support for them being
classified as likely foreground stars.

\end{description}

\begin{figure}
 \centering
 \includegraphics[angle=-90,width=0.6\textwidth]{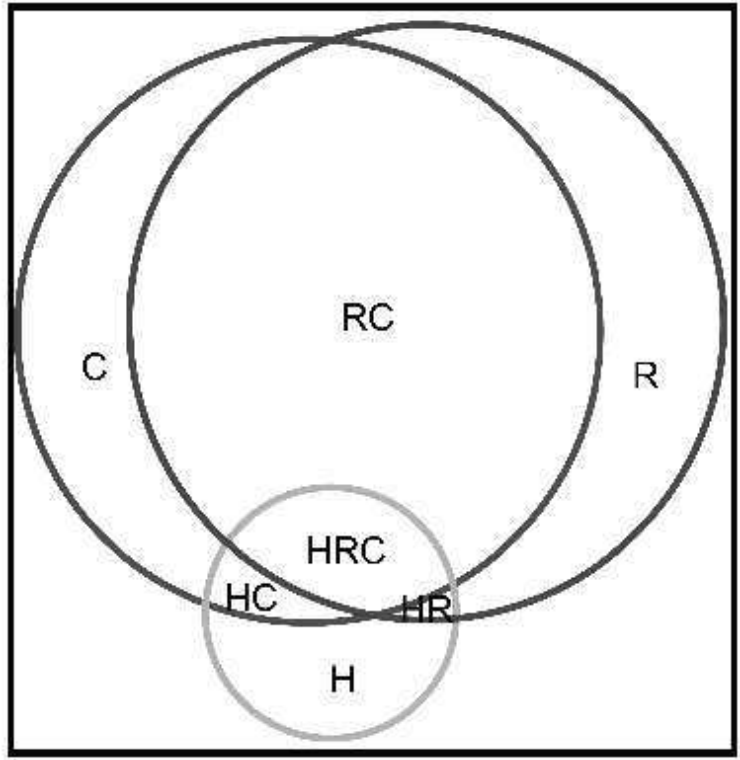}
\caption{A cartoon illustration of the source classification in \S
\ref{classification.sec}: $R$--Reliable X-ray sources with optical and
NIR stellar identifications; $C$--Clustered X-ray sources with
off-axis angle $\Theta \le 5.0 \arcmin$; $H$--Hard X-ray sources with
$\langle E \rangle \ge 3.0$ keV. The area of each circle is
proportional to the number of sources in the class. Group $U$
(unreliable X-ray detections) and $F$ (foreground field stars) are not
shown here.
\label{fig:source_class}}
\end{figure}

The classes $R$, $C$, and $H$ are not exclusive and many sources fall
in two or three categories as illustrated in
Figure~\ref{fig:source_class}.  Thirty-eight sources are not
classified because they are either too faint or poorly characterized.

\citet{Moffat73} listed 15 bright stars in Pismis 24 as their
photometry targets and concluded 12 of them are members.
Spectroscopic observation confirmed that seven of these, and four
additional stars, are massive OB stars \citep{Massey01}. Our X-ray
sample with infrared counterparts increases the known cluster
population 50-fold, from $\sim 15$ to $\sim 750$, providing the
first opportunity to investigate the full stellar population of
the region.

The $H-K$ vs.\ $J-H$ color-color diagram of 142 {\it Chandra} sources
with high-quality 2MASS photometry \footnote{The 2MASS sources plotted
here and in the $J-H$ vs.\ $J$ diagram have signal-to-noise ratios
greater than 6 (photometric quality higher than ``$C$'') and are free
of confusion flags and warning of contamination flags. See Explanatory
Supplement to the 2MASS All Sky Data Release (Cutri et~al.\ 2003;
\url{http://www.ipac.caltech.edu/2mass/releases/allsky/doc/}) for
details.} is shown in Figure~\ref{fig:ccd}.  The typical uncertainty
in color is $\sim 0.06$ mag.  The locus for main-sequence stars of
different spectral types from early O to late M \citep{Bessell88} is
plotted as the thick line. Stars that are reddened by normal
interstellar extinction occupy the color space between the two
reddening vectors of an O dwarf and an M giant (dashed lines). The
foreground A8 subgiant (\# 773) clearly reveals itself, lying on the
locus with almost no reddening. The stars to the right of the
reddening band are $K$-band excess sources. It is clearly seen that
the X-ray observation selects far more Class III infrared sources
(stars that have simple blackbody spectral-energy distributions, with
little or no inner accretion disks) than Class II sources \citep[stars with
$K$-band excess circumstellar disks;][]{Lada87,Wolk96}.

The 2MASS $J-H$ vs.\ $J$ color-magnitude diagram
(Figure~\ref{fig:cmd}) gives both the rough mass distribution and the
absorption distribution of the ACIS sources with 2MASS
counterparts. Zero age main sequence (ZAMS) track \citep{Cox00} and
two pre-MS isochrones \citep{Baraffe98,Siess00} are shown with
spectral types, stellar masses, and reddening vectors for estimation
of the dereddened masses and the associated absorption.  We assume 1
Myr for the age of Pis 24 cluster and a distance of $2.56$ kpc when
estimating stellar properties from the color magnitude diagram.  We
remind the reader that the estimated mass and the amount of absorption
for a given star also relies on the assumption of distance and cluster
age besides the uncertainties in the star's color and magnitude.  For
example, if we adopt a younger cluster age (0.3 Myr), the pre-MS
isochrone will shift upwards and yield a lower dereddened mass and
less amount of absorption as shown in Figure~\ref{fig:cmd}. Note that
the inferred properties for stars with spectral type later than B3 are
less dependent on the age and distance assumptions. In addition we
show the $H-K$ vs.\ $K$ color-magnitude diagram for the same sample of
stars accounting for some small $K$-band excess. Known Pis 24 optical
members detected in X-ray are located along the reddening vectors from
the locus of O stars with a typical $A_V\simeq 5$ mag.  Not
surprisingly, the foreground star \#773 resembles a very massive and
luminous object at the distance of NGC 6357. Besides the foreground
star, WR 93 is the most luminous object.

Most of the other ACIS sources occupy infrared color domains that are
consistent with being high-mass stars and pre-MS intermediate-mass
stars ($M \ge 2M_{\odot}$) with $5<A_V<15$ at 2.5 kpc.  The
highly-reddened bright 2MASS sources are possibly obscured massive
stars.  The new high-mass stars double the known high-mass population
of the cluster (\S 6.2).  The intermediate-mass stars, mostly Class
III objects, are newly identified members of the cluster (\S 6.3).
The ratio between Class III and Class II objects for the high-mass and
intermediate-mass cluster members is $\sim 15:1$ based on the
$\sim$140 stars with mass $M \ge 2M_{\odot}$.  Although the
uncertainties of the estimated stellar masses are not negligible
\citep[as much as a factor of 2; see the discussion of Orion stars
in][]{Preibisch05,Luhman99,Hillenbrand04}, this is a quantitative
measurement for the massive end of the cluster members and implies
that the fraction of stars possessing inner disks among the
intermediate-mass and high-mass stars is low even in such a young
stellar cluster. Those intermediate-mass stars with significant IR
excess are likely the accreting pre-MS stars known as Herbig Ae/Be
stars \citep[HAeBes;][]{Herbig60,The94}. X-ray properties of the
intermediate mass stars with infrared excess will be examined in
\S~\ref{ximf.sec}. A handful of objects are heavily obscured with $A_V
> 20$. Due to their brightness in K-band ($K_s \sim 9-13$), these
objects, some of which exhibit K-band excess, are easily accessible
for ground based follow-up infrared spectroscopic or photometric
study. The remaining $\sim 170$ ACIS members without 2MASS
counterparts but seen with SIRIUS and/or HST are likely lower mass
stars. The close separations to the high/intermediate mass stars
suggest they perhaps are companions, but this could well be a spatial
projection effect in such a crowded cluster.  Similar to those
unobscured ACIS sources without ONIR counterparts, they are either too
faint or too close to bright stars to be cataloged in 2MASS.

\subsection{EGGs and Protostars}

Two-thirds of the IR counterparts -- those with low photometric
quality -- do not appear in Figure~\ref{fig:ccd}.  Although candidate
protostars may be present, the large uncertainty in their IR color and
the upper limits in $K$-band do not allow us to confidently classify
class II/I objects. Precise infrared photometry will be the key to
establishing whether they are true embedded protostellar objects.

When compared to an archival {\it HST} WFPC2 image
(Figure~\ref{fig:hst}; Healy et~al.\ 2004b), the X-ray source \#461
(CXOU J172445.74-341106.9) is spatially coincident with the tip of an
evaporating clump extending from the wall of the molecular cloud,
which is known as an evaporating gaseous globule
\citep[EGG;][]{Hester96,Hester05}. It has only 4.6 net counts but the
X-ray photons are spatially concentrated; it is a reliable Table~1
source. The X-ray emission appears to be moderately absorbed with
$\medE=2.5$ keV. The colors of its infrared counterpart show large NIR
excess ($J-H=1.06$, $H-K=1.42$). The 2MASS counterpart was previously
detected in the \citet{Persi86} infrared observation (=IRS 4) and its
$JHKLMNQ$ spectral energy distribution (SED) gives a 1--20 $\mu m$
luminosity of 630 $L_{\odot}$ \citep{Persi86}. This luminosity and its
infrared colors are consistent with that of a protostar with spectral
type B0--B2 \citep{Persi86,Bohigas04}. In addition, it also coincides
with the compact radio peak of a 6 cm VLA observation \citep[=VLA
A;][]{Felli90}.

CXOU J172445.74-341106.9 may be the first detection of X-ray emission
from an EGG, which represents one important evolutionary stage of star
formation. \citet{Linsky03} identified deeply embedded hard X-ray
sources in the ``pillars of creation'' of the Eagle Nebula \citep[M
16;][]{Hester96} but were unable to find any such sources associated
with the EGGs. The non-detection of X-ray emission in M 16 EGGs may be
related to their high threshold of detection ($\ge$6 counts), heavy
extinction, or the spectral type and evolutionary phases of the
embedded stars. Although the uncertainty in the X-ray luminosity is
potentially very large for such a faint source, the estimated $L_h
\sim 30.1$ ergs s$^{-1}$ from PIMMS is somewhat consistent with other
X-ray detected high-mass protostars (NGC 2024 IRS 2b, O8--B2, Skinner
et al.\ [2003]; S106 IRS4, O7--B0, Giardino et al.\ [2004]). It is
likely that the actual X-ray source is the embedded protostar instead
of the gaseous globule.

No X-ray detection is associated with the HST jet candidate, Pismis 24
\# 14 in \citet{DeMarco06}. We examined the X-ray image at the HST
position and no photons can be found within a 90\% PSF contour.

\section{Properties of the Stellar Cluster and Its Environment}
\label{XLF.sec}

\subsection{X-ray Luminosity Function}

We construct the observed hard band (2--8 keV) XLF for all X-ray
sources, excluding sources classified as $U$ and $F$. Based on the
fact that the extragalactic contamination is small, we only exclude
ACIS \# 1 that is very likely an AGN. We do not attempt to study the
observed total band (0.5--8.0 keV) XLF or absorption corrected total band
luminosities ($L_{t,c}$; 0.5--8.0 keV), since we do not have reliable
measurements of $N_H$ for faint sources. In the case of heavily
absorbed X-ray sources, the unknown soft component can introduce a large
uncertainty in the observed total band X-ray luminosity.

For sources with more than 20 net counts, the hard band (2.0--8.0 keV)
flux $F_h$ and absorption-corrected hard band flux $F_{h,c}$ (and
therefore the luminosities $L_h$ and $L_{h,c}$) are available from
XSPEC spectral fitting.  For fainter sources, we define the following
procedure to estimate $F_h$ and also $F_{h,c}$ consistently. First, if
the faint X-ray source has a 2MASS counterpart with reliable
photometry, we use the $A_V$ derived from the color magnitude diagram
to obtain the $N_H$, assuming a 1 Myr age for the cluster members
$N_H/A_V=1.6\times 10^{21}$ cm$^{-2}$ mag$^{-1}$ \citep{Vuong03}. We
note that adopting a younger age or a smaller distance will shift the
pre-MS isochrone upwards in the color magnitude diagram, causing
generally lower values for the derived $A_V$. However for these faint
sources that are likely low-mass cluster members, the decrease in
$A_V$ is very small, $\sim 1$ mag (Figure~7).  Otherwise, as shown in
\citet{Feigelson05}, empirical conversion from median photon energy
$\langle E \rangle$ provides us with an effective approximate $\log
N_H$ value for sources with only a few counts: $\log N_H=9.96+13.62
\times \langle E \rangle -3.86 \times \langle E \rangle ^2$ cm$^{-2}$
for relatively soft sources ($1.0 \le \langle E \rangle \le 1.7$ keV);
$\log N_H=21.22+0.44 \times \langle E \rangle$ cm$^{-2}$ for harder
ones ($\langle E \rangle > 1.7$ keV). We estimate that the
uncertainties in $\log N_H$ are $\sim \pm 0.3$ and can be as large as
$\pm 0.8$ for soft sources that have $\langle E \rangle \le 1.3$
keV. We assign plasma temperature $kT$ based on $\langle E \rangle$ to
reflect the observed trend of increasing temperature with increasing
median energy: typically $kT=1$ keV for $\langle E \rangle < 1.3$ keV,
$kT=2$ keV for $1.3 \le \langle E \rangle \le 2.0$ keV, $kT=3$ keV for
$2.0 \le \langle E \rangle \le 4.0$ keV, and $kT=5$ keV for $\langle E
\rangle \ge 4.0$ keV \citep[see Figure~8 in][]{Feigelson05}.  With the
$N_H$ and $kT$ parameters set and ACIS-I count rates known for each
source, we run PIMMS simulations to derive the observed (absorbed) and
absorption corrected flux in the hard band.

The XLFs for the lightly obscured ($\medE < 3.0$ keV) and heavily
obscured ($\medE \ge 3.0$ keV) stellar populations in NGC 6357 are
plotted in Figure~\ref{fig:XLF}. For comparison, XLFs derived for the
lightly obscured COUP Orion Nebula population \citep{Feigelson05} and
for the Cep B unobscured and obscured populations \citep{Getman06} are
also shown. The shapes of the high luminosity portions of the lightly
obscured and obscured COUP XLFs (see Figure 3 in Feigelson et
al. 2005) are very similar, but the weaker sources in the obscured
COUP sample are missing. Therefore we adopt the lightly obscured COUP
XLF as the template in this population calibration exercise. At the
high $L_X$ end, the NGC 6357 XLF is clearly consistent with the power
law relation as seen in COUP and Cep B, allowing us to scale to the
COUP XLF and make the first reliable estimate of Pismis 24's total
cluster population.

At $\log L_h\sim 30.3$ ergs s$^{-1}$, the NGC 6357 XLF begins to
flatten and decline dramatically.  We examine the completeness in the
source detection to determine whether the decline is due to a real
deficiency of cluster members in this luminosity range or is simply an
expected effect when reaching the faint source detection limit. If the
observed X-ray flux from a cluster member with such a luminosity is
well above our sensitivity limit, the deficiency of stars in this
luminosity range would imply a different X-ray luminosity distribution
for the population.  Given that sources with 10 net counts in the
full-band at any off-axis angle are confidently detected, we adopt two
different methods to estimate what is the corresponding flux limit to
10 net counts.  First, assuming a 2 keV thermal plasma, PIMMS
simulation gives $\log F_h= -14.6$ ergs s$^{-1}$ cm$^{-2}$ for a
count rate of $10/38$ counts ks$^{-1}$. Independently, using all
sources with more than 20 net counts that have flux estimates from 
spectral fitting, we derive an empirical
conversion factor between flux and net counts for this observation,
and the 10 net counts also converts to $\log F_h \sim -14.6$ ergs
s$^{-1}$ cm$^{-2}$. Therefore the two approaches give the same
$F_h$, which converts to $\log L_h \sim 30.2$ ergs
s$^{-1}$. The decline in the NGC 6357 XLF below $\log L_h\sim
30.3$ ergs s$^{-1}$ is thus consistent with incomplete detection of
the less luminous cluster members. We do not have sufficient
sensitivity to search for a difference in XLF shape compared to COUP
below $\log L_h \sim 30$ ergs s$^{-1}$, as seen in Cep B by
\citet{Getman06}.

The scaling factor to match the COUP XLF to the NGC 6357 XLF is $\sim 5-6$
at higher values of $\log L_h$ for the lightly obscured
population. Using the XLF constructed with $L_{h,c}$ gives the same
scaling factor. The XLF from $L_h$ for the obscured stellar population
in NGC 6357 is comparable to the COUP XLF, while the absorption
corrected XLF seems to be consistently $\sim 5$ times higher than the COUP
XLF for the high luminosity bins. Altogether, this implies that the X-ray
emitting stellar population in NGC 6357 is roughly $5-10$
times\footnote{If the true distance to the cluster is a smaller value, e.g.\
1.7 kpc, all the derived luminosities, including $\log L_h$ and $\log
L_{h,c}$, will be systematically lower by 0.35 dex and the estimated
population will be $\sim$3 times the Orion Nebular Cluster.} richer than
the lightly obscured Orion Nebula Cluster \citep[839 unobscured cool
stars;][]{Feigelson05}. Therefore the X-ray emitting population in
Pismis 24 is $\sim 5000$. Given that the ONC has $\simeq 2000$ known members
from the deep optical study by \citet{Hillenbrand97} and COUP detects
more than half of the stellar population in X-rays, the total stellar
population in Pismis 24 is estimated to be two times larger than the
population detected in X-rays, $\sim 10,000$ members. It is not
unexpected, assuming the standard IMF, to find massive stars with $>100
M_{\odot}$ in a cluster with such a large population. A note of
caution here is that, even in such a young star-forming region, the
stellar population in consideration is not perfectly coeval. It
perhaps includes a younger population triggered to form by the
ionizing front from the massive stars, but the small population of the
second generation stars has little impact on the IMF of this rich
cluster.

\subsection{Spatial Distribution of the Stellar Cluster}
\label{spatial.sec}

Aside from several dozen known and new OB stars (see \S
\ref{ximf.sec}), the locations of over 700 new lower-mass cluster
members are now determined with sub-arcsec precision.
Figure~\ref{fig:radec} shows the spatial distribution of the cluster
members.  The large scale optical nebulosity contour from the DSS
R-band image, which also traces the mid-IR ring-like morphology, is
shown together with the outlines of the ionization front in yellow and
the CO ``South-Eastern Complex'' in green \citep[S.E.C. in the CO
emission map in][]{Massi97}.  The geometry of the cluster appears
spherically symmetric: the source density in the cluster core near the
ACIS-I aimpoint is the highest, and declines radially with no apparent
discontinuity. There seem to be fewer X-ray detections northwards of
the optical nebula, although there is neither an enhancement nor a
deficit of stars at the bright photodissociation region
(PDR). Although the reduced sensitivity at the chip gaps between the
ACIS-I CCDs is not taken into account, we doubt that the number of
missing sources in the chip gaps is large enough to alter the spatial
distribution of the stellar cluster.

Projected stellar surface density (number of stars arcmin$^{-2}$)
contour maps for the unobscured and obscured populations are shown in
Figure~\ref{fig:stellar_density}, in logarithmic scale.  The stellar
density is smoothed using a $0.5^{\prime}$ radius kernel. As expected,
the highest concentration of stars is at the core of the cluster,
where the massive O stars are located.  The distribution of stars is
rather spherically symmetric centered at the O3 stars, except for a
density enhancement centered at the tip of the S.E.C.  (centered at
RA=17:24:48.3, Dec=-34:15:05).  Figure~\ref{fig:offaxis_hist} further
elucidates the structure of the stellar distribution by showing the
radial profile of the stellar surface density. The stellar
distribution appears to consist of two components: a compact cluster
within the central 1.5 pc with a log-normal distribution, and a
cluster halo extending to 8 pc with an approximately exponential
distribution. The number of X-ray stars in the central region is 333
while the extended region contains 446 stars. Because these stars are
poorly studied, we can only speculate that the stars in the central
region represent the coeval dense cluster core of Pismis 24, while the
apparent extended wing is a mixture of evaporating cluster members and
newly formed stars distributed throughout the molecular cloud
enveloping the cluster.

Figure~\ref{fig:offaxis_hist} also compares the radial profile to the
profile of the Orion Nebula Cluster within $\sim$1.5 pc for Chandra
stars \citep{Getman05a,Feigelson05}. The ONC is only extensively
studied within a projected radius of $\sim$2.5 pc from $\theta ^1$ Ori
C (see Hillenbrand \& Hartmann 1998 for a review of ONC structure and
dynamics). It is strongly concentrated, with a central density $>200$
times that of the widespread cloud stellar population distributed at
radii $>2.5$ pc. The ONC could possess an extended halo of stars as we
see in the more massive Pismis 24 cluster but it would be confused
with the surrounding stellar population. Thus we limit the comparison
of the two X-ray selected clusters to radii $<$1.5 pc.

The stellar density of the ONC core appears to be $\sim$5 times higher than
the X-ray population in the Pismis 24 cluster core, but this can be
largely attributed to the differencein the number of faint stars
detected: the ONC sample from the 800~ksec COUP observation extends
into the brown dwarf regime, while the Pismis 24 sample becomes
incomplete around $M \le 3M_{\odot}$ (\S 5.1). Extrapolating to brown
dwarf masses using the COUP mass distribution (Getman et al. 2005a),
we estimate that the NGC 6357 cluster central density would be
$\sim$20 times higher than the currently observed one if it were as
sensitive as the COUP observation. We thus conclude that the NGC 6357
cluster central density is probably $\sim$4 times higher than the ONC
central density and similar to the central density of the NGC 3603
super star cluster (see footnote 1 in Hillenbrand \& Hartmann 1998).

Figure~\ref{fig:radec_zoomin} zooms in to the core region of the
cluster using the SIRIUS K image as background. The ONIR identified
X-ray sources (circles) and unidentified X-ray sources (crosses) are
shown with the known massive members highlighted in white. Only a few
of the unidentified sources appear isolated, without infrared sources
nearby. Most are clustered in the $15^{\prime \prime}$ neighborhood of
the two O3 stars; 24 are located $\sim 1.5-2^{\prime \prime}$ from
near-IR sources, therefore the available NIR images may not be able to
resolve them. These unidentified sources very likely are real low-mass
members concentrated around the massive stars.

\subsection{The Morphology of the Ring-like Nebula}
\label{morphology.sec}

A puzzling feature arises when we compare the core of the massive
cluster centered on the two O3 stars to the large ring-like nebula
with a large cavity that makes up the wider view of NGC 6357
(Figure~1): the cluster is significantly off-center to the north. The
``diamond ring'' morphology of NGC 6357 is apparent in the H$\alpha$
and mid-IR images, with the bright point along the edge brightened
torus. In contrast, the Rosette Nebula has a similar annular nebula
morphology, but the massive OB stars in NGC 2244 exciting the nebula
are located right in the center. The hollow HII morphology is
attributed to strong OB stellar winds excavating molecular materials
and the depletion of gas by newly formed stars
\citep{Dorland86,McCullough00,Townsley03}. Therefore the nebular shape
is generally accepted as spherical or cylindrical. In NGC 6357, the
most massive members are so close to the northern edge of the nebula
that it seems implausible that the cavity is created by the strong
winds of the current generation of hot stars. In general, supernovae
explosions and strong stellar winds can shape the natal cloud by
creating wind-blown ``bubbles''--cavities.  However, there is no
evidence for a recent supernova: neither diffuse X-ray emission from a
hot tenuous plasma nor non-thermal radio emission is detected in this
region.  We further note that the presence of WR 93, an evolved star,
is circumstantial evidence that some population of massive stars in
NGC 6357 is old enough to produce supernovae.

It is possible that a previous-generation of hot stars created the
large cavity and formation of the current ionizing population was
triggered at the densest edge of the excavated cavity.  Combined with
different viewing angles, this can result in diverse morphologies such
as an egg, or a diamond ring, rather than a symmetrical
annulus. H$\alpha$ images of quite a number of regions where the stars
are displaced with respect to the center of an evacuated cavity have
been shown in the Large Magellanic Cloud \citep{Gaustad99} and modeled
as modified Str\"{o}mgren spheres \citep{McCullough00}. The key for
testing this hypothesis is to identify the older cluster.  X-ray
sampling provides a critical test as X-ray luminosities decay only
slightly over $\sim 10$ Myr \citep{Preibisch05a}. The subcluster seen
in the surface density map $\sim 4^{\prime}$ southeast of the Pis 24
cluster could be the remnant core of the previous generation of
stars. ONIR photometric and spectroscopy study of the $\sim$30 stars
in this subcluster, permitting their placement on the HR diagram, might
reveal whether they are older than the main cluster concentrated off
the center of the nebula.

An alternative scenario is that the cavity is indeed created by the
massive Pis 24 stars that are emitting ionizing photons. In that case, the
displacement from the center could be explained with spatial
projection and an inhomogeneous molecular cloud, as first proposed by
\citet{Bohigas04}. The O stars were close to the geometrical center of
the Pismis 24 cluster and created an expanding HII bubble, which may
have encountered a much denser interstellar medium in the northern
part than in other directions. The apparent deficit of X-ray sources
to the north may be evidence for the absence of star formation in the
densest region, where a new generation of stars may emerge eventually.

\section{X-rays across the Mass Spectrum}
\label{ximf.sec}

\subsection{X-rays from known massive stars}

OB stars have been known to be X-ray sources since early observations
from the $Einstein$ satellite
\citep{Harnden79,Seward79}. Models were developed where
instabilities in radiatively-driven stellar winds from massive
stars produce shocks and heat the gas to X-ray emitting
temperatures \citep{Lucy80,Owocki88,Owocki99}. However, recent
studies indicate that more complex models are needed to account
for unexpected X-ray emission line profiles and hard, variable
continuum emission \citep[e.g.,][]{Waldron04,Stelzer05}.

We detect X-ray emission from all known early type stars ranging
from O3 to B0.5 in this 38 ks ACIS observation. Table~\ref{tbl:OB}
summarizes ACIS detections and non-detections of X-ray emission
from O and early B stars in the Pismis 24 cluster that are
classified in literature.

Two of the newly classified O3 stars Pis 24-1 (O3.5If) and Pis 24-17
\citep[O3.5IIIf;][]{Walborn02} exhibit X-ray luminosities $L_{t,c}\sim
10^{33}$ ergs s$^{-1}$. This is consistent with HD 93128 in the Carina
Nebula cluster Tr 14, which has similar spectral type \citep{Evans03}.
Their lightcurves remain constant through the observation, which is
typical for O stars, although variability in X-ray emission from O
stars like $\zeta$ Ori is well known \citep{Berghoefer94}.
Figure~\ref{fig:O3s} presents their X-ray spectra and spectral fits
(assumed abundances of 0.3 $Z_{\odot}$ unless otherwise
noted). Compared to the best fit using a single temperature plasma
model, the X-ray spectrum of Pis 24-1 (\#344) is better fit by a two
temperature plasma model ($N_H=1.3\times 10^{22}$cm$^{-2}$, $kT_1=0.5$
keV, and $kT_2=1.7$ keV). Pis 24-17 (\#420) can be adequately fitted
by a single soft component with $N_H=1.4\times 10^{22}$cm$^{-2}$ and
$kT=0.7$ keV. The fit is improved by adopting enhanced Mg ($0.6\times
Z_{\odot}$), Si ($0.6\times Z_{\odot}$), and Ar abundances ($2.7\times
Z_{\odot}$). The \sixiii\ line at 1.86 keV in Pis 24-17 is
exceptionally strong, comparable to the same line seen in the O star
IRS 2 in RCW 38 \citep{Wolk02}. The derived temperatures for soft and
hard components for these O stars are similar to the X-ray
temperatures of other single massive stars: some O stars exhibit a
simple single temperature plasma $\sim 1$ keV
\citep[e.g.,][]{MacFarlane93,Rho04,Townsley06}, while a few require an
additional high energy component
\citep[e.g.,][]{Corcoran94,Kitamoto96,Evans04,Stelzer05}.
\citet{Gagne05} shows that the magnetically channeled wind shock model
\citep{Babel97a,Babel97b} with strong line-driven winds in a single O
star can adequately reproduce both the soft and the hard compents in
the recent phase-resolved {\it Chandra} grating spectra of $\theta ^1$
Ori C (O5.5V).  In some cases, the high energy component may in fact
be an indication of binarity, as powerful winds in two massive
components shock to produce hard X-rays
\citep[e.g.,][]{Portegies-Zwart02,Albacete-Colombo03,Townsley06b}.

The brightest X-ray point source in the ACIS field of view is the
WC7+O7-9 binary WR 93 (\#747) with a $L_{t,c}=1.6\times 10^{33}$ ergs
s$^{-1}$, roughly consistent with the earlier ROSAT PSPC value
$L_X=1.25\times 10^{33}$ ergs s$^{-1}$ \citep[0.2--2.4
keV;][]{Pollock95}. It is one of the spectroscopic WR binaries that
are bright enough to be detected by previous generation X-ray
telescopes; single WC stars are much fainter \citep{Oskinova03}. Thus
the strong X-ray emission from binary WR stars is likely generated in
the region where the massive stellar winds collide. However, while
archetype colliding-wind binaries such as WR 140 (WC7+O4V) display
characteristic non-thermal emission in the radio band
\citep{Dougherty05}, WR 93 shows thermal radio emission
\citep{Abbott86}.  Also the X-ray lightcurve appears to be constant
through the entire observation; no orbital X-ray variability is seen
\footnote{No orbital period for WR 93 is reported in literature.}. The
X-ray spectral fit for WR 93 is shown in Figure~\ref{fig:O3s} with an
absorption column $N_H=7.4\times 10^{21}$ cm$^{-2}$ and a rather hard
$kT=2.4$ keV. The large absorption column densities towards the two O3
stars and the WR star derived from X-ray spectral fits match well with
their large visual extinction ($A_V\sim 5-7$) and the reddening from
$JHK$ colors.

Five of the bright Pismis 24 stars in \citet{Moffat73} -- Pismis 24-4,
7, 8, 9, 11 -- were not included in the spectral classification
observation of \citet{Massey01}. Based on their strong X-ray emission
(except for Pis 24-9 which is undetected), absorption column derived
from X-ray spectral fitting, $UBV$ brightness, and $JHK$ colors, we
suggest that they are also young OB stars and cluster members (see
Table~\ref{tbl:NewO}).

X-ray emission from OB stars was reported from ROSAT observations to
display a characteristic efficiency $L_X/L_{bol} \sim 10^{-7}$
\citep{Chlebowski89,Berghoefer97}, and this empirical ratio has been
examined in recent X-ray observations of massive stars
\citep[e.g.,][]{Corcoran99,Evans04,Stelzer05}. From the COUP
observation, the emission ratio for late O and B stars is found to
scatter around the canonical $L_X/L_{bol}$ value by three orders of
magnitude. However, as they cautioned \citep{Stelzer05}, this may not
represent the most luminous early O stars due to the absence of
spectral types earlier than O7. Therefore the Pismis 24 cluster, with
two O3 stars, is particularly valuable for assessing this relation in
a single massive cluster. Using $M_{bol}$ reported in \citet{Massey01}
\footnote{It should be noted that \citet{Heap06} have proposed a new
calibration between the spectral type and effective temperature of O
stars, based on analysis of high resolution HST, FUSE, and optical
spectra. The spectral calibration in general use is by
\citet{Vacca96}, which \citet{Massey01} used to derive bolometric
luminosity for Pismis 24 O stars. The derived $\log L_{bol}$ using the
\citet{Heap06} calibration is 0.4 dex lower than $\log L_{bol}$ using
the \citet{Vacca96} value for an O4V star, and 0.15 dex lower for an
O8V star. Here we use the new bolometric corrections from
\citet{Heap06} to derive bolometric luminosities for Pismis 24
stars. The bolometric correction value for the two O3.5 stars is not
provided by \citet{Heap06}, and we adopt $-0.6$. COUP data points are
from \citet{Stelzer05} and originally derived from
\citet{Hillenbrand97}.}, Figure~\ref{fig:LxLbol} shows the
$L_X$--$L_{bol}$ relation for our detections of cataloged O and B0
stars together with the strong wind sample and weak wind sample from
the COUP study \citep{Stelzer05}.  The $L_X/L_{bol}$ ratio for O stars
in the Pismis 24 cluster is consistent with the $\sim 10^{-7}$ value
with an order of magnitude scatter. Although there are only 9 data
points and one upper limit \citep[$L_{bol}$ for WR 93 is not well
determined due to the poorly characterized bolometric correction for
late-type WC stars;][]{Massey01}, it appears that the $L_X/L_{bol}$
correlation has less scatter for stars of the earliest spectral types
and the physics behind this relation is likely to be physically
related to stellar wind properties.

\subsection{Newly-discovered Candidate O Stars}
\label{newO.sec}

We assembled a list of candidate O stars in Table~\ref{tbl:NewO} to
facilitate future spectroscopic follow-up via two
approaches. Table~\ref{tbl:NewO}a lists the 13 sources brighter than
80 counts (to guarantee a reliable X-ray spectral fit) with 0.5--8.0
keV absorption corrected X-ray luminosities $\log L_{t,c} \ge 32.0$
ergs s$^{-1}$ as potential new high-mass stars, since pre-MS stars
rarely reach X-ray luminosities $\ge 10^{32}$ ergs s$^{-1}$
\citep{Favata03,Getman05b}. Then we examined their X-ray light curves
and observed optical and infrared properties to exclude possible AGNs.
Since certain O stars are observed to exhibit powerful flares
\citep{Feigelson02}, we do not exclude the X-ray luminous sources that
may be extraordinarily bright flaring pre-MS stars.  Their lightcurve
variability is noted in Table~\ref{tbl:NewO}.

O stars may reveal themselves as luminous in infrared but appear
relatively faint in X-rays due to heavy obscuration in their early
evolutionary stages. For example, an O9V star emitting $\log
L_{t,c}=32.0$ ergs s$^{-1}$ behind 15 mag of visual extinction would
have $K=10$ and $\sim 40$ counts in our ACIS image when observed at a
distance of 2.5 kpc (assuming a $kT\sim 0.6$ keV and without
considering a possible K-excess). We selected bright near-IR sources
($K_s \le 10.0$ mag) detected in the X-ray image to examine whether
their JHK colors and X-ray properties are suggestive of obscured early
type stars. This $K$ brightness selection is used in \citet{Hanson97}
when they select the spectroscopy sample of massive YSOs in M 17, and
we further adapt it to X-ray emitting samples. All 9 known members
with spectral type earlier than B0.5, the WR+O7 binary, and the
foreground A8IV star were recovered using this critieria.
Table~\ref{tbl:NewO}b gives 11 additional candidates that have colors
consistent with early-type stars and are not known foreground stars.
Figure~\ref{fig:cmd} and Figure~\ref{fig:cmd2} convincingly suggest
that highly reddened bright infrared source \#654 and \#694 are
indeed very obscured early O stars with some IR excess. Source \#140
is almost certainly a high-mass star since it qualifies using both the
IR and the X-ray selection criteria. As a cautionary note for this
method, the presence of $K$-band excess may mislead us to include some
accreting intermediate-mass stars as candidate O stars.

Given the uncertainties in distance, age, and absorption, readers are
cautioned that ONIR spectroscopy on these objects (Table~7) should
yield more appropriate classifications for them
\citep[e.g.,][]{Massey01,Walborn02}. The X-ray sample suggests that
the optical sample of \citet{Massey01} may have significantly
underestimated the O star population. However, using a smaller
distance of 1.7 kpc will result in a 0.35 dex decrease in the derived
X-ray luminosity. Some of the X-ray selected candidates may indeed be
lower mass flaring pre-MS stars, which are readily distinguishable
from their ONIR spectra. If a large fraction of these 24 candidate O stars
are confirmed, the OB population of Pis 24 (Table~6) is doubled to
tripled.

On the top right corner of the infrared color-magnitude diagram
(Figure~\ref{fig:cmd} and Figure~\ref{fig:cmd2}), there is another
mysterious bright source, \#19 (CXOU J172413.60-341456.7), whose color
and brightness ($J-H=2.006$, $H-K=0.991$, $K=5.334$) suggest that it
is a very luminous object subjected to significant absorption ($A_V
\sim 15-20$). It has 17 net counts in X-ray and the median photon
energy is 2.5 keV, which is consistent with being obscured and not a
foreground star. However, if the star is located at 2.5 kpc, the
intrinsic $K$-band magnitude and infrared luminosity will be
exceptionally luminous. Thus it is plausible that source \#19 is a
massive YSO or a young post main-sequence star like WR
93. Spectroscopy is highly warranted and should reveal the nature of
this unusually luminous but obscured star.

\subsection{Intermediate Mass Stars \label{HAeBe.sec}}

The large population of the Pismis 24 cluster offers an exceptionally
rich sample of intermediate mass stars. Of particular interest are
those with infrared excess, possibly due to the presence of
protoplanetary disks.  We note that the estimated stellar masses from
the infrared color-magnitude diagrams rely on the uncertain age and
distance, which are determined with improved precision in recent
literature.  The ambiguity is more severe among the early B stars
(Figure~\ref{fig:cmd}; for example, the inferred stellar mass for a
14~$M_{\odot}$ star from the 1 Myr isochrone can be as low as
4~$M_{\odot}$ when estimated from a 0.3 Myr isochrone. However, the
X-ray selected sample is important for future follow-up for this
poorly studied region.  We list in Table~\ref{tbl:HAeBe1} 100 X-ray
selected candidate intermediate-mass stars, which have dereddened
spectral types A0$-$B0 estimated from the NIR color-magnitude diagram
(Figure~\ref{fig:cmd} and Figure~\ref{fig:cmd2}) assuming 1 Myr for
age and 2.56 kpc for distance.  Optical spectroscopy is needed to give
accurate spectral classifications.  Among them, four stars exhibit
significant $K$-band excesses in the NIR color-color diagram
(Figure~\ref{fig:ccd}), which are good candidates for HAeBe
stars. $H_\alpha$ spectroscopy on this sample might further reveal the
existence of accretion. Their X-ray luminosities are on the order of
$\log L_X \sim 30.0-31.5$ ergs s$^{-1}$, consistent with previous
ROSAT PSPC (0.1-2.4 keV) observation of HAeBe stars
\citep{Zinnecker94}.

One interesting result has been mentioned in \S 4.5 -- the available
IR photometry data suggest that the ratio between the number of Class
II and Class III pre-MS stars with mass $>2M_{\odot}$ is low, $\sim
1:15$. Over the mass range $2-16 M_{\odot}$, the ratio between the
number of Class II and Class III pre-MS stars is even lower, $\sim 4
\%$. It is important to note that such a ratio is derived from X-ray
selected stars, without any prior knowledge of disk indicators. The
low fraction of the cluster members possessing disks is rather
surprising since apparently these X-ray emitting pre-MS stars are very
young (in the company of O3 stars). However, the low disk frequency is
consistent with the previous studies in NGC 6611 by
\citet{Hillenbrand93}. They find that optically thick circumstellar
disks are already rare among the intermediate-mass pre-MS stars with
ages less than 1 Myr and suggest the disk lifetimes are much shorter
for the massive stars than those of solar type stars. It seems also
consistent with the suggestion that disk evolution happens rapidly in
clusters based on recent {\em Spitzer} observations
\citep[e.g.,][]{Hartmann05,SiciliaAguilar06,Lada06}. The drastic
radiation environment may play a bigger role here in the dissipation
of disks, rather than mass loss via accretion. Further {\em Spitzer}
study on this statistically significant sample will provide
constraints on the timescale of the circumstellar disk dissipation and
the formation of planetary systems.

\subsection{Flaring pre-MS Stars \label{pre-MS.sec}}

There are 31 sources identified as highly variable ($P_{KS}
\le 0.005$; variability flag ``$c$'' in Table~\ref{tbl:primary} and
Table~\ref{tbl:tentative}); their lightcurves exhibit flaring
activity as most frequently seen in pre-MS stars when magnetic
reconnection events occur in their coronae. Eleven of them have more
than 80 net counts (0.5--8.0 keV), so we can reliably derive their X-ray
luminosities through spectral fits. Seven of the flaring
sources exhibit luminosities above $\log L_{t,c} \ge 32.0$ ergs
s$^{-1}$, which is exceptionally high for pre-MS stars in their flaring phase
\citep{Feigelson99,Favata03}.

One extraordinarily intense flare is seen in source \#672 (CXOU
J172457.87-341203.9), comparable to some of the strongest X-ray
flares known in pre-MS stars \citep{Grosso04,Favata05,Getman06}. The
time-energy diagram, binned lightcurve, and X-ray spectrum for \#672
are shown in Figure~\ref{fig:flare}. In contrast to the typical
impulsive fast rise phase of 2 hr observed in the X-ray flares of
pre-MS stars, this flare shows a $\sim 4$ hr rising phase to reach its
highest flux and still remains at the peak level at the end of our
observation. This is similar to the powerful flare seen in the
non-accreting pre-MS star LkH$\alpha$ 312 in the Orion B cloud
\citep{Grosso04}. The count rate is $\sim 15$ times higher during
flaring compared to the quiescent level before flaring.  The spectrum
is hard ($\medE=3.0$ keV), as seen in many pre-MS stars in COUP during
flares \citep{Favata05}. The spectral fitting gives a rather hard
$kT>10$ keV, corresponding to thermal plasma of $>120$ MK. The
infrared countpart for this X-ray source shows $K=10.2$ and its
location in the color-magnitude diagram suggests a high-mass star. The
SIRIUS images were visually examined and there is a possibility that
the flare originates from a poorly resolved low mass companion to the
high-mass star. If a future spectrum confirms that it is an early type
star, the flaring behavior is similar to that observed in the Orion
Nebula O9.5 star $\theta ^2$ Ori A \citep{Feigelson02}, either from an
extraordinary magnetic reconnection flare in a low mass companion, or
from non-standard wind shocks in the massive stellar winds
\citep{Feigelson02, Stelzer05}.

\subsection{X-ray Selected Deeply Embedded Population}
\label{embedded.sec}

In Table~\ref{tbl:Embedded} we assemble a sample of 16 heavily
obscured objects. These are sources that have $\langle E \rangle \ge
3.0$ keV (with $\ge 20$ net counts to be reliable) or $\log N_{H,X}
\ge 22.5$ cm$^{-2}$ derived from spectral fitting. This subsample of
the hardest sources (classified $H$ in Table~5) likely contains deeply
embedded protostars in addition to older stars heavily obscured by
molecular material.  Their spatial distribution is shown in
Figure~\ref{fig:embedded}, overlaid on a DSS optical image. Several of
them appear to be clustered in the optical/infrared dark column
(S.E.C. in the CO map) clearly seen in visible and 2MASS images
(Figure~1). Others are widely distributed along the rim of the
ringlike nebula. A plausible explanation is that the nebulosity traces
the rim of a bowl-shape shell facing us, thus the column density in
our line-of-sight to those objects is much higher. This geometric
configuration is consistent with the previous conclusion that
G353.2+0.9 is a blister HII region viewed face-on \citep{Massi97}.

\section{Diffuse X-ray Emission \label{diffuse.sec}}

With two of the most massive stars known and a luminous WR star in the
field, it is reasonable to expect diffuse X-ray emission in NGC 6357,
from strong stellar winds and parsec-scale wind-wind collisions
\citep{Townsley03}. A more recent study of X-ray diffuse emission in
the 30 Doradus star-forming complex is reported in
\citet{Townsley06b}, which provides our technical procedures to
extract diffuse emission.  From the point source locations and
extraction regions (Table~1 and 2), we can mask the point sources and
extract the spectrum of possible remaining diffuse emission. Two
regions of different scales are used for extraction, which yield
consistent measurements of the emission.  They are outlined by the
white and the cyan polygons in Figure~\ref{fig:ACIS_I}b.  The
background spectrum was scaled by the effective area and by the ratio
of the geometric areas to be appropriate for the region of diffuse
emission being fit; see Section 5.1 of Townsley et al.\ (2003) for
details. The spectra can be fit using a soft optically thin thermal
plasma with $kT\sim 0.6$ keV and $\log N_H \sim 22.1$ cm$^{-2}$.  The
corresponding observed soft band X-ray luminosity is $\log L_s\sim
32.4$ ergs s$^{-1}$ ($\log L_{s,c}\sim 33.6$ ergs s$^{-1}$ if
corrected for absorption). This is comparable to the observed diffuse
X-ray emission in M 17 and the Rosette Nebula \citep{Townsley03}.

However, the detected emission does not necessarily arise from
wind-generated plasma. Under the assumption that the gas temperature
within the cluster is relatively constant and hot ($>10^7$ K),
\citet{Stevens03} show that the luminosity level of the expected
diffuse emission in a stellar cluster can be estimated using a scaling
factor (equation 10), which \citet{Getman06} adopt to evaluate the
nature of X-ray diffuse emission in the Cepheus B region. We follow
the same method to obtain a crude estimate of diffuse emission from
the massive star winds
\footnote{The WC+O star WR 93 is not in the cluster core region, thus
we do not include it in the calculation. For reader's reference, the
mass loss rate and the terminal wind velocity for WR 93 is given in \S
2}.  Two O3 stars dominates the cluster core and two parameters are
needed to predict the X-ray luminosity: the total stellar mass loss
rate $\dot{M_*}$, and the mean weighted terminal velocity for the
stars in the cluster $v_*$. For the O3 star Pis 24-1,
\citet{Benaglia01} derived a terminal wind velocity of $v_{\infty}\sim
2295$ km s$^{-1}$ and an upper limit for the mass loss rate of
$\dot{M}\le 4.5\times 10 ^{-6}M_{\odot}$ yr$^{-1}$. No measurements
for Pis 24-17 is found in the literature, and we assume $v_{\infty}$
and $\dot{M}$ are comparable to Pismis 24-1.  Adopting these wind
parameters, and an unrealistically optimistic 100\% efficiency for
conversion from stellar wind to thermalization of the cluster, and
$\sim 2$ pc for the radius of the cluster core, we estimate the
diffuse X-ray luminosity $L_X<10^{31}$ ergs s$^{-1}$ (equation 8 and
10). This is apparently insufficient to account for the observed
luminosity.

An alternative explanation is that the diffuse emission is mainly the
contribution from unresolved X-ray emission from thousands of low mass
pre-MS members in Pismis 24.  Assuming that Pismis 24 and the ONC
have identical IMFs with identical XLFs, the only difference here is
that Pismis 24 cluster has 5 times the total known population of the ONC (see
\S 5.1). We focus on the lightly obscured populations since they are
less affected by the different amount of absorption along the two
sightlines. Scaling from COUP \citep{Feigelson05}, the total expected
soft-band X-ray luminosity is $\log L_{s,c}\sim 33.8$ ergs s$^{-1}$
from the entire Pismis 24 population and the actual observed soft
luminosity from resolved cluster members is $\log L_s\sim 33.2$ ergs
s$^{-1}$ (since absorption correction to individual stars is
infeasible, we consider this to be a lower limit to the intrinsic
soft-band luminosity $L_{s,c}$). Thus the integrated X-ray emission
from the unresolved X-ray emitting pre-MS population in Pismis 24 is
estimated to be $\log L_{s,c}\le 33.7$ ergs s$^{-1}$, after deducting
the observed $L_s$ from the expected $L_{s,c}$.  This is fully
consistent with the absorption-corrected soft diffuse X-ray luminosity
$\log L_{s,c}\sim 33.6$ ergs s$^{-1}$.  Therefore we suggest that most
of the diffuse emission is probably the combined contribution from
individually undetected X-ray-faint pre-MS cluster members.

We further note that X-ray emission is clearly present on the ACIS-S
chips. Several point sources are detected, with very large PSFs due to
large off-axis angles. The X-ray emission captured at the lower edge
of the S-array field corresponds to the location of the \hii\ region
G353.2+0.7 (see Figure~2b and Figure~\ref{fig:dss2mass}f). But due to
the poor PSF, no further conclusion can be drawn on whether the
emission is diffuse in nature.  Nevertheless, this detection confirms the
ROSAT detection of G353.2+0.7 and implies that it contains lots of
X-ray-emitting young stars.

\section{Summary \label{summary.sec}}

This 38 ks ACIS observation of the NGC 6357 field has provided the first
high spatial resolution X-ray image of this massive star-forming
complex. We summarize the main results of our study as follows:

1. We detect 779 X-ray sources with a limiting X-ray sensitivity
   of $\sim 10^{30}$ ergs s$^{-1}$. There are 445 ACIS sources matched
   to 2MASS point sources with $\le 2^{\prime\prime}$
   separation. Further visual inspection of an archival HST image and
   SIRIUS $JHK$ images yield a total of 616 ONIR counterparts. We
   estimate the extragalactic contaminants via a careful simulation of
   detection efficiency of the background AGN population, and conclude
   that no more than 4\% of our detections are AGNs. Similarly we
   adopt the Besan\c{c}on stellar population sythesis model of the Galactic
   disk and suggest 11 stars as foreground objects based on their
   X-ray softness and NIR color. Excluding $\sim 20$ X-ray sources
   without ONIR identifications that are possible AGNs, the rest of the
   $\sim 140$ X-ray sources without ONIR counterparts are likely new
   Pismis 24 members that are clustered around the massive stars and/or
   deeply embedded in the cloud. The X-ray detected population provides
   the first deep probe of the rich population of this massive
   cluster, increasing the number of known members from optical study
   by a factor of $\sim 50$.

\begin{figure}
 \centering
 \includegraphics[width=0.8\textwidth]{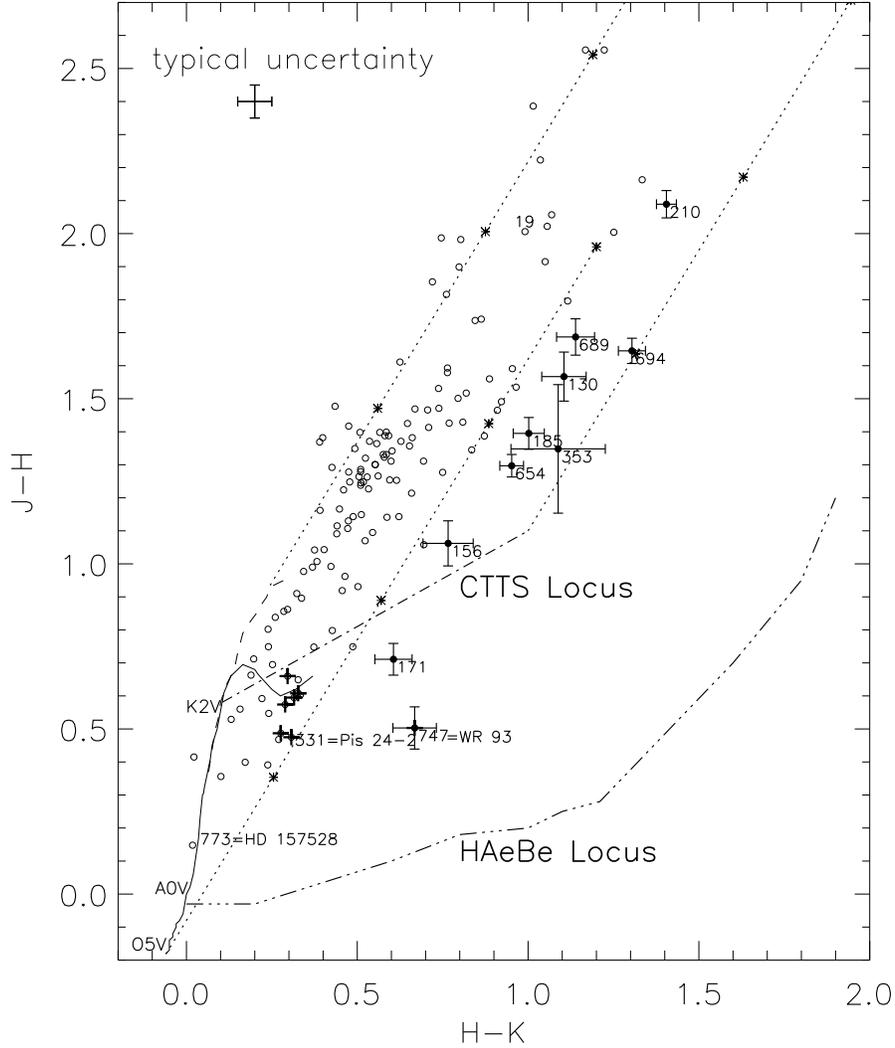}
\caption{Color-color diagram of X-ray stars with high quality 2MASS
counterparts. Open circles and filled circles represent sources
without and with significant $K$-band excess ($E(H-K) > 1\sigma$
uncertainty in $H-K$ color), respectively. Crosses mark the known
members of the Pismis 24 cluster from \citet{Massey01}. ACIS source
numbers and error bars are given for the counterparts with significant
$K$-band excess. The interesting source \#19, the Pis24 member \#331,
and the foreground star \#773 are also labeled to guide the
reader. The solid and long-dash lines denote the locus of main
sequence stars and giants from \citet{Bessell88}, respectively. The
dash dotted line is the locus for T Tauri stars from \citet{Meyer97}
and the dash triple-dotted line is the locus for HAeBe stars from
\citet{Lada92}. The dash lines represent the standard reddening vector
for $A_V=20$ mag, with asterisks marking every $A_V=5$ mag. Many
additional X-ray stars with lower quality 2MASS photometry are
omitted.
\label{fig:ccd}}
\end{figure}

\begin{figure}
 \centering
\includegraphics[width=.8\textwidth]{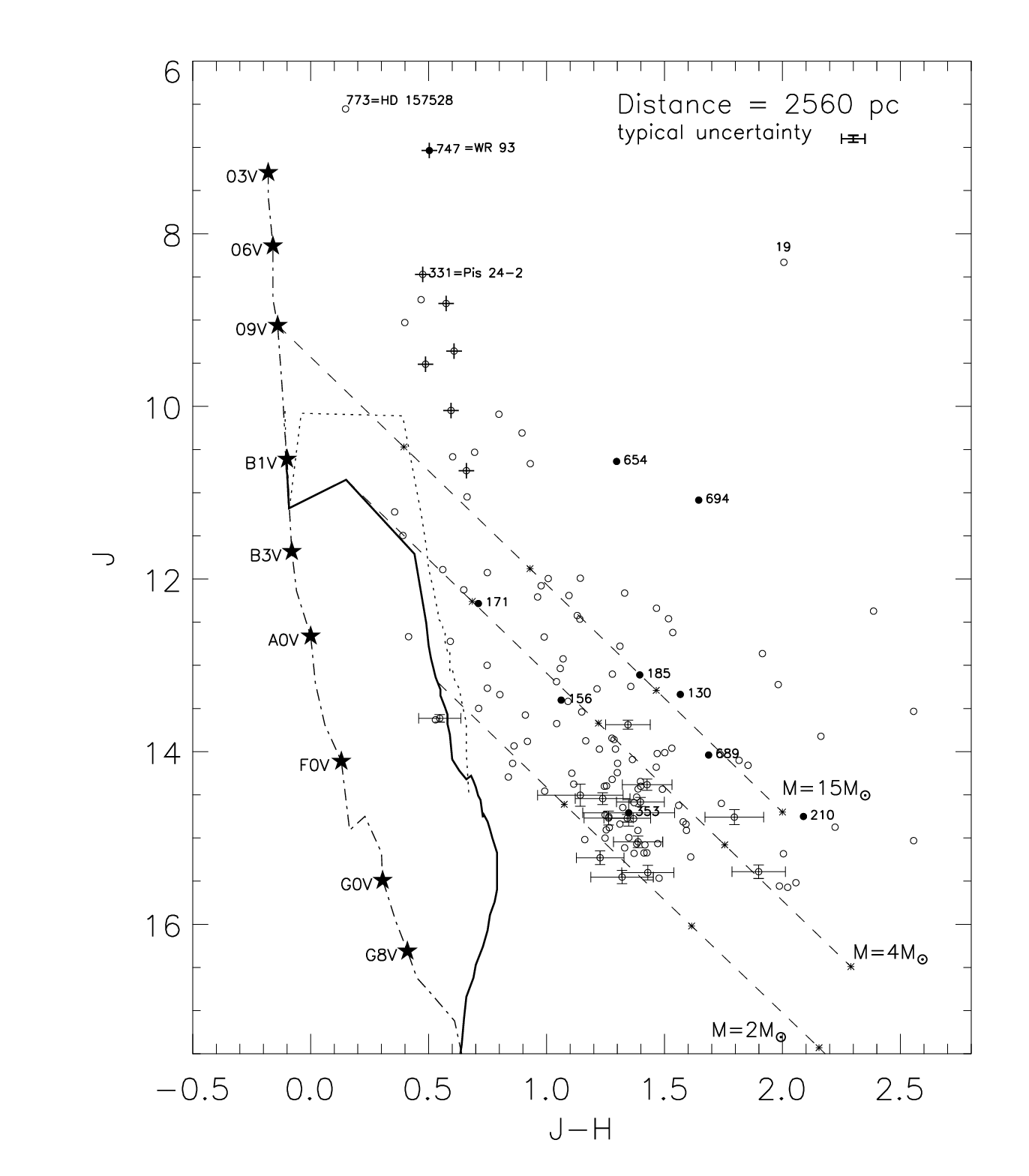}
\caption{Infrared $J-H$ vs. $J$ color-magnitude diagram. The solid
line is the 1 Myr isochrone for pre-MS stars from \citet{Baraffe98}
and \citet{Siess00}. A 0.3 Myr isochrone \citep{Siess00} is also shown
(dotted line) to demonstrate how the estimated masses differ if we
adopt a younger age for the cluster. The dash dotted line marks the
location of Zero Age Main Sequence (ZAMS) stars. Dashed lines
represent the standard reddening vector with asterisks marking every
$A_V=5$ mag; the corresponding stellar masses are marked. ACIS numbers
for sources with significant $K$-excess are shown. The interesting
source \#19, the Pis24 member \#331, and the foreground star \#773 are
labeled to guide the reader. Error bars are shown for all sources with
color uncertainties $> 0.1$ mag. This diagram uses the same source
sample and symbols as Figure~6.
\label{fig:cmd}}
\end{figure}

\begin{figure}
 \centering
\includegraphics[width=1.0\textwidth]{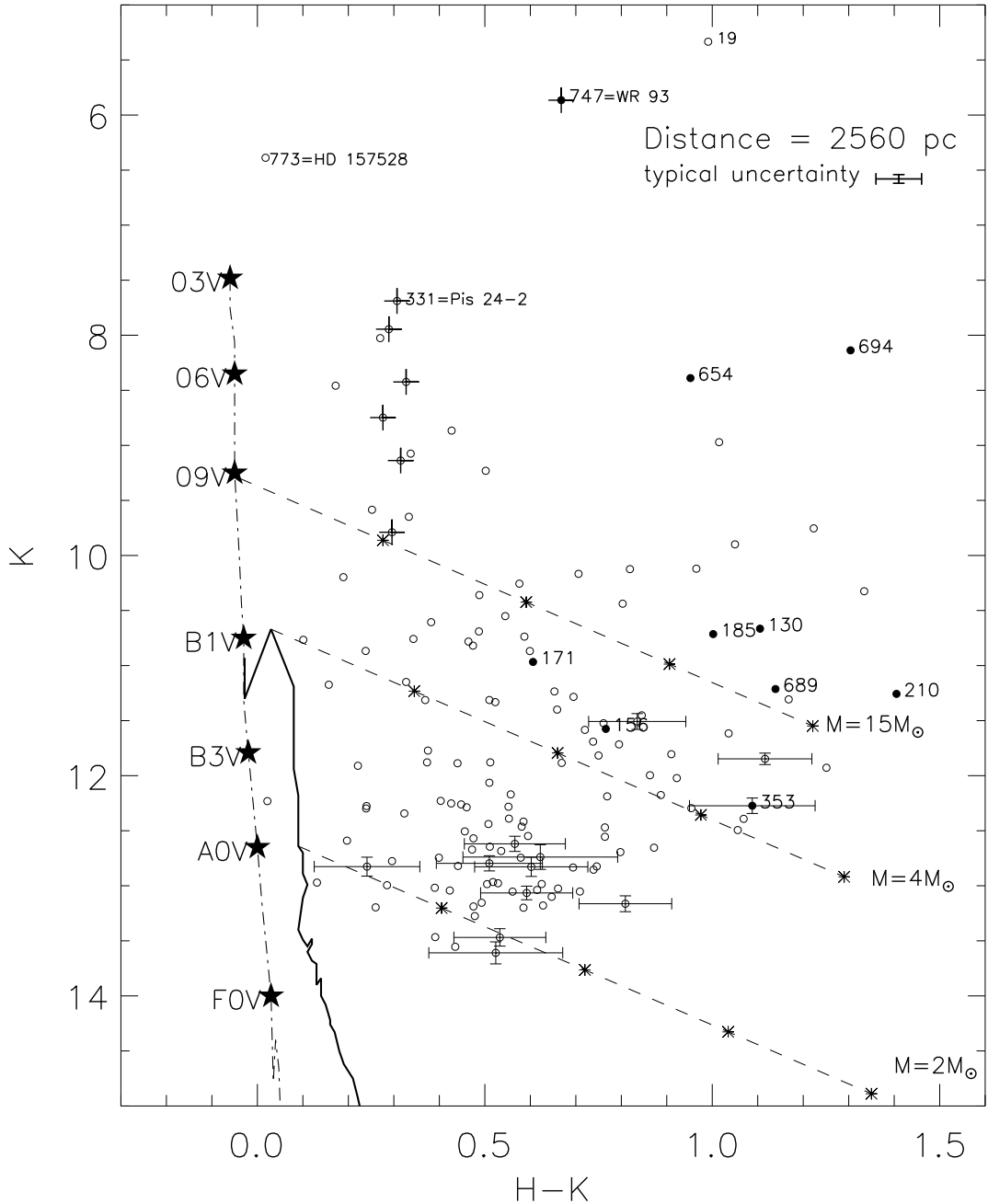}
\caption{Infrared $H-K$ vs. $K$ color-magnitude diagram. The solid
line is the 1 Myr isochrone from \citet{Baraffe98} and
\citet{Siess00}. The dash dotted line marks the location of ZAMS
stars. Dashed lines represent the standard reddening vector with
asterisks marking every $A_V=5$ mag; the corresponding stellar masses
are marked. ACIS numbers for sources with significant $K$-excess are
shown. The interesting source \#19, the Pis24 member \#331, and the
foreground star \#773 are labeled to guide the reader. Error bars are
shown for all sources with color uncertainties $> 0.1$ mag.  This
diagram uses the same source sample and symbols as Figure~6.
\label{fig:cmd2}}
\end{figure}

2. On the NIR color-color diagram (Figure~\ref{fig:ccd}), it is clear
   that ACIS detects more Class III sources than Class II/I
   sources. In the color-magnitude plot (Figure~\ref{fig:cmd}), the
   locations of most of the ACIS sources are consistent with being
   high mass stars and pre-MS intermediate mass stars ($M \ge
   2M_{\odot}$), subjected to a visual extinction of $5<A_V<15$ at 2.5
   kpc. The new high mass stars, if spectroscopically confirmed,
   double the known high mass population of the cluster. Dozens of
   intermediate mass stars are newly recognized members of the
   cluster. The ratio between Class III and Class II objects in the
   intermediate- to high-mass range is $\sim 15:1$. This is an
   important new sample for stellar and disk studies of intermediate
   mass pre-MS stars. The brightness of several K-excess sources makes
   them extremely favorable for ground based follow-up. We find a very
   luminous X-ray emitting infrared source subjected to significant
   absorption, which may be a new obscured massive YSO or a Wolf-Rayet
   star.

3. We determine the spatial distribution of over 700 new members to
   sub-arcsecond precision for the first time. The cluster is roughly
   spherically symmetric, with the highest stellar density around the
   massive O3 stars. A density enhancement centered at the tip of the
   molecular filament deserves further investigation
   (Figure~\ref{fig:stellar_density}). A radial profile of the stellar
   surface density shows that the physical size of the cluster spans
   over 2 pc in radius and extends beyond 6 pc (Figure~\ref{fig:offaxis_hist}).

\begin{figure}
 \centering
\plottwo{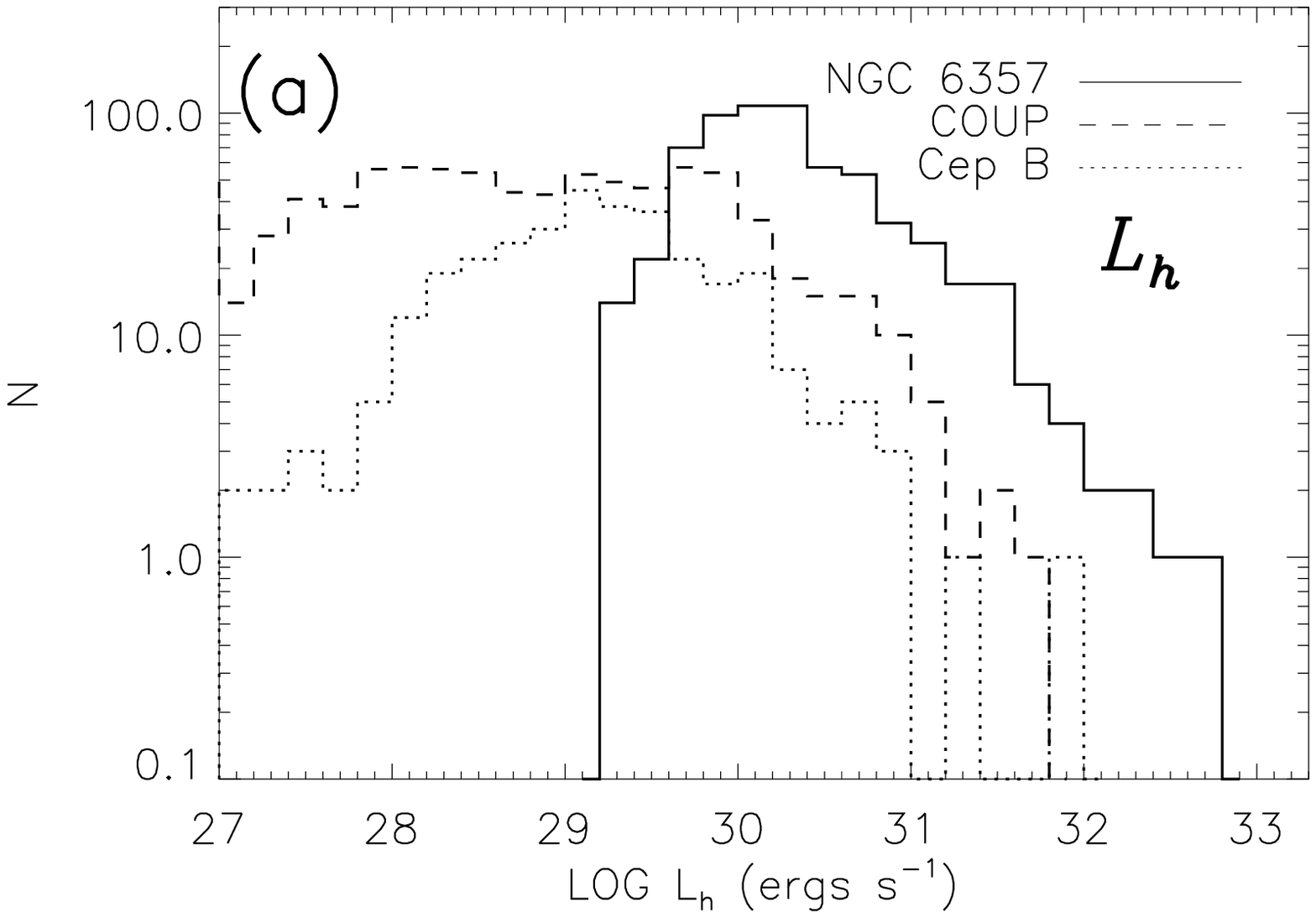}{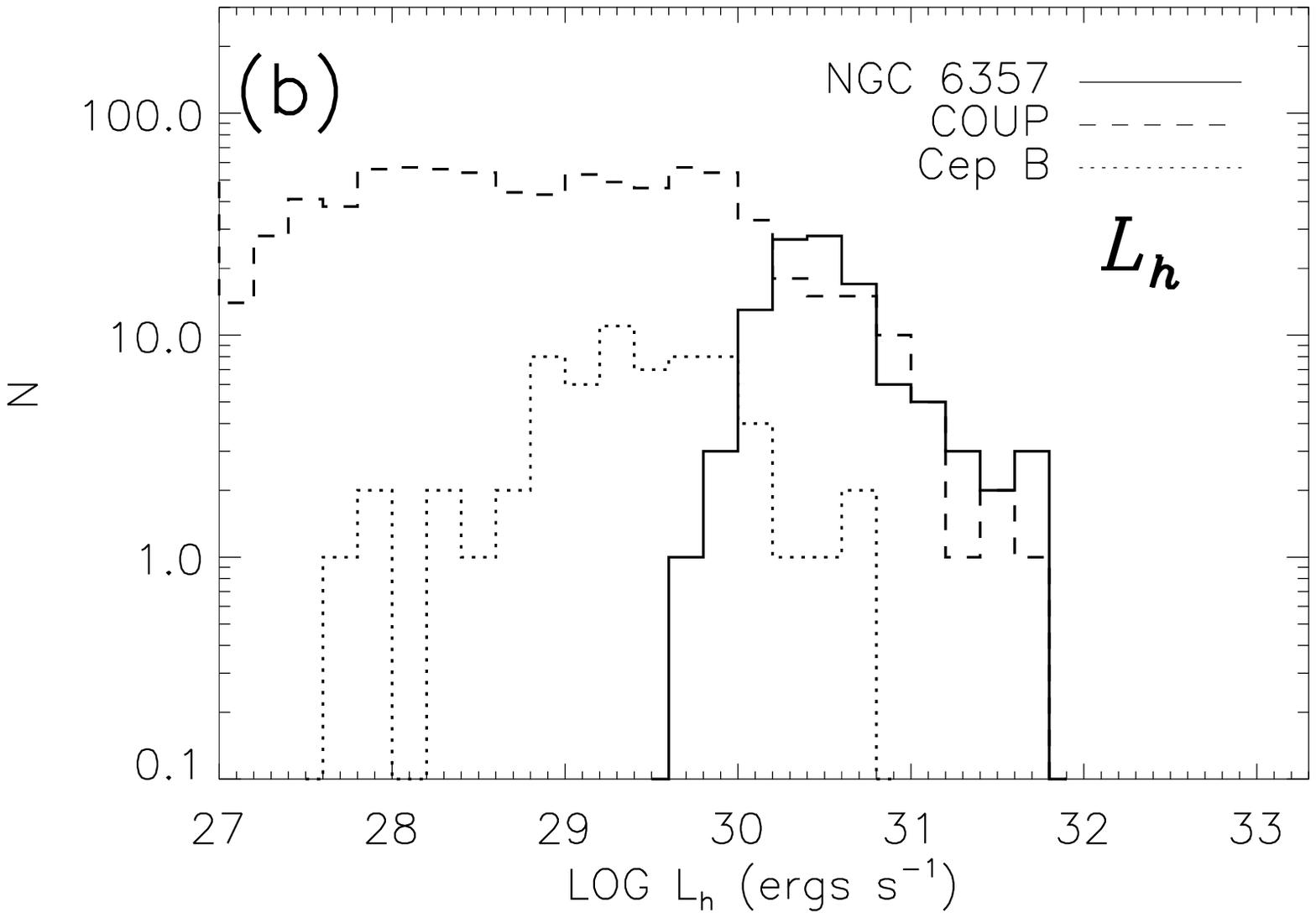}
\plottwo{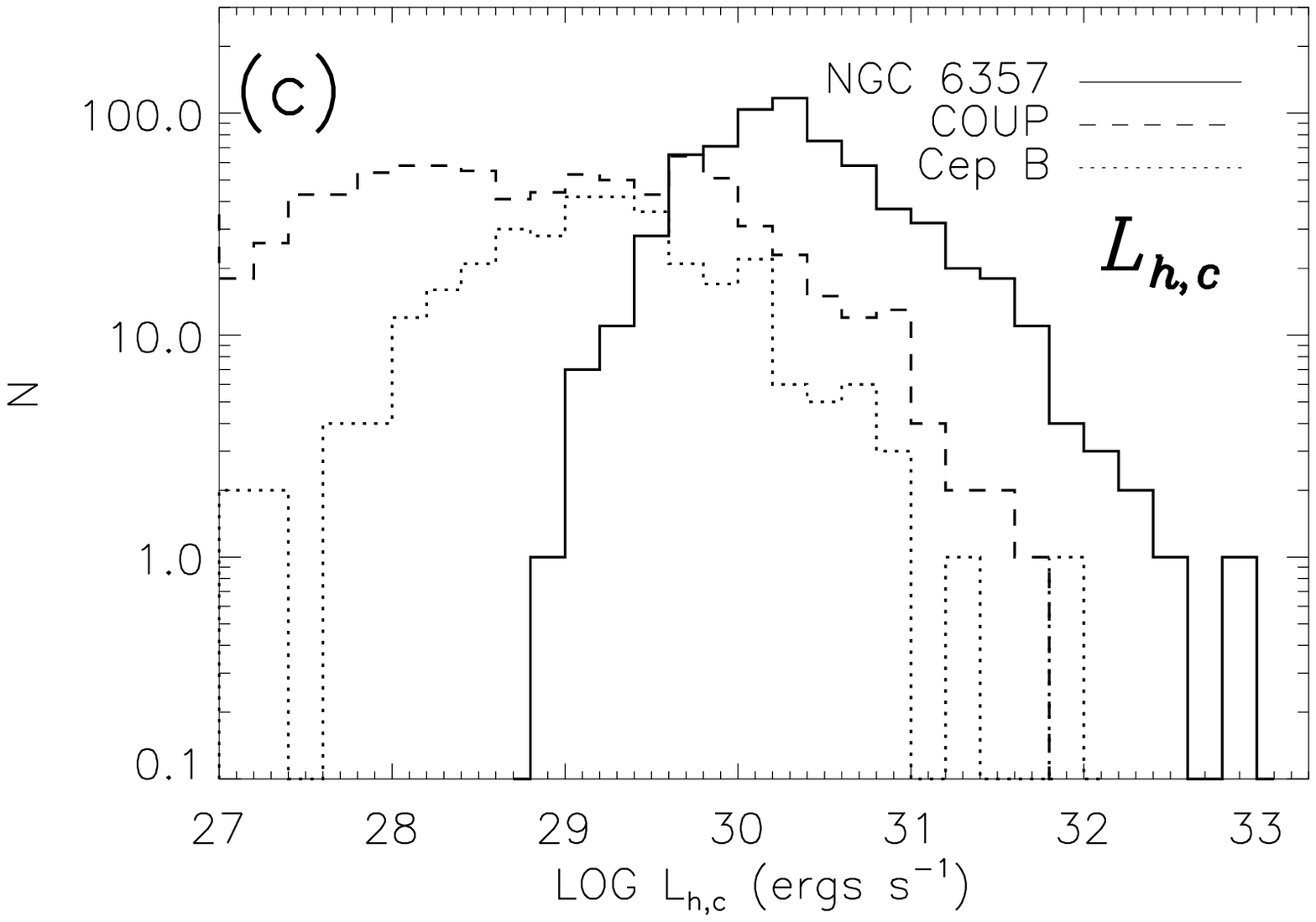}{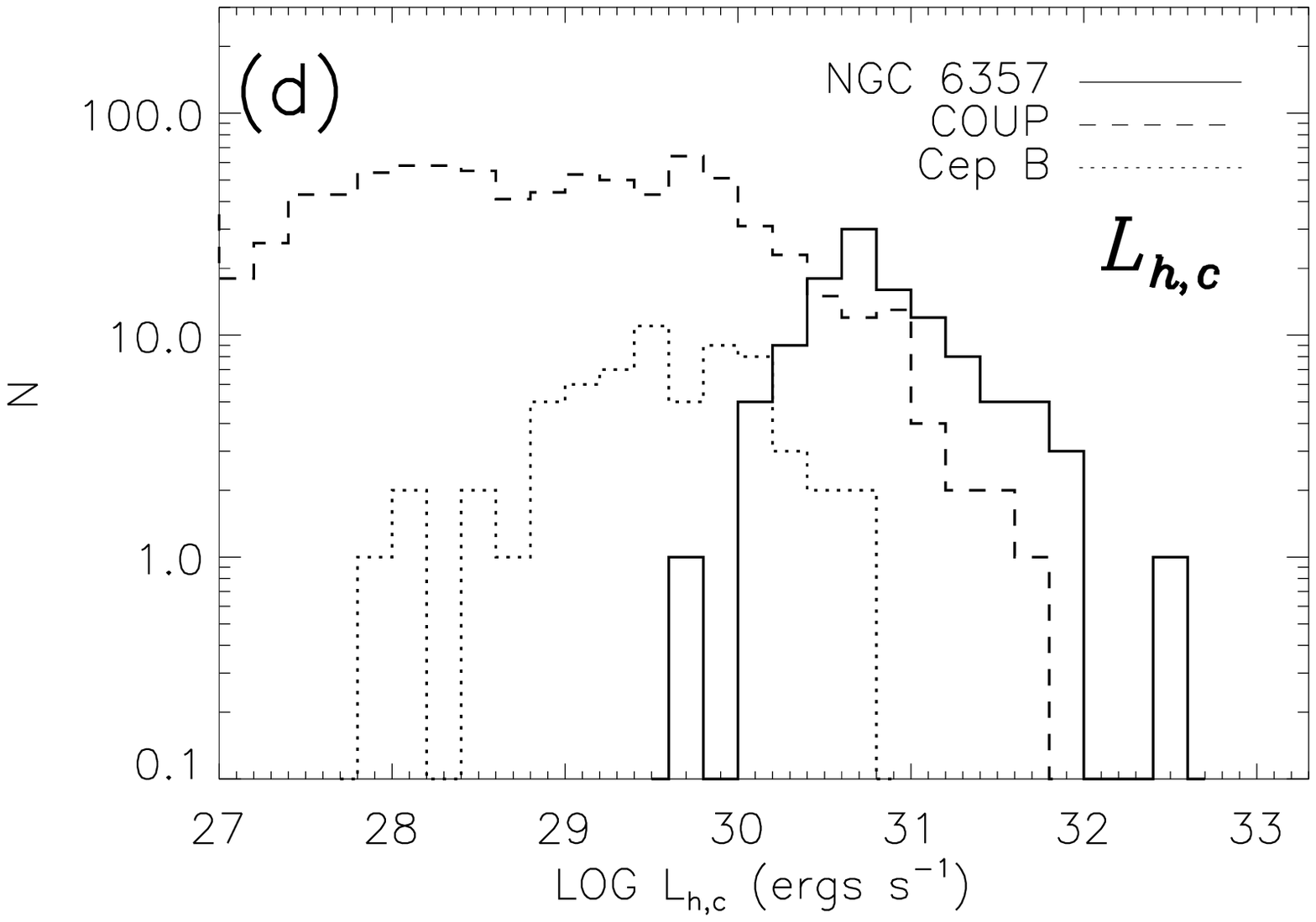}
\caption{(a): X-ray luminosity function (XLF) constructed from the
observed (not corrected for absorption) hard band (2.0--8.0 keV) X-ray
luminosity $L_h$ for the lightly obscured population in NGC 6357
(solid line), Orion \citep[dashed line,][]{Feigelson05}, and Cep B
\citep[dotted line,][]{Getman06}. (b): XLF using $L_h$ for the heavily
obscured population in NGC 6357 and Cep B compared to the XLF for the
lightly obscured population in Orion. (c): XLF using absorption
corrected hard band (2.0--8.0 keV) X-ray luminosity $L_{h,c}$ for the
lightly obscured population in NGC 6357, Orion, and Cep B. (d): XLF
using $L_{h,c}$ for the heavily obscured population in NGC 6357 and
Cep B compared to the XLF for the lightly obscured population in Orion.
 \label{fig:XLF}}
\end{figure}

\begin{figure}
 \centering
 \includegraphics[width=1.0\textwidth]{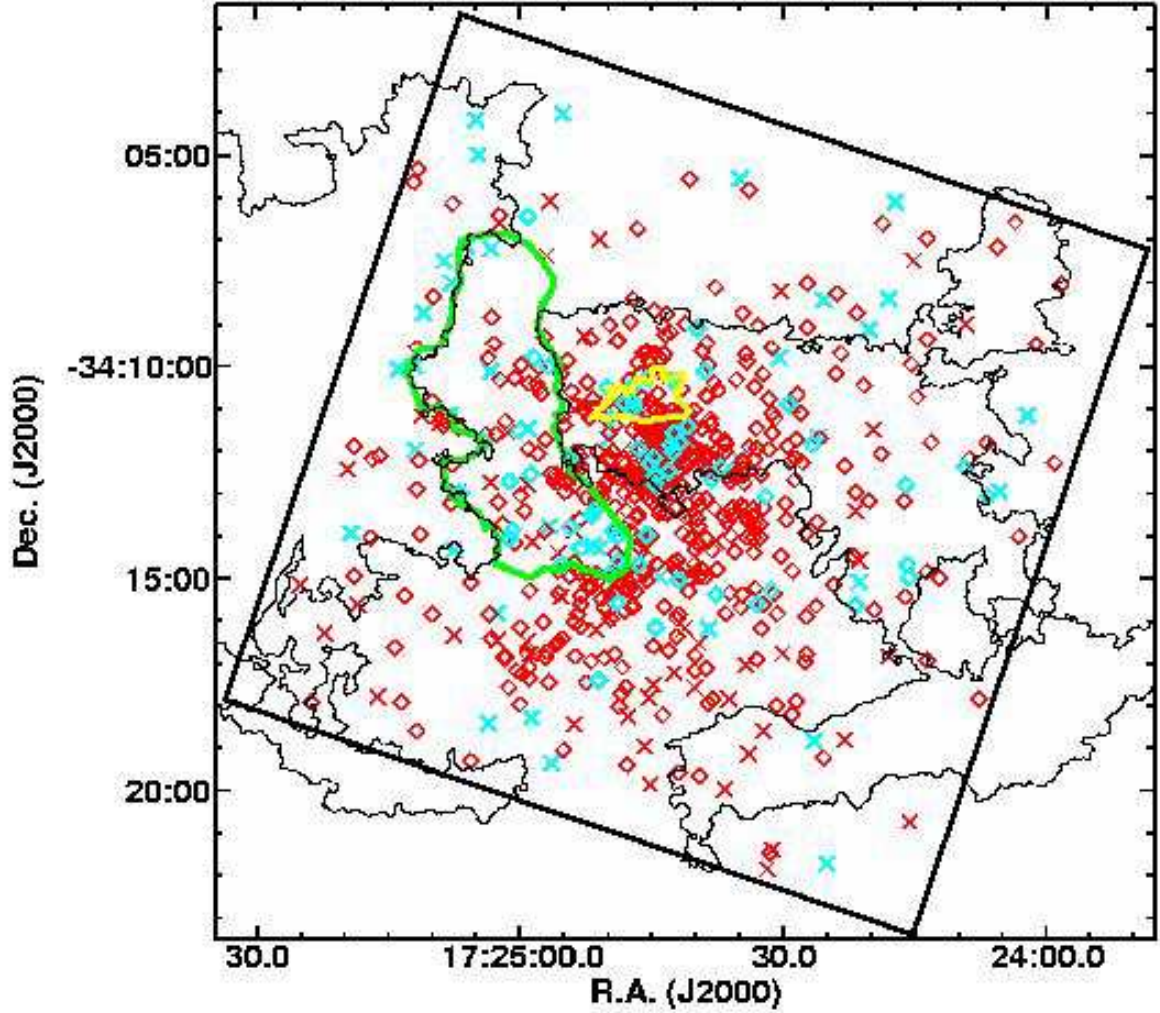}
\caption{The spatial distribution of the X-ray stellar
sources. Members with ONIR counterparts are plotted as diamonds, and
likely members without ONIR identifications are shown with
crosses. Obscured sources ($\medE \ge 3.0$ keV) are plotted in blue
and less obscured sources ($\medE < 3.0$ keV) are plotted in red. The
large scale optical nebulosity contour from the DSS image traces the
mid-IR ring-like morphology; it is shown in black together with the
outlines of the ionization front in yellow and the CO ``South-Eastern
Complex'' in green.
 \label{fig:radec}}
\end{figure}

\begin{figure}
 \centering
 \plottwo{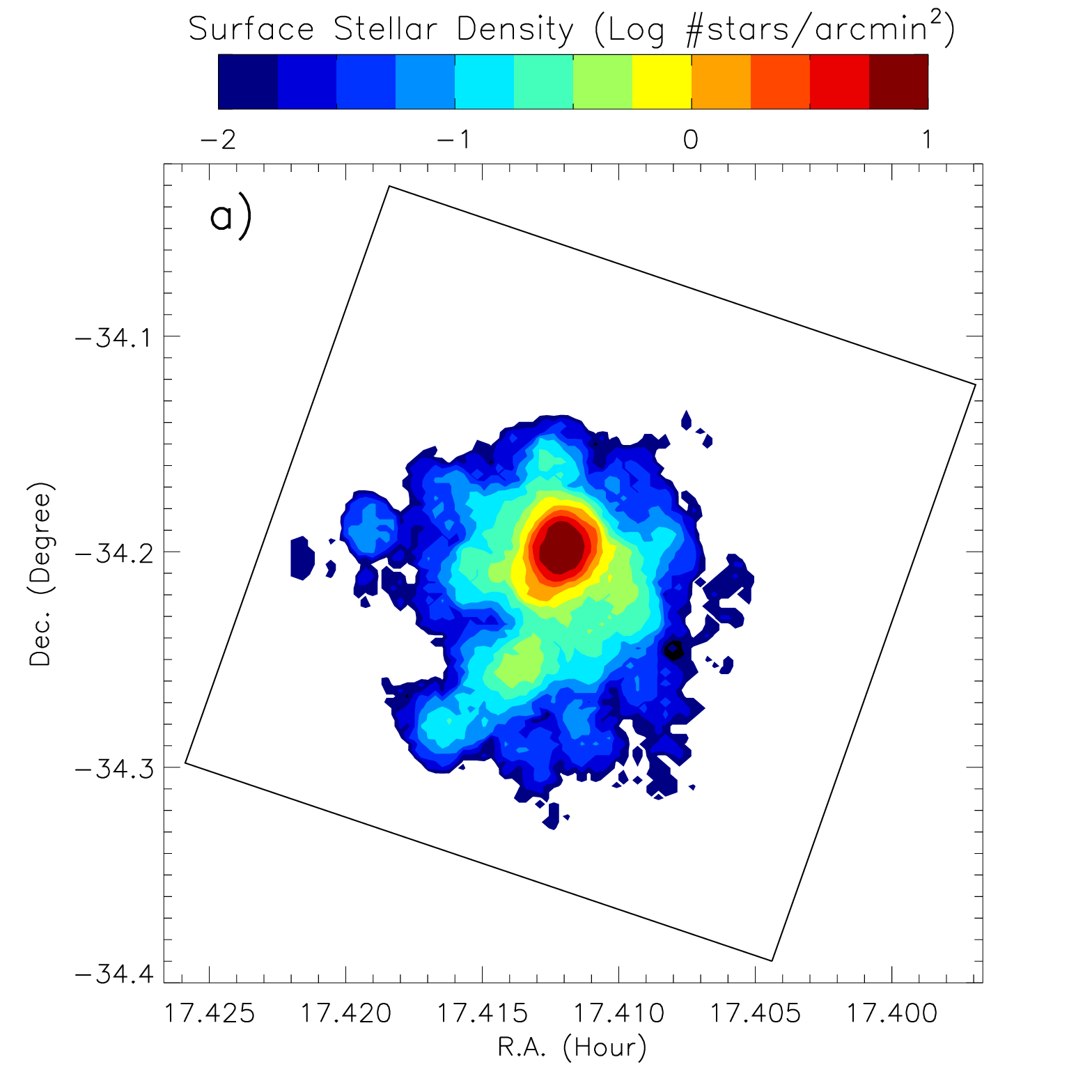}{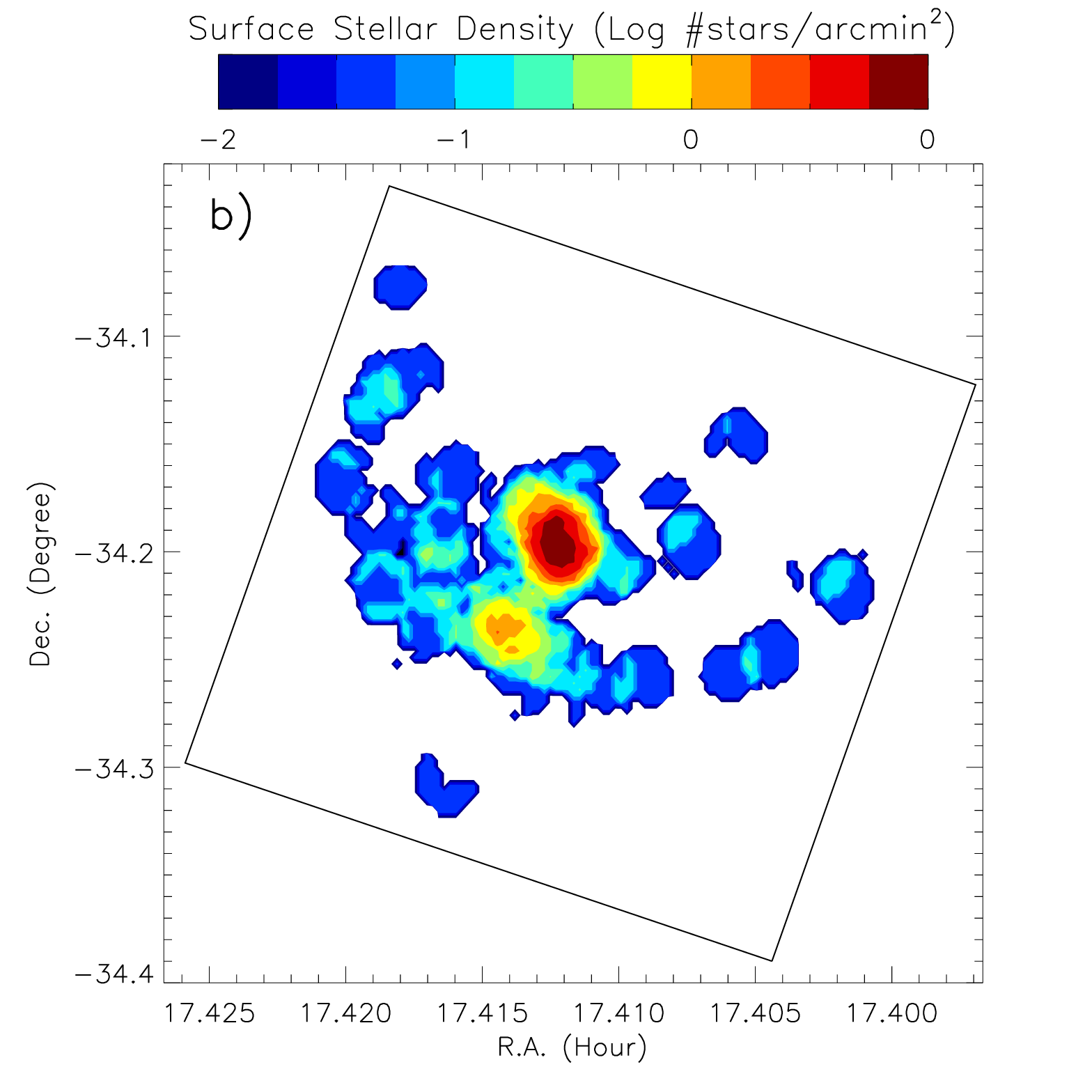}
\caption{Projected stellar surface density (number of stars
arcmin$^{-2}$) maps for (a) the unobscured population and (b) the
obscured population, shown in logarithmic scale using a $0^{\prime}.5$
radius smoothing kernel.  The highest concentration of stars coincides
with the core of the cluster, where the massive O stars are located. A
density enhancement, or a subcluster is seen centered at RA=17.414
(hours), Dec=-34.24 (degrees) in (b). There is a small overdensity
around the WR star, at RA=17.419 (hours), Dec=-34.19 (degrees) in (a).
\label{fig:stellar_density}}
\end{figure}

\begin{figure}
 \centering
\includegraphics[width=0.7\textwidth]{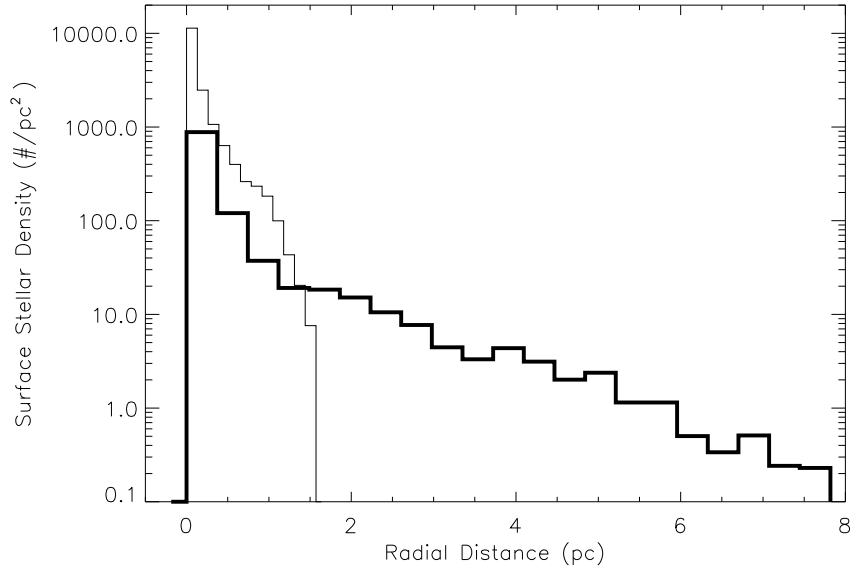}
\caption{Radial density profiles for X-ray stars in the Orion Nebula
Cluster (COUP sample, Getman et~al.\ 2005; thin line) and for Pismis 24 (thick
line). COUP sample is spatially complete in the central 1 pc of Orion
Nebula Cluster.
 \label{fig:offaxis_hist}}
\end{figure}

\begin{figure}
 \centering
\includegraphics[width=0.7\textwidth]{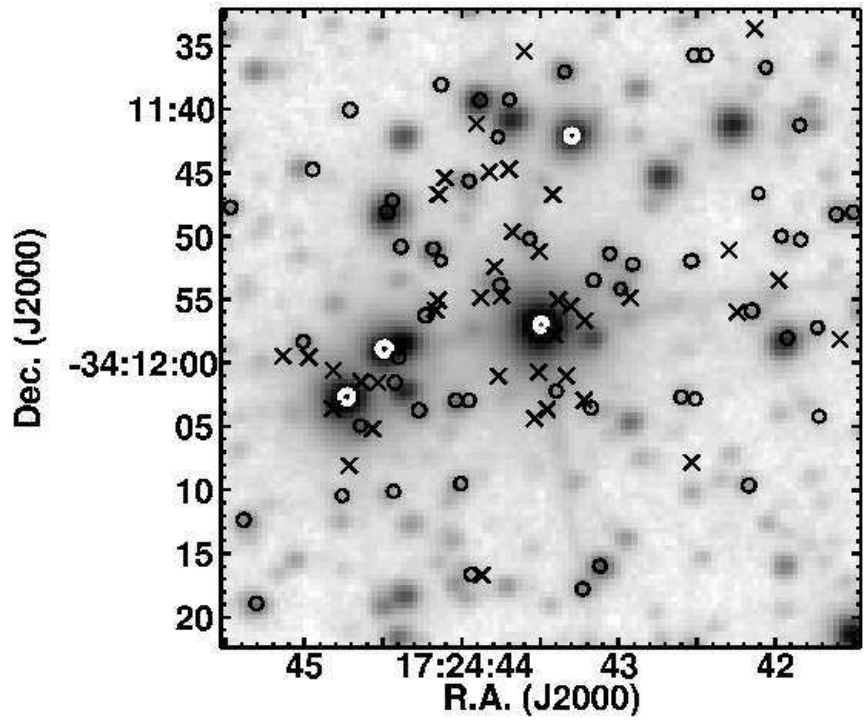}
 \caption{The distribution of X-ray sources overlaid on the central
$50^{\prime\prime}$ SIRIUS K image. The ONIR identified sources (circles), and
unidentified sources (crosses) are shown, with the known massive
members highlighted in white.
 \label{fig:radec_zoomin}}
\end{figure}

\begin{figure}
 \centering

 \includegraphics[angle=-90,width=0.35\textwidth]{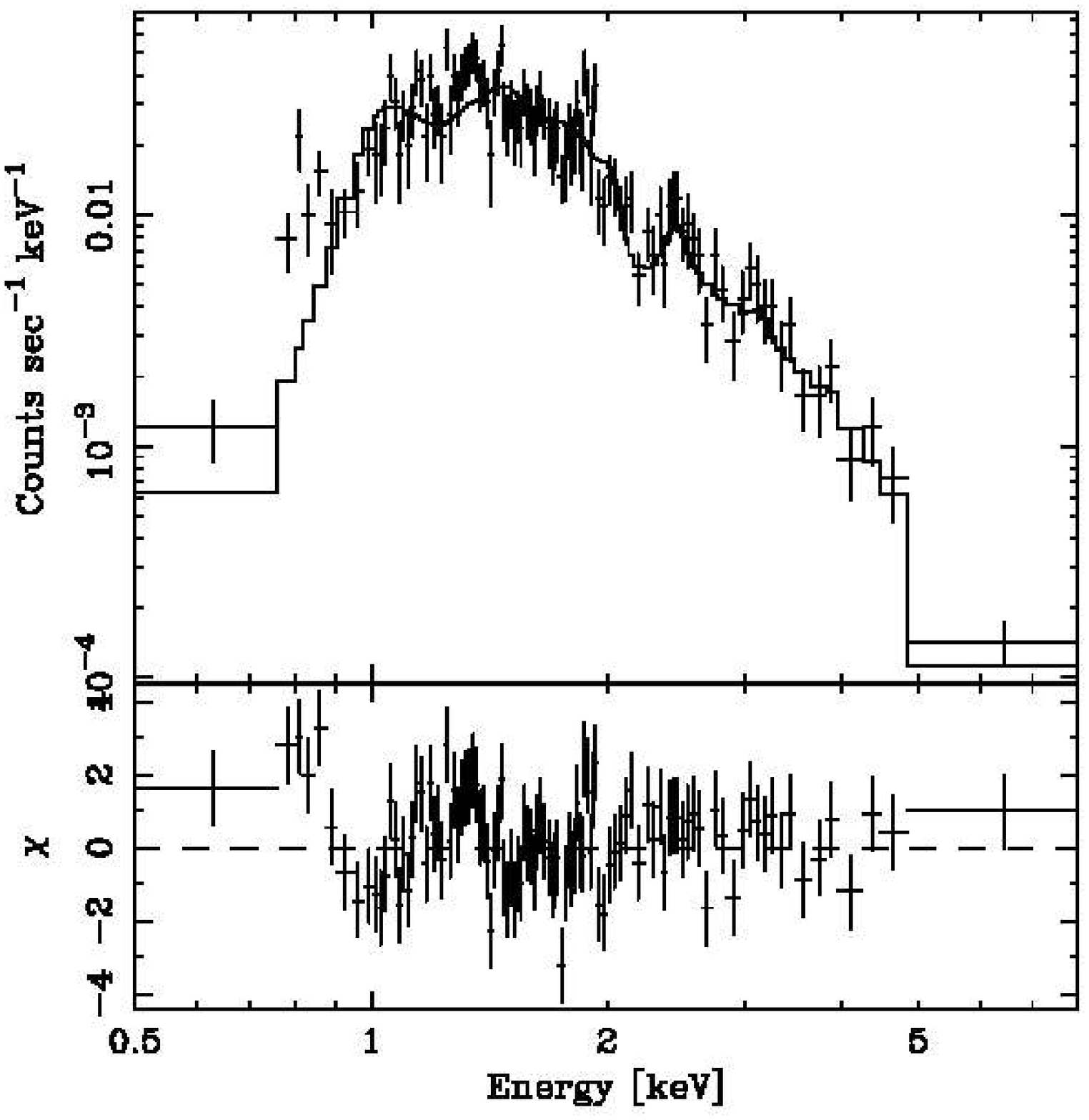}
 \includegraphics[angle=-90,width=0.35\textwidth]{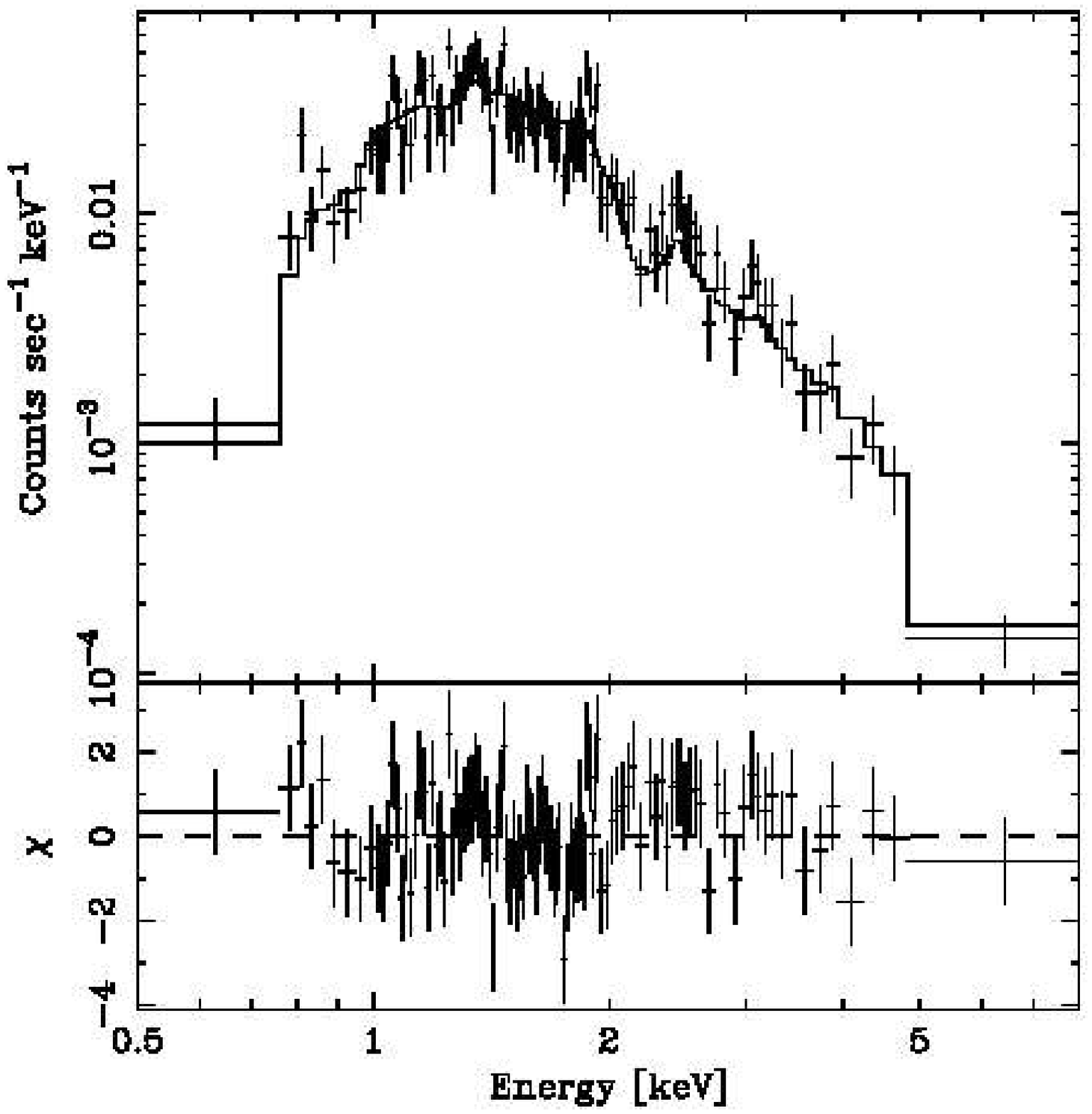}
 \includegraphics[angle=-90,width=0.35\textwidth]{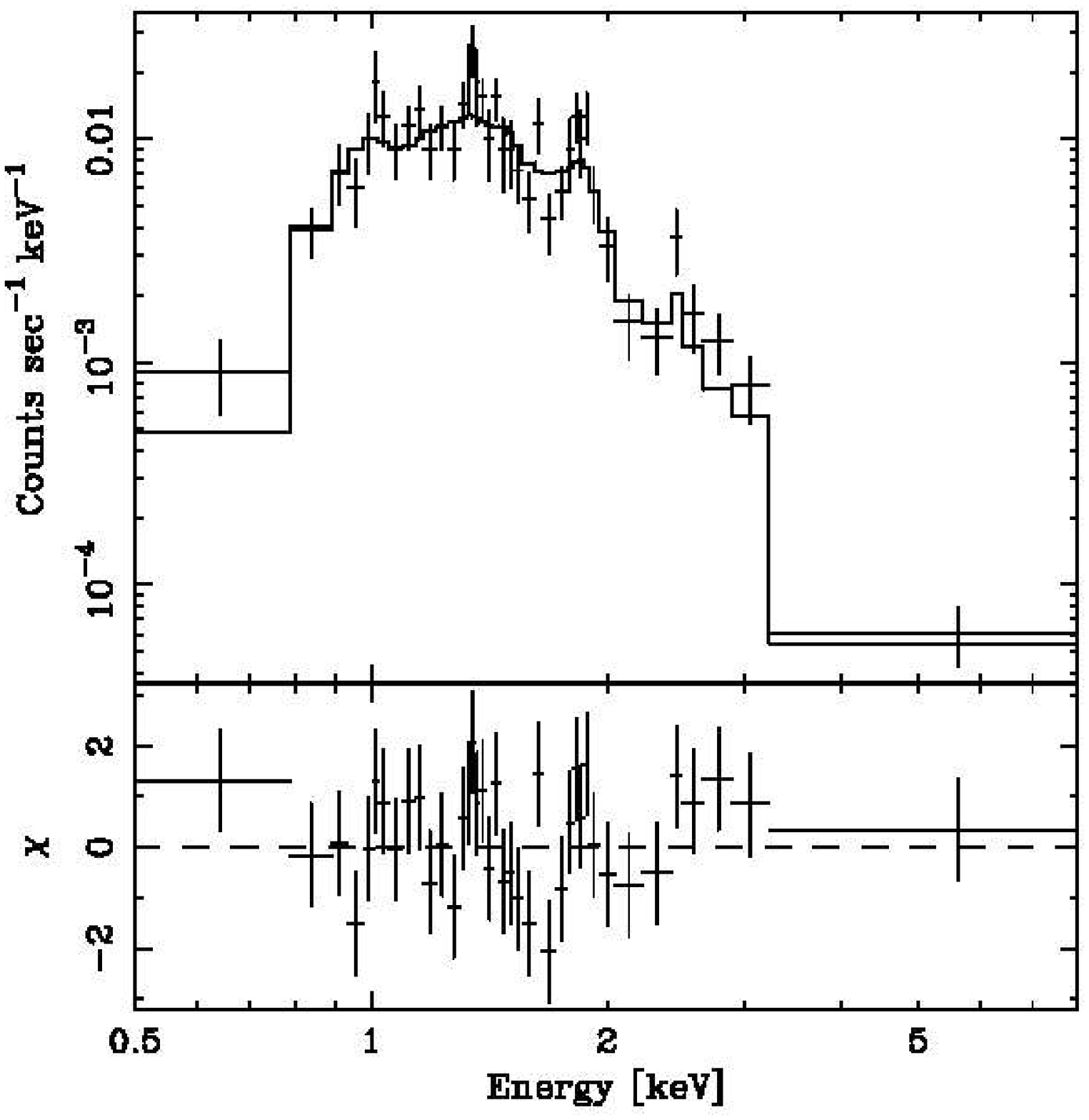}
 \includegraphics[angle=-90,width=0.35\textwidth]{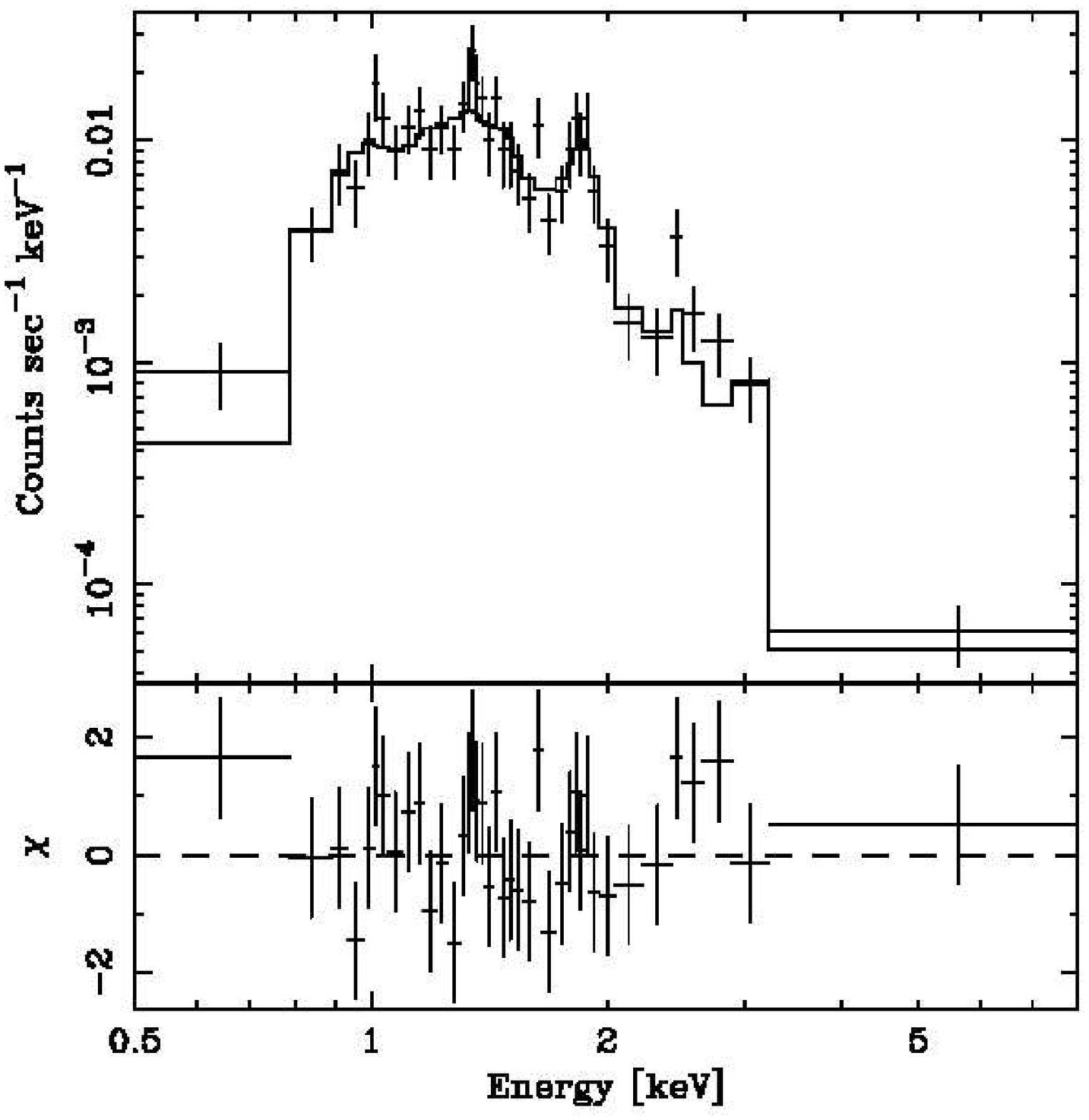}
 \includegraphics[angle=-90,width=0.35\textwidth]{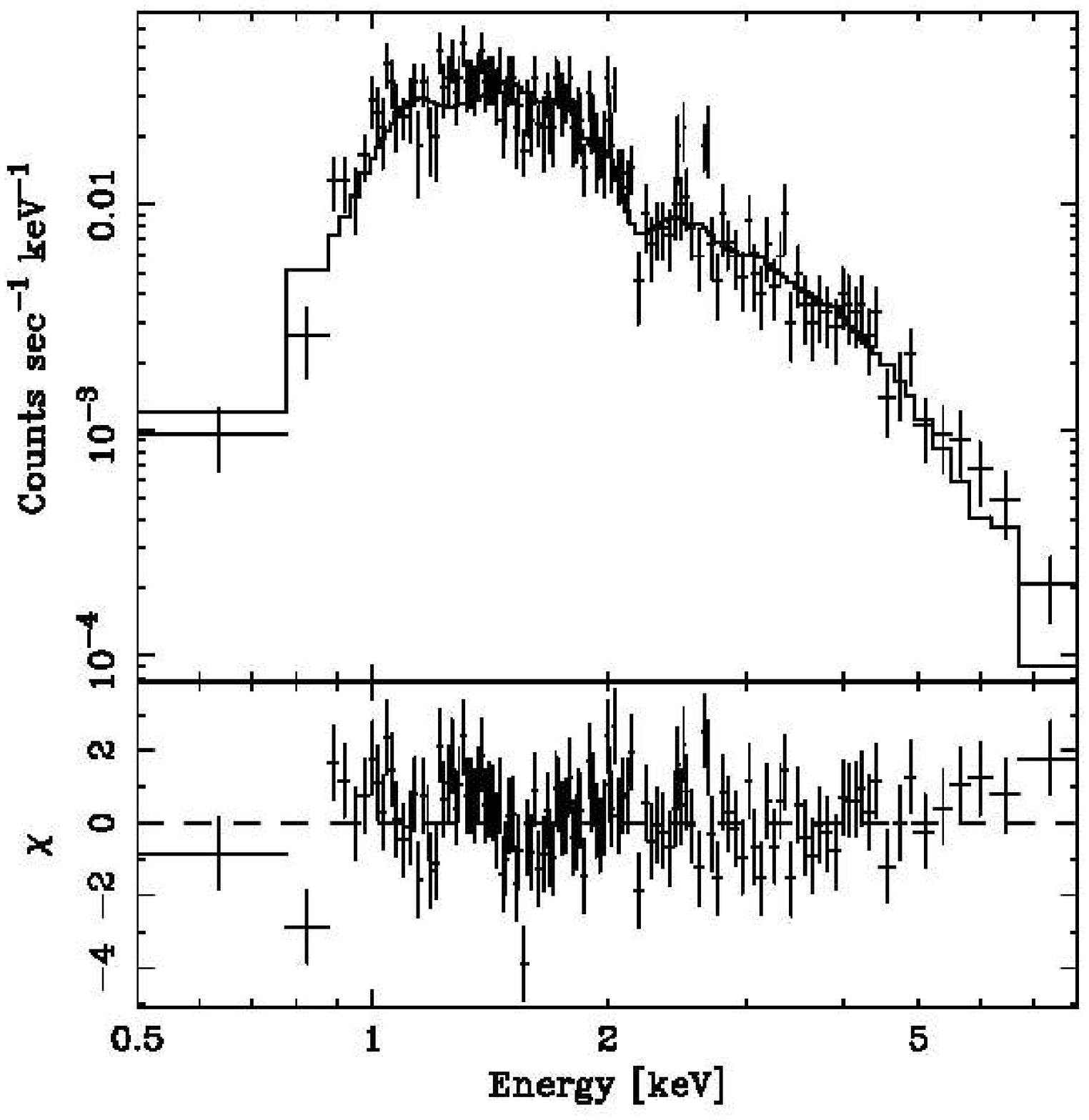}
 \includegraphics[angle=-90,width=0.35\textwidth]{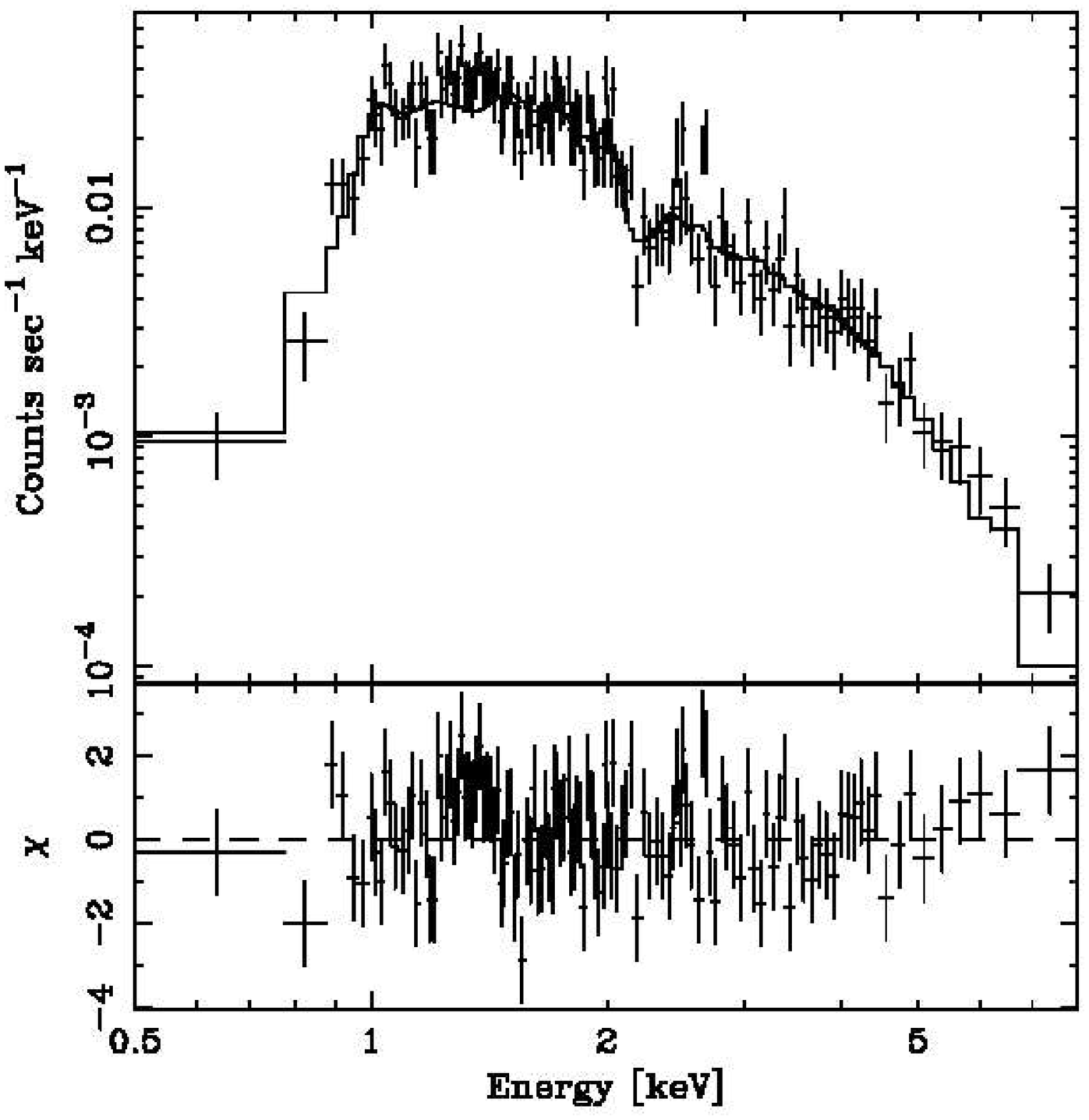}
\caption{X-ray spectra of the O3 stars and the brightest source
WR93. The left three panels give {\it apec} model fits for Pis24-1
($N_H=1.1\times 10^{22}$ cm$^{-2}$ and $kT=1.2$ keV), Pis24-17
($N_H=1.2\times 10^{22}$ cm$^{-2}$ and $kT=0.7$ keV), and WR 93
($N_H=6.9\times 10^{21}$ cm$^{-2}$ and $kT=2.3$ keV). Top right shows
an improved fit with a two temperature thermal plasma fit for Pis24-1
($N_H=1.3\times 10^{22}$ cm$^{-2}$, $kT_1=0.5$ keV and $kT_2=1.7$
keV). Middle right gives a slightly improved fit for Pis24-17 with a
variable abundance thermal plasma fit ($N_H=1.4\times 10^{22}$
cm$^{-2}$ and $kT=0.7$ keV). Bottom right presents an improved fit for
WR 93 with the {\it vapec} model ($N_H=7.4\times 10^{21}$ cm$^{-2}$
and $kT=2.4$ keV).
 \label{fig:O3s}}
\end{figure}

\begin{figure}
 \centering
\includegraphics[width=1.0\textwidth]{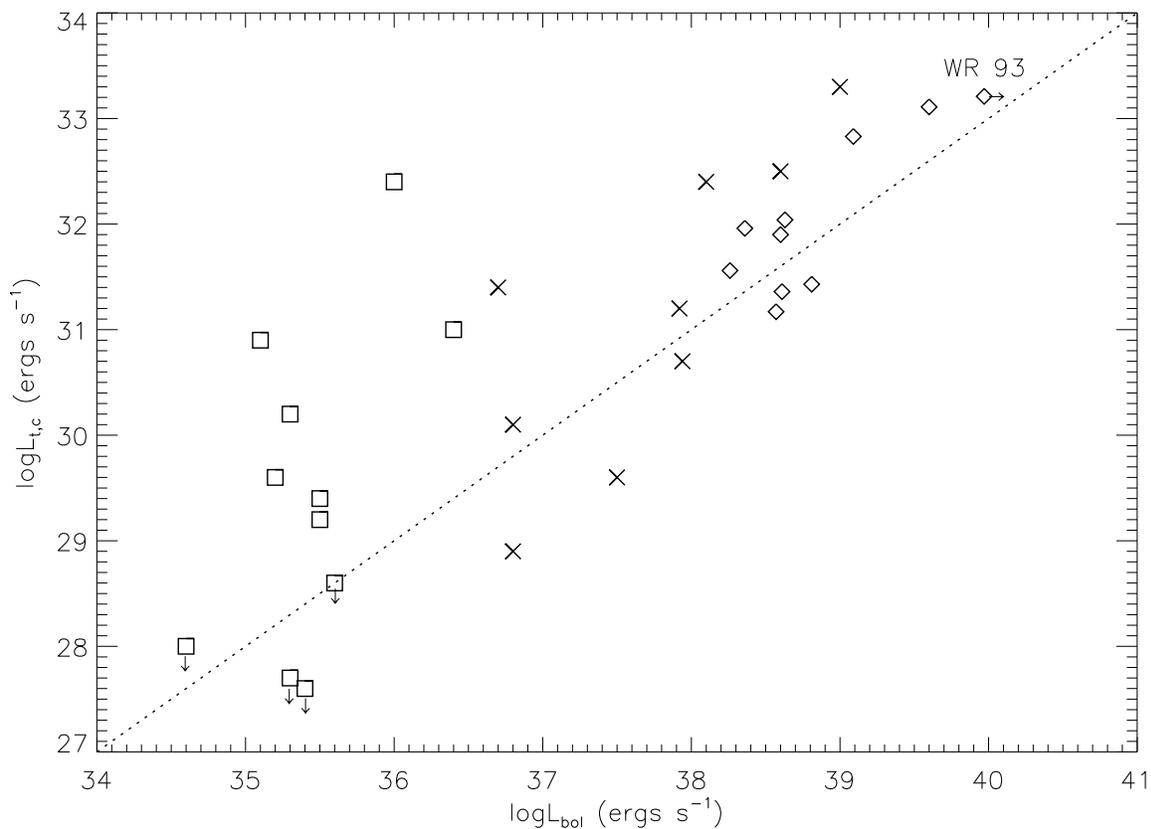}
\caption{The $L_X$ vs. $L_{bol}$ relation for the X-ray detected OB
stars in Pismis 24 (diamonds) compared to COUP massive stars with
strong winds (crosses) and weak winds (squares).  Four COUP stars with
weak winds have upper limits in $L_X$ reported in
\citet{Stelzer05}. The $L_{bol}$ for WR 93 is a lower limit due to
uncertain bolometric correction for late-type WC stars. The dashed
line represents the canonical relation $\log (L_X/L_{bol})\sim -7$.
 \label{fig:LxLbol}}
\end{figure}

\begin{figure}
 \centering
 \epsscale{0.5}
 \plotone{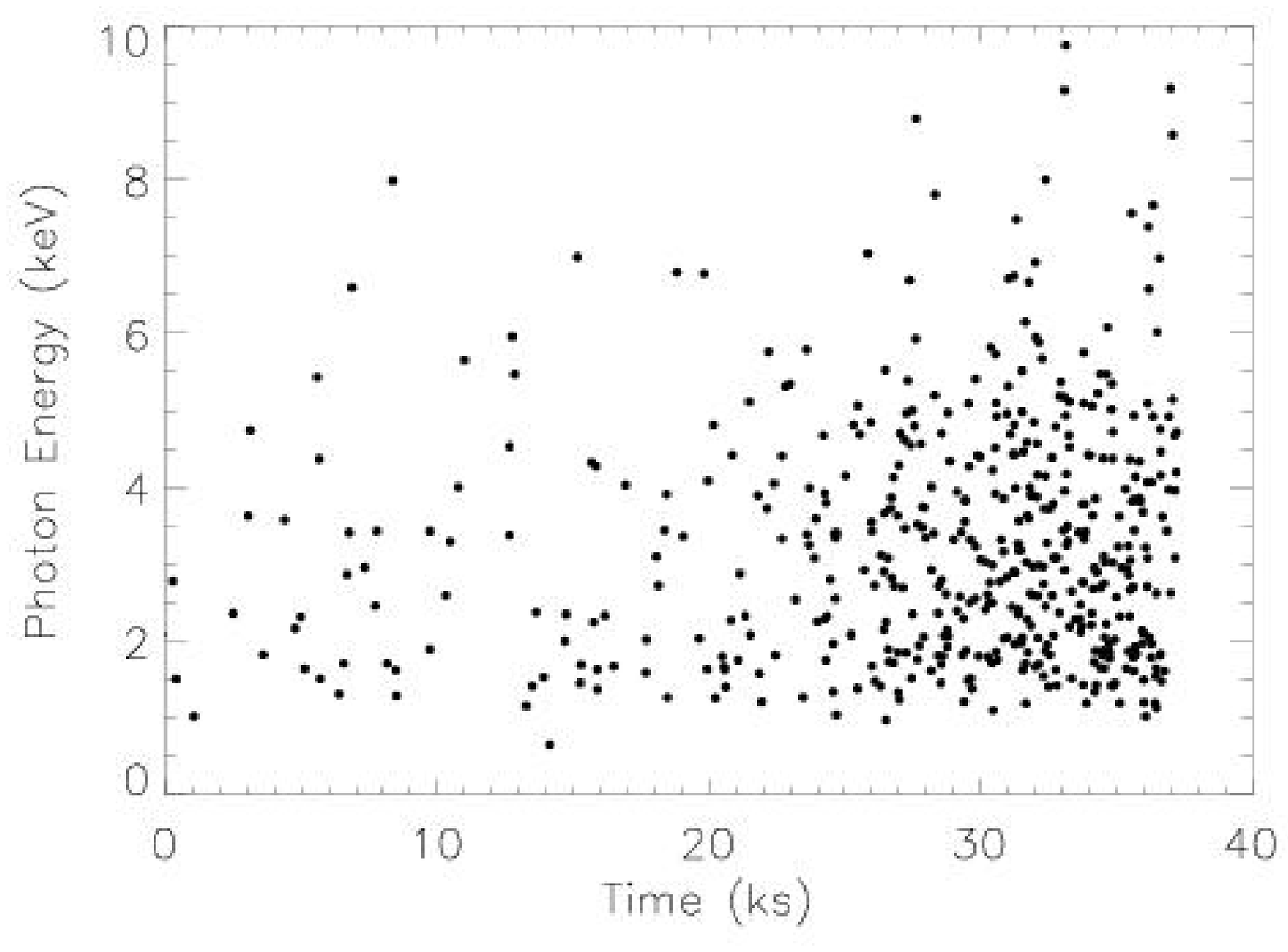}
 \plotone{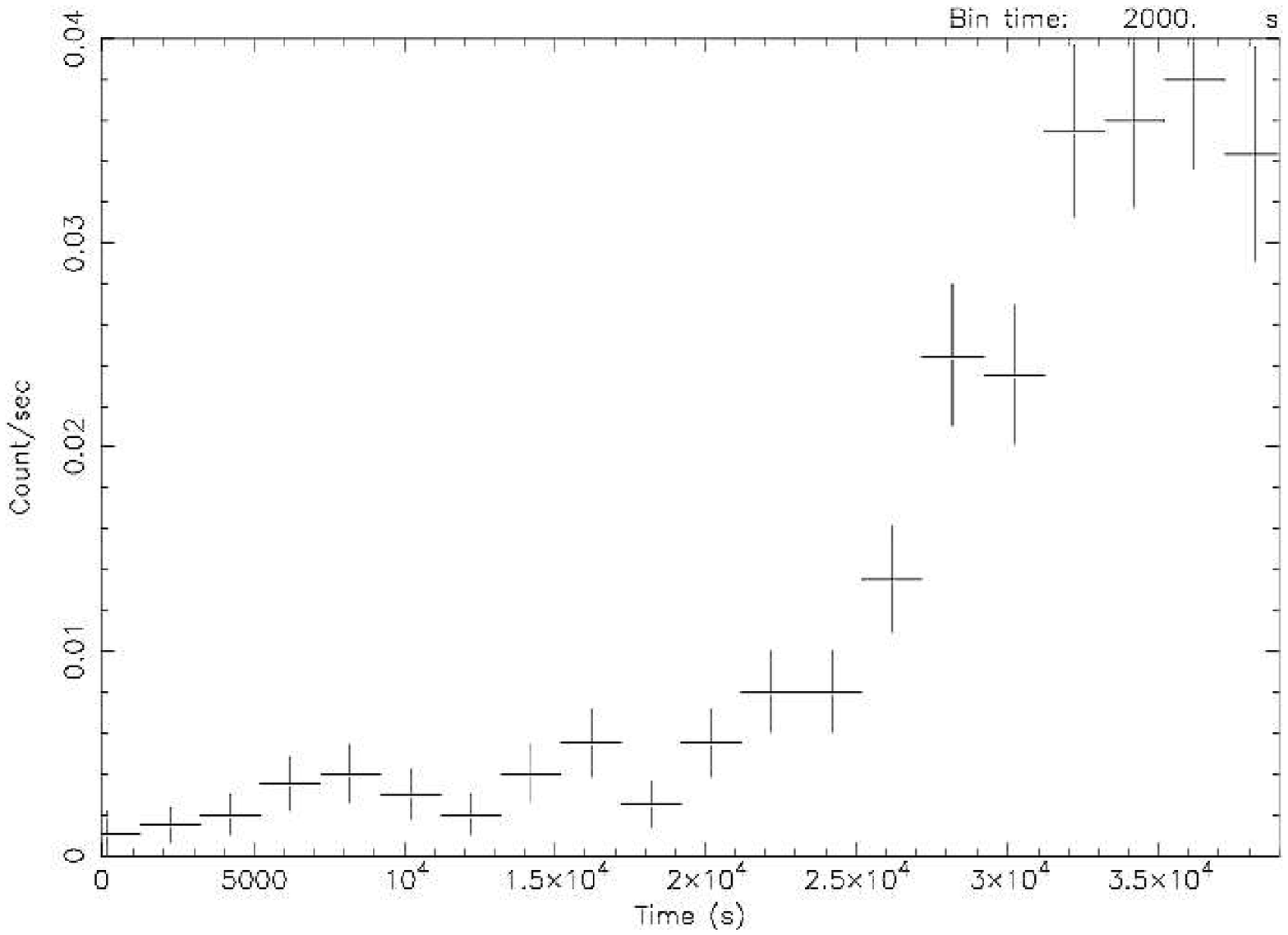}
 \plotone{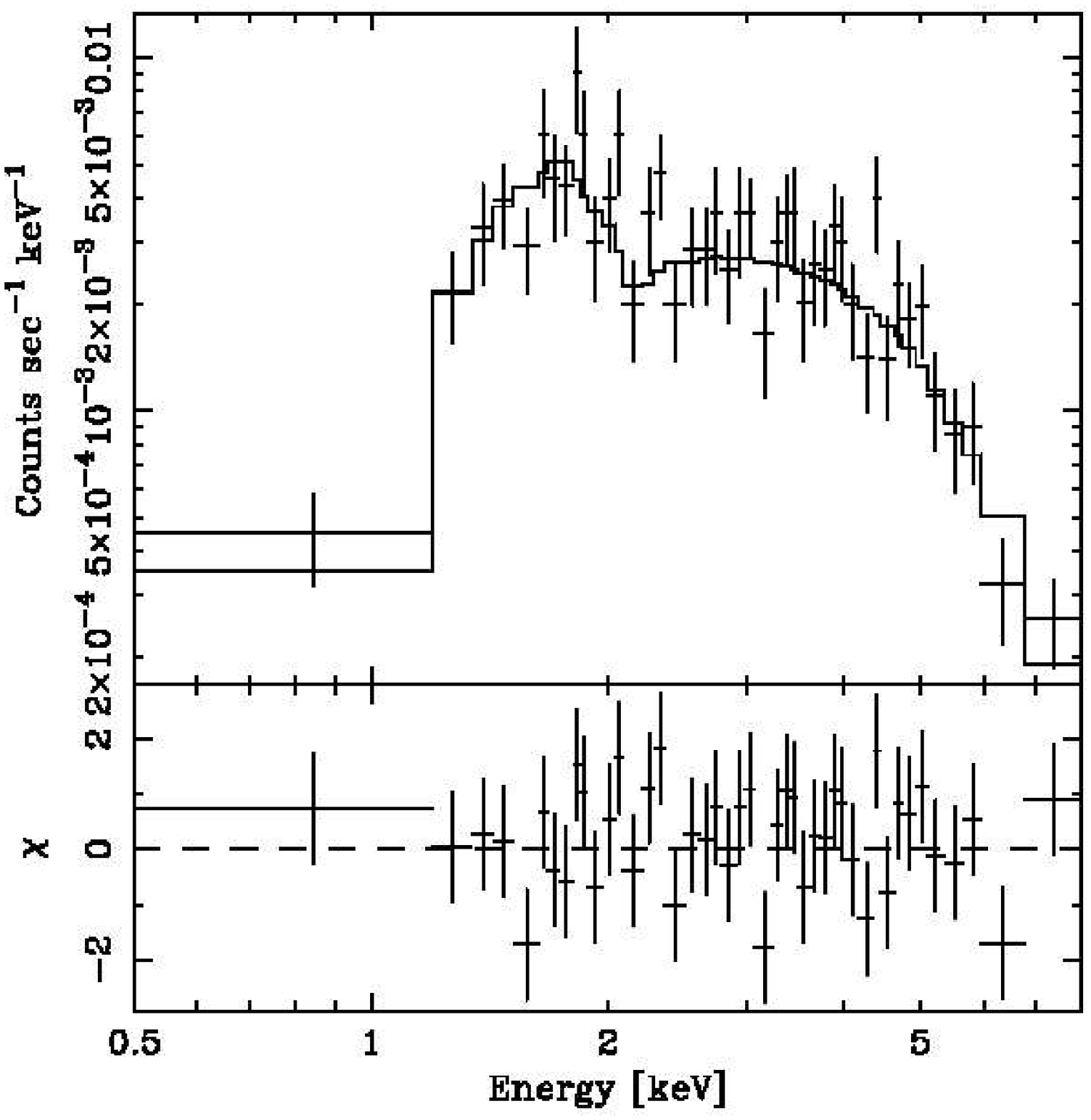}
\epsscale{1.0}
\caption{Photon arrival diagram, lightcurve (2~ks bins) and X-ray
spectrum of source \#672 (CXOU J172457.87-341203.9), with a powerful
flare detected in this ACIS observation.  The best spectral fit is
obtained with $N_H=1.7\times 10^{22}$ cm$^{-2}$ and $kT>10$ keV.
 \label{fig:flare}}
\end{figure}

\begin{figure}
 \centering
 \includegraphics[width=0.9\textwidth]{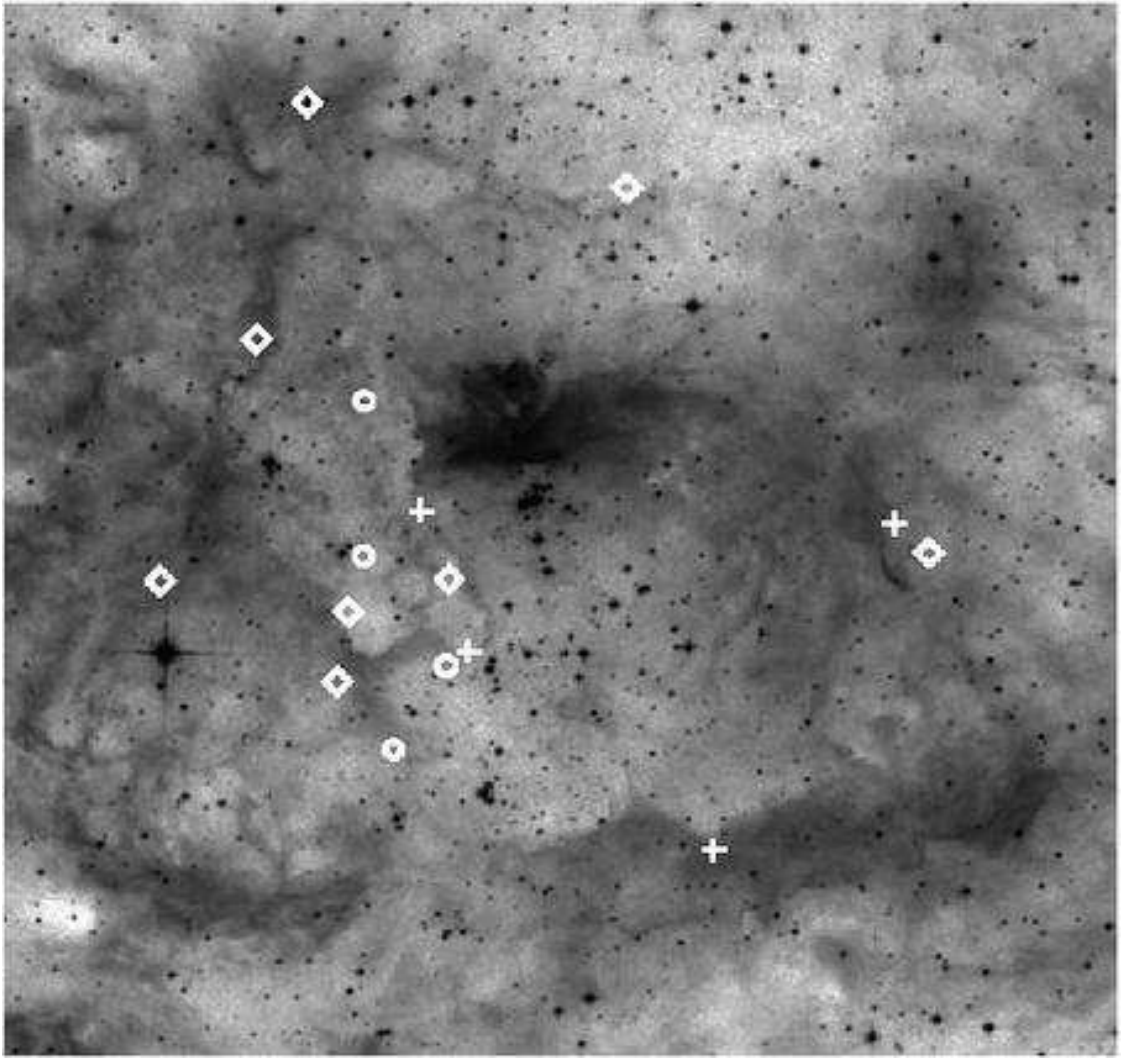}
\caption{The spatial distribution of deeply embedded sources.  Sources
that have $\langle E \rangle \ge 3.0$ keV are plotted with crosses and
sources that have $\log N_{H,X} \ge 22.5$ are plotted with
circles. Diamonds represent sources that satisfy both criteria. The
background DSS image is inverted so that brighter regions
appear darker.
 \label{fig:embedded}}
\end{figure}

4. XLFs are derived for lightly obscured and heavily obscured stellar
   populations in NGC 6357 (Figure~\ref{fig:XLF}). The high X-ray
   luminosity end ($\log L_X \ge 30.3$ ergs s$^{-1}$) of the NGC 6357
   XLF is clearly consistent with a power law relation as seen in COUP
   and Cep B, allowing a scaling comparison with the COUP XLF. The
   first estimate of the total cluster population of Pismis 24 is a
   few times the known Orion population. The presence of two O3 stars is
   consistent with the standard IMF.

5. We detect all 10 known OB stars with spectral type earlier than B1
   in the Pismis 24 cluster, including the WC7+O7 binary WR 93. Their
   $L_X/L_{bol}$ ratios are consistent with the canonical $10^{-7}$
   value with less than an order of magnitude scatter. The O3 stars
   Pis 24-1 (O3.5If) and Pis 24-17 (O3.5IIIf) exhibit X-ray
   luminosities $L_X\sim 10^{33}$ ergs s$^{-1}$. The X-ray spectrum of
   Pis 24-1 shows a soft emission component ($kT\sim 0.5$ keV) and a
   harder component ($kT\sim 1.7$ keV), while Pis 24-17 exhibits an
   unusally strong \sixiii\ line with a soft $0.7$ keV thermal
   plasma. The WC7+O7 binary WR93 is the brightest X-ray source in the
   field ($L_X\simeq 2 \times 10^{33}$). Its hard spectrum suggests
   that the X-ray emission is generated from colliding winds. The
   lightcurves of these massive stars are constant through the
   observation.  We assembled a list of candidate O stars from sources
   with high X-ray and infrared luminosity to facilitate future
   spectroscopy follow-up. Four candidate HAeBe stars are also
   identified.

6. We report the detection of X-ray emission from an EGG at the tip of
   a molecular pillar, which was previously found to be an IR source
   and the peak of the radio continuum. It is in the early
   evolutionary of stages of star formation. The spectral type
   estimated from its SED is B0--B2. The non-detection of EGGs in the
   Eagle Nebula may be the result of lower stellar mass and earlier
   evolutionary phases. We assemble a sample of 16 deeply embedded
   objects.

7. Several flaring sources exhibit luminosities above $\log L_{t,c}
   \ge 32.0$ ergs s$^{-1}$, which is exceptionally high for pre-MS
   stars. One powerful flare seen in ACIS \#672 (CXOU
   J172457.87-341203.9) is comparable to the strongest X-ray flares
   known in pre-MS stars ($L_{t,c}\sim 32.5$ ergs s$^{-1}$). Its
   lightcurve rises 15 times above the quiescent level during the
   flare, with a slow rise phase that is distinct from the fast rise
   flares seen pre-MS stars.

8. Soft unresolved X-ray emission in the NGC 6357 region is
   present. However, the luminosity derived from spectral fitting is
   consistent with the estimated level of integrated emission from the
   unresolved pre-MS stars.

We sincerely thank the anonymous referee for his/her careful review
and many constructive suggestions to improve the clarity of our
paper. We are grateful to Steve Strom for an enlightening discussion
on disk fractions among massive stars.  We thank Takahiro Nagayama,
Takahiro Kawadu, and the IRSF/SIRIUS team for taking the SIRIUS images
of NGC 6357.  Support for this work was provided to Gordon Garmire,
the ACIS Principal Investigator, by the National Aeronautics and Space
Administration (NASA) through NASA Contract NAS8-38252 and Chandra
Contract SV4-74018 issued by the Chandra X-ray Observatory Center,
which is operated by the Smithsonian Astrophysical Observatory for and
on behalf of NASA under contract NAS8-03060. This publication makes
use of data products from the Two Micron All Sky Survey, which is a
joint project of the University of Massachusetts and the Infrared
Processing and Analysis Center/California Institute of Technology,
funded by NASA and the National Science Foundation.  This research has
made use of the SIMBAD database and the VizieR catalogue access tool,
operated at CDS, Strasbourg, France. This research made use of data
products from the Midcourse Space Experiment.  Processing of the data
was funded by the Ballistic Missile Defense Organization with
additional support from NASA Office of Space Science.  This research
has also made use of the NASA/ IPAC Infrared Science Archive, which is
operated by the Jet Propulsion Laboratory, California Institute of
Technology, under contract with NASA.

{\it Facility}: \facility{CXO (ACIS)}

\onecolumn

\clearpage
\pagestyle{empty}
\begin{deluxetable}{rcrrrrrrrrrrrrrccccc}
\centering
\rotate
\tabletypesize{\tiny} \tablewidth{0pt}
\tablecolumns{20}

\tablecaption{ Main {\it Chandra} Catalog:  Basic Source Properties \label{tbl:primary}}

\tablehead{
\multicolumn{2}{c}{Source} &
  &
\multicolumn{4}{c}{Position} &
  &
\multicolumn{5}{c}{Extracted Counts\tablenotemark{a}} &
  &
\multicolumn{6}{c}{Characteristics} \\
\cline{1-2} \cline{4-7} \cline{9-13} \cline{15-20}

\colhead{Seq} & \colhead{CXOU J} &
  &
\colhead{$\alpha_{\rm J2000}$} & \colhead{$\delta_{\rm J2000}$} & \colhead{Err} & \colhead{$\theta$\tablenotemark{b}} &
  &
\colhead{Net} & \colhead{$\Delta$Net} & \colhead{Bkgd} & \colhead{Net} & \colhead{PSF} &
  &
\colhead{Signif} & \colhead{$\log P_B$\tablenotemark{c}} & \colhead{Anom\tablenotemark{d}} & \colhead{Var\tablenotemark{e}} &\colhead{EffExp} & \colhead{Med E}
\\

\colhead{\#} & \colhead{} &
  &
\colhead{(deg)} & \colhead{(deg)} & \colhead{(\arcsec)} & \colhead{(\arcmin)} &
  &
\colhead{Full} & \colhead{Full} & \colhead{Full} & \colhead{Hard} &
\colhead{Frac} &
  &
\colhead{} & \colhead{} & \colhead{} & \colhead{} & \colhead{(ks)} & \colhead{(keV)} \\

\colhead{(1)} & \colhead{(2)} &
  &
\colhead{(3)} & \colhead{(4)} & \colhead{(5)} & \colhead{(6)} &
  &
\colhead{(7)} & \colhead{(8)} & \colhead{(9)} & \colhead{(10)} & \colhead{(11)} &
  &
\colhead{(12)} & \colhead{(13)} & \colhead{(14)} & \colhead{(15)} & \colhead{(16)} & \colhead{(17)} }

\startdata
  1 & 172353.98-340850.8 &  &  260.974932 & -34.147448 &  0.4 & 10.7 &  &   267.2 &  17.7 &  22.8 &   160.0 & 0.91 &  &  14.7 & $<$-5 & .... & a &   32.7 & 2.4 \\
  2 & 172358.42-340802.6 &  &  260.993436 & -34.134056 &  0.5 & 10.2 &  &   107.0 &  11.5 &  12.0 &    71.7 & 0.91 &  &   8.9 & $<$-5 & .... & b &   33.1 & 2.6 \\
  3 & 172359.11-341217.6 &  &  260.996328 & -34.204895 &  0.8 &  9.1 &  &    23.2 &   6.6 &  12.8 &    12.4 & 0.91 &  &   3.2 & $<$-5 & .... & a &   33.0 & 2.0 \\
  4 & 172401.20-340928.0 &  &  261.005039 & -34.157783 &  0.9 &  9.1 &  &    13.7 &   5.3 &   8.3 &     7.2 & 0.89 &  &   2.4 & -4.3  & .... & a &   33.7 & 2.1 \\
  5 & 172402.27-341110.4 &  &  261.009476 & -34.186237 &  0.8 &  8.5 &  &    19.4 &   5.8 &   7.6 &    20.0 & 0.90 &  &   3.1 & $<$-5 & .... & a &   34.2 & 3.7 \\
  6 & 172403.19-341402.2 &  &  261.013313 & -34.233962 &  1.1 &  8.5 &  &    16.6 &   5.7 &   9.4 &     8.0 & 0.90 &  &   2.7 & $<$-5 & .... & a &   34.2 & 1.9 \\
  8 & 172405.42-341257.5 &  &  261.022606 & -34.215981 &  0.4 &  7.9 &  &    63.2 &   8.9 &   6.8 &    58.3 & 0.89 &  &   6.7 & $<$-5 & .... & a &   34.7 & 3.9 \\
  9 & 172405.64-340710.0 &  &  261.023532 & -34.119459 &  0.7 &  9.3 &  &    38.2 &   7.3 &   7.8 &     1.4 & 0.90 &  &   4.8 & $<$-5 & .... & a &   33.6 & 1.1 \\
 10 & 172406.70-341305.0 &  &  261.027933 & -34.218067 &  0.8 &  7.6 &  &    11.0 &   4.8 &   7.0 &     9.3 & 0.90 &  &   2.1 & -3.5  & .... & a &   34.9 & 3.8 \\
 12 & 172407.65-341751.9 &  &  261.031884 & -34.297766 &  0.8 &  9.3 &  &    27.8 &   7.1 &  13.2 &    13.7 & 0.90 &  &   3.7 & $<$-5 & .... & a &   31.6 & 2.1 \\
\enddata

\tablenotetext{a}{ Full band = 0.5--8~keV; Hard band = 2--8 keV.}
\tablenotetext{b}{ Off-axis angle.}
\tablenotetext{c}{ $P_B =$ probability extracted counts (full band) are solely from background.}
\tablenotetext{d}{ Source anomalies:  g = fractional time that source was on a detector (FRACEXPO from {\em mkarf}) is $<0.9$ ; e = source on field edge; p = source piled up; s = source on readout streak. Note that a reduced PSF fraction (significantly below 90$\%$) indicates that the source is in a crowded region.}
\tablenotetext{e}{ Source variability:  a = no evidence for variability; b = possibly variable; c = definitely variable.  No test is performed for sources with fewer than 4 total full-band counts.  No value is reported for sources in chip gaps or on field edges. }

\tablecomments{See \S 3.3 for a detailed description of the columns.  The full
table of 665 {\it Chandra} sources is available in the electronic
edition of the Journal.   }

\end{deluxetable}
\clearpage
\pagestyle{empty}
\begin{deluxetable}{rcrrrrrrrrrrrrrccccc}
\centering
\rotate
\tabletypesize{\tiny} \tablewidth{0pt}
\tablecolumns{20}

\tablecaption{ Tentative Source Properties \label{tbl:tentative}}
\tablehead{
\multicolumn{2}{c}{Source} &
  &
\multicolumn{4}{c}{Position} &
  &
\multicolumn{5}{c}{Extracted Counts\tablenotemark{a}} &
  &
\multicolumn{6}{c}{Characteristics} \\
\cline{1-2} \cline{4-7} \cline{9-13} \cline{15-20}

\colhead{Seq} & \colhead{CXOU J} &
  &
\colhead{$\alpha_{\rm J2000}$} & \colhead{$\delta_{\rm J2000}$} & \colhead{Err} & \colhead{$\theta$\tablenotemark{b}} &
  &
\colhead{Net} & \colhead{$\Delta$Net} & \colhead{Bkgd} & \colhead{Net} & \colhead{PSF} &
  &
\colhead{Signif} & \colhead{$\log P_B$\tablenotemark{c}} & \colhead{Anom\tablenotemark{d}} & \colhead{Var\tablenotemark{e}} &\colhead{EffExp} & \colhead{Med E}
\\

\colhead{\#} & \colhead{} &
  &
\colhead{(deg)} & \colhead{(deg)} & \colhead{(\arcsec)} & \colhead{(\arcmin)} &
  &
\colhead{Full} & \colhead{Full} & \colhead{Full} & \colhead{Hard} &
\colhead{Frac} &
  &
\colhead{} & \colhead{} & \colhead{} & \colhead{} & \colhead{(ks)} & \colhead{(keV)} \\

\colhead{(1)} & \colhead{(2)} &
  &
\colhead{(3)} & \colhead{(4)} & \colhead{(5)} & \colhead{(6)} &
  &
\colhead{(7)} & \colhead{(8)} & \colhead{(9)} & \colhead{(10)} & \colhead{(11)} &
  &
\colhead{(12)} & \colhead{(13)} & \colhead{(14)} & \colhead{(15)} & \colhead{(16)} & \colhead{(17)} }

\startdata
7 & 172403.64-340634.7 &  &  261.015178 & -34.109657 &  1.4 & 10.0 &  &    10.4 &   5.1 &   9.6 &     2.6 & 0.91 &  &   1.8 & -2.7 & .... & a &   32.4 & 1.6 \\
11 & 172406.84-341149.7 &  &  261.028513 & -34.197157 &  1.2 &  7.5 &  &     6.4 &   3.7 &   3.6 &     4.0 & 0.90 &  &   1.5 & -2.4 & .... & a &   33.0 & 2.0 \\
16 & 172412.29-341500.3 &  &  261.051238 & -34.250107 &  1.0 &  7.0 &  &     6.8 &   3.9 &   4.2 &     3.7 & 0.90 &  &   1.5 & -2.4 & .... & a &   33.1 & 2.1 \\
23 & 172415.22-340730.0 &  &  261.063419 & -34.125011 &  0.9 &  7.5 &  &     5.2 &   3.6 &   3.8 &     3.8 & 0.90 &  &   1.3 & -1.8 & .... & a &   32.2 & 2.6 \\
24 & 172415.43-340958.2 &  &  261.064294 & -34.166176 &  0.9 &  6.2 &  &     3.7 &   3.0 &   2.3 &     0.5 & 0.90 &  &   1.0 & -1.5 & g... & \nodata &   33.6 & 1.4 \\
25 & 172415.57-342044.7 &  &  261.064909 & -34.345752 &  1.5 & 10.3 &  &     7.9 &   5.3 &  13.1 &     1.2 & 0.91 &  &   1.4 & -1.6 & .... & a &   28.1 & 1.4 \\
27 & 172415.91-341441.2 &  &  261.066326 & -34.244799 &  0.8 &  6.2 &  &     6.3 &   3.5 &   2.7 &     6.2 & 0.90 &  &   1.5 & -2.7 & .... & a &   34.9 & 3.5 \\
32 & 172418.05-340824.5 &  &  261.075226 & -34.140162 &  0.9 &  6.5 &  &     4.5 &   3.2 &   2.5 &     5.9 & 0.89 &  &   1.2 & -1.8 & .... & a &   35.3 & 4.5 \\
33 & 172418.06-341650.6 &  &  261.075253 & -34.280743 &  1.0 &  7.0 &  &     6.5 &   3.7 &   3.5 &     2.2 & 0.89 &  &   1.5 & -2.5 & .... & a &   34.7 & 1.1 \\
37 & 172419.79-341130.3 &  &  261.082481 & -34.191754 &  1.0 &  4.9 &  &     4.8 &   3.0 &   1.2 &     2.4 & 0.90 &  &   1.3 & -2.8 & .... & a &   36.4 & 1.1 \\
\enddata

\tablenotetext{a}{ Full band = 0.5--8~keV; Hard band = 2--8 keV.}
\tablenotetext{b}{ Off-axis angle.}
\tablenotetext{c}{ $P_B =$ probability extracted counts (full band) are solely from background.}
\tablenotetext{d}{ Source anomalies:  g = fractional time that source was on a detector (FRACEXPO from {\em mkarf}) is $<0.9$ ; e = source on field edge; p = source piled up; s = source on readout streak. Note that a reduced PSF fraction (significantly below 90$\%$) indicates that the source is in a crowded region.}
\tablenotetext{e}{ Source variability:  a = no evidence for variability; b = possibly variable; c = definitely variable.  No test is performed for sources with fewer than 4 total full-band counts.  No value is reported for sources in chip gaps or on field edges.}

\tablecomments{See \S 3.3 for a detailed description of the columns.
The full table of 114 {\it Chandra} sources is available in the
electronic edition of the Journal. }
\end{deluxetable}
\clearpage
\begin{deluxetable}{rcrcccrcccccccrc}
\centering \rotate
\tabletypesize{\tiny}
\tablewidth{0pt}
\tablecolumns{16}

\tablecaption{X-ray Spectroscopy for Brighter Sources:  Thermal Plasma Fits
\label{tbl:apec}}

\tablehead{
\multicolumn{2}{c}{Source} &
  &
\multicolumn{3}{c}{Spectral Fit\tablenotemark{a}} &
  &
\multicolumn{5}{c}{X-ray Luminosities\tablenotemark{b}} &
  &
\colhead{Goodness} &
  &
\colhead{Notes\tablenotemark{c}} \\
\cline{1-2} \cline{4-6} \cline{8-12}

\colhead{Seq} & \colhead{CXOU J} &
  &
\colhead{$\log N_H$} & \colhead{$kT$} & \colhead{$\log EM$} &
  &
\colhead{$\log L_s$} & \colhead{$\log L_h$} & \colhead{$\log L_{h,c}$} & \colhead{$\log L_t$} & \colhead{$\log L_{t,c}$} &
  &
\colhead{of Fit} &
  &
\colhead{}  \\

\colhead{\#} & \colhead{} &
  &
\colhead{(cm$^{-2}$)} & \colhead{(keV)} & \colhead{(cm$^{-3}$)} &
  &
\multicolumn{5}{c}{(ergs s$^{-1}$)} &
  &
\colhead{$\chi_{\nu}^2$} &
  &
\colhead{} \\

\colhead{(1)} & \colhead{(2)} &
  &
\colhead{(3)} & \colhead{(4)} & \colhead{(5)} &
  &
\colhead{(6)} & \colhead{(7)} & \colhead{(8)} &\colhead{(9)} & \colhead{(10)} &
  &
\colhead{(11)} &
  &
\colhead{(12)}
}


\startdata
1 & 172353.98-340850.8 &  &  {\tiny $-0.08$} 22.0 {\tiny $+0.10$} &  $>10.0$  & {\tiny $-0.05$} 55.0 {\tiny $+0.05$} &  &   31.15 &   31.97 &  32.00 &  32.03 &  32.18 &  & 0.92 & & S; AGN? \\
2 & 172358.42-340802.6 &  &  {\tiny $-0.2$} 22.3 {\tiny $+0.2$} & {\tiny $-1.1$} 3.0 {\tiny $+2.8$} & {\tiny $-0.2$} 54.9 {\tiny $+0.3$} &  & 30.68 & 31.54 & 31.65 & 31.60 & 31.97 &  & 0.93 & & 2MASS\\
51 & 172423.93-340816.2 &  &  {\tiny $-0.3$} 21.4 {\tiny $+0.3$} & {\tiny $-3.7$} 7.1 {\tiny $+2.9$} & {\tiny $-0.09$} 54.3 {\tiny $+0.13$} &  & 30.76 & 31.22 & 31.23 & 31.35 & 31.44 &  & 1.07 & & USNO 0558-0532169+2MASS\\
77 & 172428.94-341450.6 &  &  {\tiny $-0.10$} 21.5 {\tiny $+0.09$} & {\tiny $-0.8$} 4.1 {\tiny $+1.2$} & {\tiny $-0.05$} 54.9 {\tiny $+0.05$} &  & 31.40 & 31.76 & 31.77 & 31.91 & 32.04 &  & 1.06 & & Pis24-15 \\
78 & 172429.03-341813.5 &  &  {\tiny $-0.2$} 21.9 {\tiny $+0.2$} & {\tiny $-0.8$} 2.7 {\tiny $+1.3$} & {\tiny $-0.2$} 54.7 {\tiny $+0.2$} &  & 30.88 & 31.32 & 31.37 & 31.46 & 31.72 &  & 0.27 & &  USNO 0556-0531216+2MASS\\
112 & 172433.04-341654.5 &  &  {\tiny $-0.1$} 22.1 {\tiny $+0.1$} & {\tiny $-1.1$} 3.6 {\tiny $+2.0$} & {\tiny $-0.1$} 54.8 {\tiny $+0.2$} &  & 30.84 & 31.54 & 31.60 & 31.62 & 31.88 &  & 0.46 & & USNO 0557-0528440+2MASS\\
140 & 172434.79-341318.0 &  &  {\tiny $-0.09$} 21.9 {\tiny $+0.08$} & {\tiny $-1.9$} 7.0 {\tiny $+3.0$} & {\tiny $-0.05$} 55.0 {\tiny $+0.06$} &  & 31.27 & 31.96 & 32.00 & 32.04 & 32.20 & & 0.71 &  & Pis24-11\\
158 & 172436.63-341550.7 &  &  {\tiny $-0.1$} 22.0 {\tiny $+0.1$} & {\tiny $-0.8$} 2.9 {\tiny $+1.2$} & {\tiny $-0.1$} 55.0 {\tiny $+0.2$} &  & 31.05 & 31.63 & 31.69 & 31.73 & 32.01 & & 0.80 &  & USNO 0557-0528485+2MASS\\
188 & 172438.73-341202.9 &  &  {\tiny $-0.1$} 22.0 {\tiny $+0.1$} & {\tiny $-0.9$} 3.3 {\tiny $+1.3$} & {\tiny $-0.1$} 54.8 {\tiny $+0.1$} &  & 30.96 & 31.54 & 31.60 & 31.64 & 31.89 & & 0.66 &  & USNO 0557-0528507+2MASS\\
196 & 172438.98-341220.6 &  &  {\tiny $-0.1$} 22.0 {\tiny $+0.1$} & {\tiny $-0.6$} 2.2 {\tiny $+1.0$} & {\tiny $-0.2$} 54.7 {\tiny $+0.2$} &  & 30.76 & 31.21 & 31.28 & 31.34 & 31.68 & & 0.47 &  & USNO 0557-0528516+2MASS \\
203 & 172439.51-341317.9 &  &  {\tiny $-0.2$} 21.8 {\tiny $+0.2$} & {\tiny $-1.7$} 4.8 {\tiny $+4.9$} & {\tiny $-0.1$} 54.5 {\tiny $+0.1$} &  & 30.79 & 31.34 & 31.37 & 31.45 & 31.61 & & 0.69 &  & USNO 0557-0528522+2MASS\\
227 & 172440.67-341403.8 &  &  {\tiny $-0.08$} 22.1 {\tiny $+0.08$} & {\tiny $-2.8$} 7.6 {\tiny $+2.4$} & {\tiny $-0.07$} 54.9 {\tiny $+0.09$} &  & 30.90 & 31.81 & 31.87 & 31.86 & 32.06 & & 0.76 &  & 2MASS\\
243 & 172441.23-341138.6 &  &  {\tiny $-0.2$} 22.0 {\tiny $+0.2$} & {\tiny $-1.0$} 3.3 {\tiny $+2.3$} & {\tiny $-0.2$} 54.8 {\tiny $+0.2$} &  & 30.91 & 31.50 & 31.55 & 31.60 & 31.85 & & 1.39 &  & USNO 0558-0532303+2MASS\\
270 & 172441.92-341158.0 &  &  {\tiny $-0.1$} 22.0 {\tiny $+0.1$} & {\tiny $-0.5$} 2.2 {\tiny $+0.6$} & {\tiny $-0.1$} 55.0 {\tiny $+0.2$} &  & 31.05 & 31.50 & 31.56 & 31.63 & 31.96 & & 1.31 &  & USNO 0558-0532308+2MASS \\
284 & 172442.14-341155.9 &  &  {\tiny $-0.2$} 22.0 {\tiny $+0.2$} & {\tiny $-3.2$} 6.9 {\tiny $+3.1$} & {\tiny $-0.10$} 54.8 {\tiny $+0.17$} &  & 30.92 & 31.69 & 31.74 & 31.76 & 31.94 & & 0.60 &  & . \\
290 & 172442.28-341128.0 &  &  {\tiny $-0.1$} 21.6 {\tiny $+0.1$} & {\tiny $-3.3$} 8.0 {\tiny $+2.0$} & {\tiny $-0.06$} 54.6 {\tiny $+0.08$} &  & 30.97 & 31.55 & 31.57 & 31.65 & 31.76 & & 0.61 &  & 2MASS \\
319 & 172443.11-341215.9 &  &  {\tiny $-0.2$} 21.8 {\tiny $+0.2$} & {\tiny $-2.3$} 5.6 {\tiny $+4.4$} & {\tiny $-0.1$} 54.6 {\tiny $+0.1$} &  & 30.87 & 31.46 & 31.49 & 31.56 & 31.71 & & 1.10 &  & USNO 0557-0528600+2MASS\\
331 & 172443.28-341243.9 &  &   21.3  &  1.4  &  54.5  &  & 31.08 & 30.76 & 30.78 & 31.25 & 31.43 & & 2.61 &  &  P; Pis24-2+TYC-7383-421-1\\
335 & 172443.34-341137.0 &  &  {\tiny $-0.1$} 22.0 {\tiny $+0.2$} & {\tiny $-1.4$} 3.6 {\tiny $+2.5$} & {\tiny $-0.1$} 54.7 {\tiny $+0.2$} &  & 30.82 & 31.49 & 31.54 & 31.57 & 31.83 & & 1.02 &  & . \\
344 & 172443.49-341156.9 &  &  {\tiny $-0.04$} 22.1 {\tiny $+0.03$} & {\tiny $-0.07$} 1.2 {\tiny $+0.07$} & {\tiny $-0.05$} 56.2 {\tiny $+0.05$} &  & 32.19 & 32.26 & 32.36 & 32.53 & 33.11 & & 1.57 &  & 2T; Pis24-1+TYC-7383-7-1 \\
372 & 172443.88-341139.3 &  &  {\tiny $-0.1$} 21.8 {\tiny $+0.1$} & {\tiny $-1.6$} 5.2 {\tiny $+3.9$} & {\tiny $-0.08$} 54.8 {\tiny $+0.09$} &  & 31.07 & 31.65 & 31.68 & 31.75 & 31.91 & & 1.40 &  & 2.$^{\prime\prime}7$ from Pis24-19 \\
379 & 172443.95-341145.6 &  &  {\tiny $-0.03$} 22.0 {\tiny $+0.06$} & $>10.0$  & {\tiny $-0.02$} 55.5 {\tiny $+0.05$} &  &  31.55 &  32.43 &  32.47 & 32.48 &   32.65 & & 0.81 &  & S; 2MASS \\
399 & 172444.37-341039.8 &  &  {\tiny $-0.09$} 22.0 {\tiny $+0.09$} & {\tiny $-1.1$} 4.1 {\tiny $+2.0$} & {\tiny $-0.09$} 54.9 {\tiny $+0.11$} &  & 31.04 & 31.74 & 31.79 & 31.82 & 32.05 & & 0.95 &  & USNO 0558-0532335+2MASS\\
420 & 172444.72-341202.6 &  &  {\tiny $-0.03$} 22.1 {\tiny $+0.03$} & {\tiny $-0.06$} 0.7 {\tiny $+0.06$} & {\tiny $-0.08$} 55.9 {\tiny $+0.13$} &  & 31.73 & 31.42 & 31.55 & 31.90 & 32.83 & & 1.16 &  & VT; Pis24-17 \\
464 & 172445.78-340939.8 &  &  {\tiny $-0.4$} 21.1 {\tiny $+0.3$} & {\tiny $-0.3$} 2.0 {\tiny $+0.5$} & {\tiny $-0.08$} 54.4 {\tiny $+0.08$} &  & 31.03 & 30.91 & 30.92 & 31.27 & 31.36 & & 1.16 &  & Pis24-13 \\
471 & 172446.01-341407.3 &  &  {\tiny $-0.07$} 21.9 {\tiny $+0.07$} & {\tiny $-1.3$} 5.7 {\tiny $+2.4$} & {\tiny $-0.06$} 55.2 {\tiny $+0.06$} &  & 31.35 & 32.06 & 32.10 & 32.13 & 32.32 & & 0.73 &  & 2MASS \\
472 & 172446.02-340942.3 &  &  {\tiny $-0.1$} 22.0 {\tiny $+0.1$} & {\tiny $-0.7$} 2.6 {\tiny $+1.2$} & {\tiny $-0.1$} 54.7 {\tiny $+0.2$} &  & 30.78 & 31.27 & 31.33 & 31.39 & 31.69 & & 0.36 &  & . \\
493 & 172446.68-341233.4 &  &  {\tiny $-0.1$} 22.0 {\tiny $+0.1$} & {\tiny $-1.2$} 3.5 {\tiny $+2.6$} & {\tiny $-0.1$} 54.7 {\tiny $+0.2$} &  & 30.88 & 31.46 & 31.51 & 31.56 & 31.80 & & 0.70 &  & USNO 0557-0528655+2MASS \\
519 & 172447.56-341048.6 &  &  {\tiny $-0.07$} 22.1 {\tiny $+0.06$} & {\tiny $-2.2$} 7.5 {\tiny $+2.5$} & {\tiny $-0.05$} 55.2 {\tiny $+0.06$} &  & 31.27 & 32.13 & 32.18 & 32.19 & 32.38 & & 0.93 &  & 2MASS \\
530 & 172447.95-341735.3 &  &  {\tiny $-0.4$} 21.4 {\tiny $+0.3$} & {\tiny $-0.9$} 3.1 {\tiny $+1.6$} & {\tiny $-0.1$} 54.4 {\tiny $+0.1$} &  & 30.98 & 31.17 & 31.18 & 31.38 & 31.49 & & 0.33 &  & 2MASS \\
577 & 172450.15-341243.4 &  &  {\tiny $-0.07$} 22.0 {\tiny $+0.08$} & {\tiny $-1.0$} 4.1 {\tiny $+1.2$} & {\tiny $-0.07$} 55.1 {\tiny $+0.09$} &  & 31.26 & 31.94 & 31.99 & 32.02 & 32.25 & & 1.29 &  & USNO 0557-0528716+2MASS\\
612 & 172453.10-341525.6 &  &  {\tiny $-0.1$} 22.2 {\tiny $+0.1$} & {\tiny $-0.4$} 1.8 {\tiny $+0.7$} & {\tiny $-0.2$} 54.9 {\tiny $+0.2$} &  & 30.81 & 31.32 & 31.42 & 31.44 & 31.90 & & 0.83 &  & USNO 0557-0528750+2MASS \\
634 & 172454.31-341209.4 &  &  {\tiny $-0.06$} 22.3 {\tiny $+0.08$} & $>10.0$  & {\tiny $-0.05$} 54.8 {\tiny $+0.09$} &  & 30.53   & 31.72 & 31.80 & 31.75 &  31.98 & & 1.65 &  & S; 2MASS \\
649 & 172455.55-341631.3 &  &  {\tiny $-0.10$} 22.1 {\tiny $+0.10$} & {\tiny $-1.3$} 4.3 {\tiny $+2.5$} & {\tiny $-0.1$} 55.1 {\tiny $+0.1$} &  & 31.08 & 31.86 & 31.92 & 31.93 & 32.18 & & 0.59 &  & USNO 0557-0528794+2MASS\\
672 & 172457.87-341203.9 &  &  {\tiny $-0.04$} 22.3 {\tiny $+0.04$} & $>10.0$  & {\tiny $-0.03$} 55.3 {\tiny $+0.04$} &  & 31.17 &  32.27 & 32.34 & 32.30  &   32.52 & & 0.98 &  & S; 2MASS\\
692 & 172459.74-340958.7 &  &  {\tiny $-0.1$} 22.5 {\tiny $+0.1$} & {\tiny $-0.6$} 2.2 {\tiny $+1.1$} & {\tiny $-0.2$} 55.2 {\tiny $+0.3$} &  & 30.46 & 31.57 & 31.77 & 31.60 & 32.17 & & 0.49 &  & . \\
746 & 172508.76-341115.2 &  &   22.0  &  1.9  &  55.2  &  & 31.31 & 31.65 & 31.72 & 31.82 & 32.18 & & 2.19 &  & P; MSX6C G353.2261+00.8295 \\
747 & 172508.83-341112.5 &  &  {\tiny $-0.03$} 21.8 {\tiny $+0.03$} & {\tiny $-0.1$} 2.3 {\tiny $+0.2$} & {\tiny $-0.03$} 56.2 {\tiny $+0.02$} &  & 32.45 & 32.78 & 32.82 & 32.95 & 33.21 & & 1.35 &  & VT; WR 93+TYC7383-162-1 \\
753 & 172510.93-340843.2 &  &  {\tiny $-0.07$} 22.5 {\tiny $+0.10$} &  $>10.0$  & {\tiny $-0.06$} 54.9 {\tiny $+0.14$} &  & 30.31  & 31.78 &  31.90 & 31.80 &   32.08 & & 0.98 &  & S \\
770 & 172516.76-341211.3 &  &  {\tiny $-0.1$} 22.0 {\tiny $+0.1$} & {\tiny $-1.1$} 3.8 {\tiny $+2.2$} & {\tiny $-0.1$} 54.7 {\tiny $+0.1$} &  & 30.92 & 31.52 & 31.57 & 31.62 & 31.84 & & 0.74 &  & USNO 0557-0529090+2MASS \\
\enddata

\tablenotetext{a}{ All fits were ``wabs(apec)'' in XSPEC and assumed $0.3 \times$solar abundances. }

\tablenotetext{b}{ X-ray luminosities:  s = soft band (0.5--2 keV); h = hard
band (2--8 keV); t = total band (0.5--8 keV).  Absorption-corrected luminosities
are subscripted with a $c$.  Luminosities were calculated assuming a distance
of 2.56~kpc.}

\tablenotetext{c}{ 2T means a two-temperature fit was also performed;
VT means a variable abundance fit was also performed; see \S 6.1 for
fit parameters.  S means the spectral fit is only used as a spline fit
for a rough estimate of luminosity; actual fit values are
unreliable. P means the fit does not formally satisfy the convergence
criterion, nevertheless the fit is acceptable to estimate the
luminosity. The confidence intervals are missing since no error
calculation is reported in {\it XSPEC}. Pis24 stars and other
cataloged names obtained from VizieR are also shown. {\it USNO}-The
USNO-B1.0 Catalog (Monet et~al.\ 2003); {it TYC}-The Tycho-2 Catalogue
(Hog et~al.\ 2000); {\it MSX6C}-MSX6C Infrared Point Source Catalog
(Egan et~al.\ 2003).}

\tablecomments{Column 1: X-ray source number. Column 2: IAU
designation. Columns 3-5: Estimated column density, plasma energy, and
the plasma emission measure. Columns 6-10: Observed and corrected for
absorption X-ray luminosities, obtained from our spectral analysis \S
3.4. Columns 11: Reduced $\chi^2$ for the spectral fit.}
\end{deluxetable}
\clearpage
\begin{deluxetable}{rcrcccrcccccrc}
\centering \rotate 
\tabletypesize{\scriptsize} \tablewidth{0pt}
\tablecolumns{14}

\tablecaption{X-ray Spectroscopy for Less Bright Sources:  Thermal Plasma Fits
\label{tbl:apec2}}

\tablehead{
\multicolumn{2}{c}{Source} &
  &
\multicolumn{3}{c}{Spectral Fit\tablenotemark{a}} &
  &
\multicolumn{5}{c}{X-ray Luminosities\tablenotemark{b}} &
  &
\colhead{Notes\tablenotemark{c}} \\ 
\cline{1-2} \cline{4-6} \cline{8-12}

\colhead{Seq} & \colhead{CXOU J} & 
  &
\colhead{$\log N_H$} & \colhead{$kT$} & \colhead{$\log EM$} &  
  &
\colhead{$\log L_s$} & \colhead{$\log L_h$} & \colhead{$\log L_{h,c}$} & \colhead{$\log L_t$} & \colhead{$\log L_{t,c}$} &
  &
\colhead{}  \\

\colhead{\#} & \colhead{} & 
  &
\colhead{(cm$^{-2}$)} & \colhead{(keV)} & \colhead{(cm$^{-3}$)} & 
  &
\multicolumn{5}{c}{(ergs s$^{-1}$)} &
  &
\colhead{} \\

\colhead{(1)} & \colhead{(2)} & 
  &
\colhead{(3)} & \colhead{(4)} & \colhead{(5)} & 
  &
\colhead{(6)} & \colhead{(7)} & \colhead{(8)} &\colhead{(9)} & \colhead{(10)} &  
  &
\colhead{(11)}
}


\startdata

3 & 172359.11-341217.6 &  &   22.0  &  4.8  &  53.9  &  & 30.01 & 30.73 & 30.78 & 30.81 & 31.02 &  & . \\
8 & 172405.42-341257.5 &  &  {\tiny $-0.1$} 22.7 {\tiny $+0.1$}  &  $>$10  &  54.7  &  & 29.59 & 31.60 & 31.76 & 31.61 & 31.89 &  & S \\
9 & 172405.64-340710.0 &  &   21.0  &  {\tiny $-0.3$} 1.8 {\tiny $+0.3$}  &  53.9  &  &  30.62 &  30.38 &  30.39 & 30.82  &  30.86 &  & . \\
12 & 172407.65-341751.9 &  &  {\tiny $-0.2$} 22.3 {\tiny $+0.1$} & {\tiny $-0.2$} 0.5 {\tiny $+0.1$} & {\tiny $-0.2$} 55.0 {\tiny $+0.2$} &  & 30.36 & 30.04 & 30.25 & 30.53 & 31.93 &  & . \\
15 & 172409.28-341224.3 &  &  {\tiny $-1.5$} 22.5 {\tiny $+0.2$} & {\tiny $-1.4$} 2.7 \phantom{{\tiny $-1.4$}} & {\tiny $-0.3$} 54.9 \phantom{{\tiny $+0.3$}} &  & 30.30 & 31.41 & 31.58 & 31.45 & 31.92 &  & . \\
19 & 172413.61-340921.8 &  &  {\tiny $-0.5$} 21.5 {\tiny $+0.4$} & {\tiny $-1.5$} 2.9 {\tiny $+14.8$} & {\tiny $-0.3$} 54.0 {\tiny $+0.3$} &  & 30.53 & 30.74 & 30.76 & 30.95 & 31.08 &  & . \\
20 & 172413.68-341451.8 &  &  {\tiny $-0.4$} 22.2 {\tiny $+0.2$} & {\tiny $-0.7$} 1.5 {\tiny $+2.6$} & {\tiny $-0.6$} 54.4 {\tiny $+0.5$} &  & 30.24 & 30.67 & 30.79 & 30.80 & 31.36 &  & . \\
29 & 172416.17-341526.6 &  &  {\tiny $-0.2$} 22.4 {\tiny $+0.2$} & {\tiny $-0.9$} 2.3 {\tiny $+4.3$} & {\tiny $-0.4$} 54.7 {\tiny $+0.4$} &  & 30.30 & 31.19 & 31.34 & 31.24 & 31.73 &  & . \\
30 & 172416.18-341311.3 &  &  {\tiny $-0.9$} 21.6 {\tiny $+0.3$} & {\tiny $-0.9$} 1.9 {\tiny $+3.5$} & {\tiny $-0.3$} 54.1 {\tiny $+0.2$} &  & 30.49 & 30.55 & 30.58 & 30.82 & 31.05 &  & . \\
35 & 172418.81-341204.7 &  &   22.1  &  1.3  &  54.4  &  & 30.43 & 30.55 & 30.65 & 30.79 & 31.37 &  & . \\

\enddata

\tablenotetext{a}{ All fits were ``wabs(apec)'' in XSPEC and assumed
$0.3 \times$solar abundances. }

\tablenotetext{b}{ X-ray luminosities: s = soft band (0.5--2 keV); h =
hard band (2--8 keV); t = total band (0.5--8 keV).
Absorption-corrected luminosities are subscripted with a $c$.
Luminosities were calculated assuming a distance of 2.56~kpc.}

\tablenotetext{c}{ S means the spectral fit is only used as a spline fit
for a rough estimate of luminosity; actual fit values are unreliable. 
Some confidence intervals are missing since no error calculation is reported in {\it XSPEC}.}

\tablecomments{Column 1: X-ray source number. Column 2: IAU
designation. Columns 3-5: Estimated column density, plasma energy, and
the plasma emission measure. Columns 6-10: Observed and corrected for
absorption X-ray luminosities.  The full table of 144 sources is
provided in the electronic edition of the Journal.}
\end{deluxetable}
\clearpage
\begin{deluxetable}{rccccccccccccccccr}
\centering
\rotate
\tabletypesize{\tiny}
\tablewidth{0pt}
\tablecolumns{18}

\tablecaption{Stellar Counterparts and Classifications \label{tbl:class}}
\tablehead{
\multicolumn{2}{c}{Source} &
  &
\multicolumn{4}{c}{Counterparts \& Class} &
  &
\multicolumn{3}{c}{2MASS Coordinates} &
  &
\multicolumn{3}{c}{2MASS Colors \& Magnitudes} &
  &
\multicolumn{2}{c}{Cautionary Flags}
\\
\cline{1-2} \cline{4-7} \cline{9-11} \cline{13-15} \cline{17-18}\\

\colhead{Seq \#} & \colhead{CXOU J} &
  &
\colhead{2MASS} & \colhead{SIRIUS} & \colhead{HST} & \colhead{Flags} & \colhead{} & \colhead{RA(J2000)} & \colhead{DEC(J2000)} & \colhead{$\Delta \phi(^{\prime\prime})$} & \colhead{} & \colhead{$J-H$} & \colhead{$H-K_s$} & \colhead{$K_s$} & \colhead{} & \colhead{QFLAG} & \colhead{Class} \\
\colhead{(1)} & \colhead{(2)} &
  &
\colhead{(3)} & \colhead{(4)} & \colhead{(5)} & \colhead{(6)}
  & & \colhead{(7)} & \colhead{(8)}
  &
\colhead{(9)} & & \colhead{(10)}
  &
\colhead{(11)} & \colhead{(12)} &
  &
\colhead{(13)} & \colhead{(14)}
}
\startdata
  1 & 172353.98-340850.8 & &$\times$ &\nodata &\nodata &\nodata& &\nodata&\nodata&\nodata& &\nodata&\nodata&\nodata& &\nodata&\nodata\\
  2 & 172358.42-340802.6 & &$\surd$ &\nodata &\nodata &R& & 17 23 58.36 & -34 08 02.5 & 0.87 & &  2.004 &  1.251 & 11.928 & &AAA& 000 \\
  3 & 172359.11-341217.6 & &$\surd$ &\nodata &\nodata &R& & 17 23 59.16 & -34 12 17.0 & 0.80 & &  0.749 &  0.487 & 10.690 & &AAA& 000 \\
  4 & 172401.20-340928.0 & &$\surd$ &\nodata &\nodata &R& & 17 24 01.32 & -34 09 27.3 & 1.58 & &  0.744 &  0.477 & 12.926 & &UAA& 000 \\
  5 & 172402.27-341110.4 & &$\times$ &\nodata &\nodata &H& &\nodata&\nodata&\nodata& &\nodata&\nodata&\nodata& &\nodata&\nodata\\
  6 & 172403.19-341402.2 & &$\surd$ &\nodata &\nodata &R& & 17 24 03.22 & -34 14 02.1 & 0.35 & &  0.990 &  0.369 & 11.313 & &AAA& 000 \\
  7 & 172403.64-340634.7 & &$\surd$ &\nodata &\nodata &R& & 17 24 03.56 & -34 06 36.2 & 1.77 & &  2.092 &  0.739 & 13.505 & &DBA& ccc \\
  8 & 172405.42-341257.5 & &$\times$ &\nodata &\nodata &H& &\nodata&\nodata&\nodata& &\nodata&\nodata&\nodata& &\nodata&\nodata\\
  9 & 172405.64-340710.0 & &$\surd$ &\nodata &\nodata &F& & 17 24 05.62 & -34 07 09.5 & 0.60 & &  0.399 &  0.172 &  8.458 & &AAA& 000 \\
 10 & 172406.70-341305.0 & &$\surd$ &\nodata &\nodata &HR& & 17 24 06.64 & -34 13 04.9 & 0.74 & &  2.319 &  1.208 & 13.316 & &UCA& 000 \\

\enddata

\tablecomments{Column 1: X-ray source number. Column 2: IAU
designation. Columns 3-5: existence of ONIR counterparts;
$\surd$-counterpart. $\times$-no counterpart, \nodata-source is out of
the field of view of the available images. Column 6: X-ray source
classifications. See \S 5.5 for descriptions of the class
designations. Column 7-9: 2MASS (J2000) coordinates and the offset
between X-ray position and near-IR position in arcsec. Column 10-12:
2MASS $J-H$,$H-K_s$, and $K_s$ measurements.  Column 13: From {\it
Explanatory Supplement to the 2MASS All Sky Data Release} (Cutri
et~al.\ 2003). 2MASS $JHK_s$ photometry quality flags with three
characters: A = very high significance detection ($>10$ SNR); B = high
significance detection ($>7$ SNR); C = moderate significance detection
($>5$ SNR); D = low significance detection; E = point spread fitting
poor; F = reliable photometric errors not available; U = upper limit
on magnitude (source not detected); X = source is detected but no
valid brightness estimate is available. Column 14: From {\it
Explanatory Supplement to the 2MASS All Sky Data Release} (Cutri
et~al.\ 2003). 2MASS $JHK_s$ confusion and contamination flags with
three characters: 0 = source unaffected by artifacts; b = bandmerge
confusion (possible multiple source); c = photometric confusion from
nearby star; d = diffraction spike confusion from nearby star; p =
persistence contamination from nearby star; s = electronic stripe from
nearby star. The full table of 779 {\it Chandra} sources is available
in the electronic edition of the Journal.}

\end{deluxetable}
\clearpage
\begin{deluxetable}{lcccccccccccccccc}
\topmargin 0.5in
\centering
\rotate
\tabletypesize{\tiny} 
\tablewidth{0pt}
\tablecolumns{17}

\tablecaption{Known O and Early B stars in Pismis 24 \label{tbl:OB}}

\tablehead{ 
\multicolumn{4}{c}{Star} &
\multicolumn{8}{c}{ONIR photometry} &
\multicolumn{4}{c}{X-ray properties} \\
\cline{1-16} \\
\colhead{ID\tablenotemark{a}} &
\colhead{SpecType\tablenotemark{b}} & \colhead{$\alpha_{\rm
J2000}$}\tablenotemark{c} & \colhead{$\delta_{\rm
J2000}$}\tablenotemark{c} & \colhead{$V$\tablenotemark{b}} &
\colhead{$U-B$\tablenotemark{b}} &
\colhead{$B-V$\tablenotemark{b}} &
\colhead{$V-J$\tablenotemark{d}} &
\colhead{$J-H$\tablenotemark{d}} &
\colhead{$H-K_s$\tablenotemark{d}} &
\colhead{$K_s$\tablenotemark{d}} &
\colhead{$A_V$\tablenotemark{e}} & \colhead{Seq \# \tablenotemark{f}}& 
\colhead{$\log L_X$\tablenotemark{g}}& \colhead{$\log N_H$\tablenotemark{g}}&
\colhead{$\log L_{bol}$\tablenotemark{b}}&
\colhead{Notes\tablenotemark{h}} }

\startdata
Pis 24-1 & O3If* & 17 24 43.49 & -34 11 57.0 & 10.43 & 0.40 & 1.45 & 3.70 & 0.55 & 0.28 & 5.89 & 6.16 & \#344 & 33.11& 22.1 & 39.60 &HDE 319718=[N78] 35\\
Pis 24-17& O3III(f*)& 17 24 44.73 & -34 12 02.7 & 11.84 & \nodata & 1.49 & 4.02 & 0.54 & 0.31 & 6.98 & 6.57 & \#420 &32.83 &22.1 & 39.09 & [N78] 57\\
Pis 24-2 & O5.5V((f))& 17 24 43.20 & -34 12 43.5 & 11.95 & 0.32 & 1.41 & 3.48 & 0.48 &  0.31 & 7.69 & 5.81 & \#331 &31.43 & 21.3& 38.81 & \nodata \\
Pis 24-13& O6.5V((f)) & 17 24 45.78 & -34 09 39.9 & 12.73 & 0.11 & 1.48 & 3.92& 0.57 & 0.29 & 7.95 & 6.45 & \#464 & 31.36 &21.1 & 38.61 & [N78] 36\\
Pis 24-16& O7.5V & 17 24 44.45 & -34 11 58.8 & 13.02 & \nodata & 1.60 & 4.57 & 0.59 &  0.49 &  7.37 & 7.35 & \#407 & 31.90& 22.0& 38.60 & [N78] 58\\
Pis 24-3 & O8V & 17 24 42.30 & -34 13 21.5 & 12.75 & 0.24 & 1.41 & 3.46 & 0.52 & 0.30 & 8.47 & 5.79 & \#293 &31.17 & 21.0& 38.57 & \nodata \\
Pis 24-15& O8V & 17 24 28.86 & -34 14 50.3 & 12.32 & 0.14 & 1.27 & 3.29 & 0.45 & 0.24 & 8.35 & 5.54 & \#77 & 32.04& 21.5& 38.63 &[N78] 46\\
Pis 24-10& O9V & 17 24 36.04 & -34 14 00.5 & 13.02 & 0.40 & 1.40 & 3.51 & 0.49 &  0.28 & 8.75 & 5.84 & \#151 & 31.96 & 22.1& 38.36 & \nodata\\
WR 93& WC7+O7-9& 17 25 08.85 & 34 11 12.5 & 10.68 & 0.70 & 1.42 & 3.65 & 0.50 & 0.67 & 5.87 & 5.86 & \#747 & 33.21 & 21.8 & $>$39.97 & HD 157504=[N78] 52\\
Pis 24-18& B0.5V& 17 24 43.29 & -34 11 41.9 & 13.97 & \nodata & 1.48 & 3.97 & 0.68 &  0.34 &  8.97 & 6.42 & \#332 & 31.56 & 22.1 & 38.26 & \nodata\\
Pis 24-12& B1V & 17 24 42.27 & -34 11 41.2 & 13.88 & 0.38 & 1.47 & 3.38 & 0.51 & 0.21 & 9.78 & 5.56 & No & \nodata &\nodata & 38.05& \nodata\\
Pis 24-19& B1V & 17 24 43.69 & -34 11 40.7 & 14.43 & \nodata & 1.39 & 3.76 & 0.59& 0.47 & 9.61 & 6.08 & No & \nodata &\nodata & 38.04& \nodata\\
\hline \\
Pis 24-8\tablenotemark{i} & O9? & 17 24 38.81 & -34 14 58.2 & 12.98 & 0.48 & 1.44 & 3.67 & 0.59 &  0.21 & 8.52 & 5.90 & \#194 & 31.29& 21.9& \nodata &\nodata\\
Pis 24-7\tablenotemark{i} & O9? & 17 24 47.81 & -34 15 16.5 & 13.46 & 0.58 & 1.68 & 4.10 & 0.61 &  0.33 & 8.42 & 6.61 & \#527 & 30.28& 22.2& \nodata & \nodata\\
Pis 24-4\tablenotemark{i} & O9-B0? & 17 24 40.39 & -34 12 05.9 & 13.93 & 0.53 & 1.43 & 3.88 & 0.60 &  0.32 & 9.14 & 6.49 & \#225 & 30.45 & 22.0& \nodata &\nodata \\
Pis 24-9\tablenotemark{i} & O9-B1? & 17 24 39.29 & -34 15 26.4 & 14.26 & 0.40 & 1.40 & 3.48 & 0.49 &  0.29 & 10.00& 5.81 & No & \nodata & \nodata& \nodata &\nodata \\
Pis 24-11\tablenotemark{i}& O9-B1? & 17 24 34.68 & -34 13 17.1 & 14.53 &0.30::& 1.57 & 3.79 & 0.66 &  0.30 & 9.79 & 7.03 & \#140 & 32.20 & 21.9& \nodata &\nodata \\

\enddata

\tablenotetext{a}{Identifications from \citet{Moffat73} and \citet{Massey01}.}
\tablenotetext{b}{Spectral types from \citet{Massey01} and \citet{Walborn02}. Optical photometry from \citet{Massey01}. Bolometric luminosities from \citet{Massey01} are recalculated using bolometric corrections from \citet{Heap06} instead of \citet{Vacca96}.}
\tablenotetext{c}{Units of RA are hours, minutes
and seconds; units of DEC are degrees, arcminutes, and
arcseconds. Coordinates are from the Catalogue of Stellar Spectral
Classifications (Skiff 2005) and \citet{Massey01}. }
\tablenotetext{d}{Infrared photometry from 2MASS.}
\tablenotetext{e}{Visual extinction from \citet{Bohigas04} ($A_V=1.39 E(V-J)$).}
\tablenotetext{f}{ACIS source number for X-ray detections.}
\tablenotetext{g}{$L_X$ is absorption corrected full-band luminosity $L_{t,c}$ (ergs s$^{-1}$). $N_H$ is X-ray derived column density (cm$^{-2}$).}
\tablenotetext{h}{Other identifications from \citet{Neckel78}.}
\tablenotetext{i}{Not included in \citet{Massey01} spectroscopic
classification sample. Tentative spectral types are inferred from
color-magnitude diagram by de-reddening 2MASS photometric colors to
the 1 Myr pre-MS isochrone.}

\end{deluxetable}
\begin{deluxetable}{lcccccccc}
\centering
\tabletypesize{\scriptsize} \tablewidth{0pt}
\tablecolumns{9}

\tablecaption{X-ray and IR luminous stars as Candidate O stars \label{tbl:NewO}}

\tablehead{
\colhead{Seq \#} &
\colhead{CXOU J } &
\colhead{$J-H$} &
\colhead{$H-K_s$} &
\colhead{$K_s$} &
\colhead{$\log L_X$}&
\colhead{$\log N_H$}&
\colhead{$kT$} &
\colhead{Notes}
}

\startdata
\multicolumn{9}{c}{(a) Stars with $\log L_X > 32.0$} \\
\cline{1-9} \\
140 & 172434.79-341318.0 & 0.660 & 0.296 & 9.789 & 32.04& 21.9& 7.0&  Pis 24-11\\
158 &172436.63-341550.7 & 1.194 & 0.504 & 12.270& 32.01& 22.0& 2.9&  V\\
227 &172440.67-341403.8 & 1.203 & 0.176 & 14.324& 32.06& 22.1& 7.6& flaring\\
379 &172443.95-341145.6 & -0.572:& 0.651:& 12.298& 32.65& 22.0& 10.0:&  HST\\
399 &172444.37-341039.8 & 1.063 & 0.957& 10.972& 32.05& 22.0& 4.1& HST\\
471 &172446.01-341407.3 & 1.636 & 0.802& 10.674& 32.32& 21.9& 5.7& ... \\
519 &172447.56-341048.6 & 1.466 & 0.706& 10.167& 32.38& 22.1& 7.5&  V;HST\\
577 &172450.15-341243.4 & 1.240 & 0.509& 10.788& 32.25& 22.0& 4.1& ... \\
649 &172455.55-341631.3 & 1.286 & 0.510& 12.064& 32.18& 22.1& 4.3&  V\\
672 &172457.87-341203.9 & 1.331 & 0.576& 10.256& 32.52& 22.3& 10.0:&  flaring\\
692 &172459.74-340958.7 & \nodata&\nodata &\nodata & 32.17& 22.5& 2.2&  faint in SIRIUS\\
746 &172508.76-341115.2 & \nodata&\nodata &\nodata & 32.18& 22.0:& 1.9:& close to WR star\\
753 &172510.93-340843.2 & \nodata&\nodata &\nodata & 32.08& 22.0 & 3.8 &  V; IR nebulosity\\
\cline{1-9} \\
\multicolumn{9}{c}{(b) Stars with $K <10$} \\
\cline{1-9} \\
9   &172405.64-340710.0 & 0.399 & 0.172 & 8.458 & 31.20 & 21.5 & 0.7 & IRAS 17207-3404\\
19  &172413.61-341656.5 & 2.006 & 0.991 & 5.334 & 31.0 & 21.9 & 2.8 & MSX6C G353.0401+00.9326 \\
120 &172433.46-341344.9 & 0.602 & 0.333 & 9.649 & 31.29 & 21.7 & 10.0:& ...\\
194 &172438.90-341459.0 & 0.587 & 0.208 & 8.519 & 31.29 & 21.9 & 1.7& Pis24-8\\
225 &172440.50-341206.3 & 0.595 & 0.315 & 9.138 & \nodata & \nodata& \nodata& Pis24-4\\
312 &172442.88-340911.6 & 0.931 & 0.502 & 9.231 & \nodata& \nodata& \nodata& ...\\
527 &172447.87-341517.0 & 0.608 & 0.327 & 8.424 & \nodata& \nodata&\nodata & Pis24-7\\
539 &172448.44-341743.7 & 0.468 & 0.270 & 8.026 & 31.14 & 21.5 & 0.9& ...\\
643 &172455.09-341111.5 & 0.798 & 0.427 & 8.866 & 32.15 & 22.4 & 0.6& ...\\
654 &172455.85-341234.0 & 1.297 & 0.952 & 8.389 & \nodata& \nodata& \nodata& ...\\
694 &172459.86-341610.8 & 1.645 & 1.304 & 8.137 & 31.34 & 22.1 & 2.5& MSX6C G353.1413+00.8083\\
\enddata
\tablecomments{Column 1: X-ray source number. Column 2: IAU designation. Columns 3-5: 2MASS colors
and $K_s$ magnitudes. Columns 6-8: X-ray luminosities (0.5--8.0 keV)
corrected for absorption, derived column densities, and thermal plasma
energy obtained from our spectral analysis. Columns 9: Related notes:
ACIS source number; V=Variable lightcurves; HST=HST counterparts
found; Other catalog names. Stars in the top panel are selected from
X-ray criteria, and those in the bottom panel are selected from
infrared criteria. Missing $\log L_X$, $\log N_H$, and $kT$ values mean that
the source has $<20$ net counts.}
\end{deluxetable}
\clearpage
\begin{deluxetable}{rcrrrrrr}
\centering
\tabletypesize{\scriptsize} \tablewidth{0pt}
\tablecolumns{8}

\tablecaption{X-ray Selected Candidate Intermediate-mass Stars\label{tbl:HAeBe1}}

\tablehead{
\colhead{Seq \#} &
\colhead{CXOU J } &
\colhead{$J$} &
\colhead{$H$} &
\colhead{$K_s$} &
\colhead{$\tilde{A_V}$\tablenotemark{a}} &
\colhead{$\tilde{M}$\tablenotemark{a}} &
\colhead{$\Delta K$\tablenotemark{b}}\\
\colhead{} &
\colhead{} &
\colhead{(mag)} &
\colhead{(mag)} &
\colhead{(mag)} &
\colhead{(mag)} &
\colhead{($M_{\odot}$)} &
\colhead{(mag)}
}

\startdata
      2 & 172358.42-340802.6 & 14.99 & 13.16 & 11.93 & 18 & 13 & \nodata \\
      3 & 172359.11-341217.6 & 11.86 & 11.16 & 10.69 &  8 & 14 & \nodata \\
      6 & 172403.19-341402.2 & 12.60 & 11.67 & 11.31 & 10 & 13 & \nodata \\
     12 & 172407.65-341751.9 & 14.32 & 13.03 & 12.46 &  8 &  3 & \nodata \\
     18 & 172413.61-340921.8 & 13.51 & 12.65 & 12.34 &  4 &  3 & \nodata \\
     20 & 172413.68-341451.8 & 13.12 & 12.13 & 11.77 &  7 &  4 & \nodata \\
     29 & 172416.17-341526.6 & 13.88 & 12.28 & 11.45 & 16 & 16 & \nodata \\
     35 & 172418.81-341204.7 & 14.80 & 13.64 & 13.04 &  6 &  2 & \nodata \\
     38 & 172420.25-341310.7 & 13.90 & 12.54 & 11.88 & 14 & 12 & \nodata \\
     45 & 172421.75-341539.4 & 15.40 & 13.55 & 12.82 & 15 &  4 & \nodata \\
    156 & 172436.45-341236.7 & 13.30 & 12.32 & 11.57 &  5 &  3 & $\surd$ \\
\enddata

\tablecomments{Column 1: X-ray source number. Columns 2: IAU
designation. Columns 3-5: 2MASS $JHK_s$-band magnitudes. Columns 6-7:
Visual extinction and mass estimated from de-reddened location along
the standard interstellar reddening vector to the 1 Myr pre-MS
isochrone in the 2MASS color-magnitude diagram, assuming a distance
$d\simeq 2.56$ kpc. Column 8: Flag of significant $K$-band excess. The
full table of 100 {\it Chandra} sources is available in the electronic
edition of the Journal.}

\tablenotetext{a}{These quantities assume a cluster age of 1 Myr and
$d\simeq 2.56$ kpc.  For early B stars, $\tilde{A_V}$ and $\tilde{M}$
(estimated visual extinction and mass) are especially dependent on the
age and distance assumptions (Figure 7).}

\tablenotetext{b}{$\surd$ marks significant $K$-band excess derived
from high quality 2MASS photometry where near-IR color excess $E(H-K)$
is larger than $\sigma (H-K)$, the uncertainty in $H-K$ color index.}

\end{deluxetable}
\clearpage
\begin{deluxetable}{rccccc}
\centering
\tabletypesize{\scriptsize} \tablewidth{0pt}
\tablecolumns{6}
\tablecaption{X-ray Selected Heavily Obscured Sources \label{tbl:Embedded}}

\tablehead{ 
\colhead{Seq \#} &
\colhead{CXOU J} &
\colhead{Criteria\tablenotemark{a}} & \colhead{$\log N_H$
(cm$^{-2}$)\tablenotemark{b}}& \colhead{$\medE$
(keV)\tablenotemark{c}} & \colhead{Notes\tablenotemark{d}} }

\startdata
8  & 172405.42-341257.5 & I;II& 22.7 & 3.9& \nodata \\
15 & 172409.28-341224.3 & I& 22.5 & 3.1& V;2MASS\\
61 & 172426.47-341850.9 & I& 22.5 & 3.3& \nodata \\
142& 172434.95-340533.3 & I;II& 23.2 & 4.7& \nodata \\
569& 172449.82-341455.6 & I& 22.4& 3.4& \nodata \\
598& 172451.90-341331.3 & I;II& 22.6& 3.1& 2MASS \\
599& 172451.95-341513.0 & II& 22.6& 2.0& 2MASS \\
634& 172454.31-341209.4 & I& 22.3& 3.1& VV;2MASS\\
664& 172456.97-341651.8 & II& 22.6& 2.1& 2MASS \\
692& 172459.74-340958.7 & II& 22.5& 2.9& \nodata \\
693& 172459.85-341302.9 & II& 22.5& 2.8& 2MASS \\
702& 172501.09-341404.6 & I;II& 22.7& 3.0& \nodata \\
713& 172502.54-341549.7 & I;II& 22.7& 4.1& \nodata \\
728& 172504.93-340410.5 & I;II& 22.6& 3.9& e\\
753& 172510.93-340843.2 & I;II& 22.5& 3.5& VV;f\\
775& 172519.23-341356.6 & I;II& 22.8& 3.7& \nodata \\
\enddata

\tablenotetext{a}{Selection criteria: I.--$\medE \ge 3.0$ keV;
II.--$N_{H,X} \ge 22.5$ cm$^{-2}$. The values shown here are rounded.}
\tablenotetext{b}{Column density derived from spectral fitting.}
\tablenotetext{c}{Median photon energy of the X-ray source.}
\tablenotetext{d}{Related notes: V=variable lightcurves; VV=flaring;
2MASS=2MASS counterparts found.}  \tablenotetext{e}{MSX6C
G353.3175+00.9075} \tablenotetext{f}{MSX6C G353.2651+00.8443}

\end{deluxetable}
\clearpage
\pagestyle{plaintop}

\clearpage
\centerline{\includegraphics[width=0.8\textwidth]{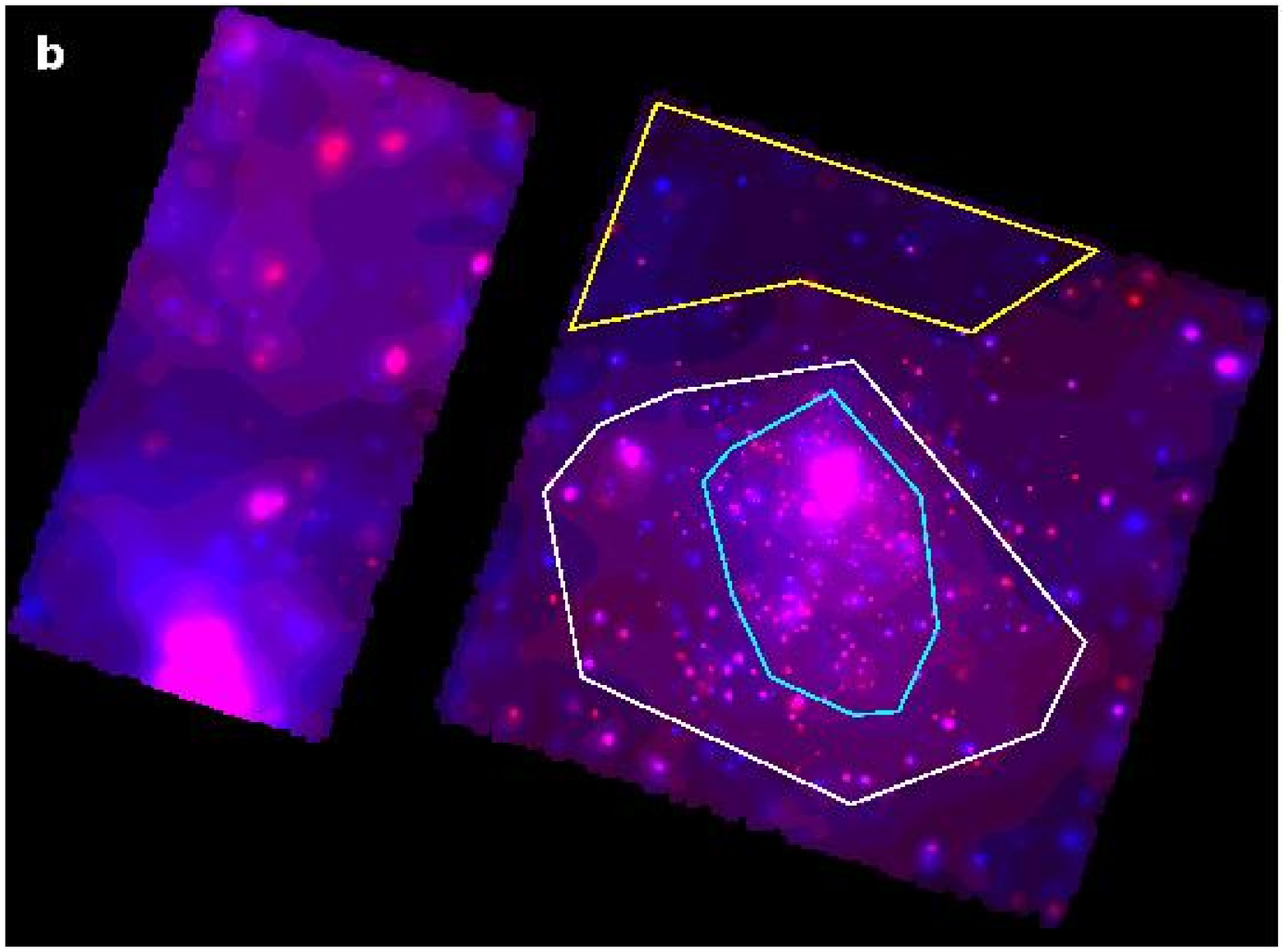}}
\centerline{Fig. 2. --- Continued.}
\clearpage
\centerline{\includegraphics[width=0.8\textwidth]{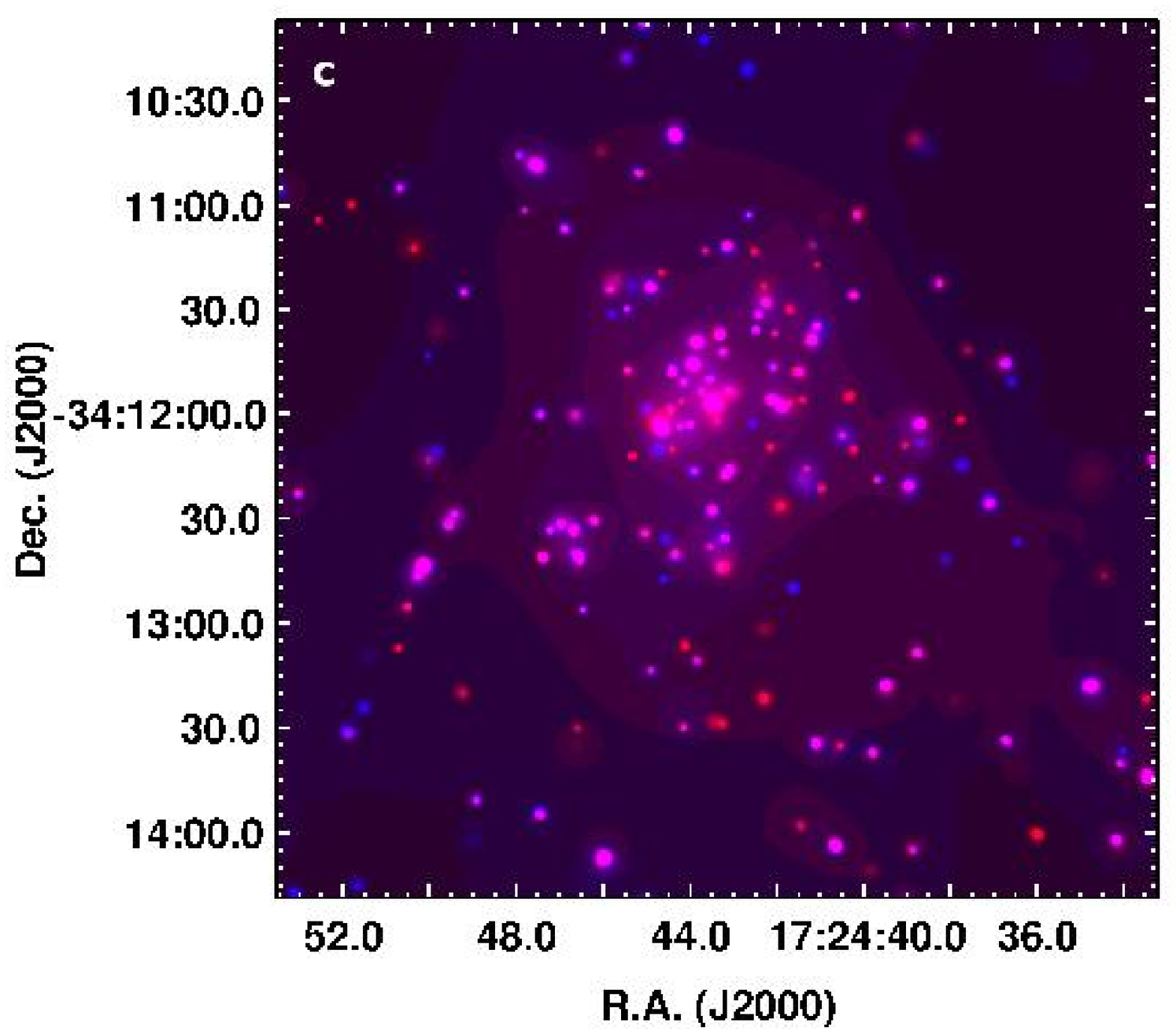}}
\centerline{Fig. 2. --- Continued.}
\clearpage
\clearpage
\clearpage
\clearpage
\end{document}